\documentclass[12pt]{article}
\usepackage{amsmath}
\usepackage{graphicx,psfrag,epsf}
\usepackage{enumerate}
\usepackage{natbib}
\usepackage{url} 

\newcommand{\blind}{1}

\usepackage{multirow}
\usepackage[T1]{fontenc}
\usepackage{float}
\usepackage{amsmath}
\usepackage{graphicx}
\usepackage{setspace}
\usepackage{lmodern}

\usepackage{caption}
\usepackage{subcaption}
\usepackage{amsmath,amssymb,amsthm,setspace,paralist}
\usepackage[colorlinks,citecolor=blue,urlcolor=blue]{hyperref}
\usepackage{epstopdf}
\usepackage{authblk}
\usepackage[normalem]{ulem}

\usepackage[top=2.7cm, bottom=3cm, left=2.7cm, right=2.7cm]{geometry}
\usepackage[space]{grffile}

\usepackage[space]{grffile}

\usepackage{algorithm2e}
\RestyleAlgo{boxruled}

\usepackage{xcite}

\usepackage{xr}
\makeatletter
\newcommand*{\addFileDependency}[1]{
\typeout{(#1)}
\@addtofilelist{#1}
\IfFileExists{#1}{}{\typeout{No file #1.}}
}
\makeatother

\newtheorem{lemma}{Lemma}
\newtheorem{theorem}{Theorem}
\newtheorem{definition}{Definition}
\newtheorem{assumption}{Assumption}

\begin{document}

	\def\spacingset#1{\renewcommand{\baselinestretch}%
		{#1}\small\normalsize} \spacingset{1}

\if1\blind
{
  \title{\bf Model-based Statistical Depth with Applications to Functional Data}
\author{Weilong Zhao$^1$, Zishen Xu$^1$, Yun Yang$^2$, Wei Wu$^1$ \\
\vspace{-6pt} 
 \textit{$^1$ Department of Statistics, Florida State University} \\
 \textit{$^2$ Department of Statistics, University of Illinois  at Urbana-Champaign}}
 \date{}
  \maketitle
} \fi

\if0\blind
{
  \bigskip
  \bigskip
  \bigskip
  \begin{center}
    {\LARGE\bf Model-based Statistical Depth with Applications to Functional Data}
\end{center}
  \medskip
} \fi


\vspace{-30pt} 

	\spacingset{1.5}
\begin{abstract}
Statistical depth, a commonly used analytic tool in non-parametric statistics, has been extensively studied for multivariate and functional observations over the past few decades. Although various forms of depth were introduced, they are mainly procedure-based whose definitions are independent of the generative model for observations. To address this problem, we introduce a generative model-based approach to define statistical depth for both multivariate and functional data. The proposed model-based depth framework permits simple computation via Monte Carlo sampling  and improves the depth estimation accuracy. When applied to functional data, the proposed depth can capture important features such as continuity, smoothness, or phase variability, depending on the defining criteria. Specifically, we view functional data as realizations from a second-order stochastic process, and define their depths through the eigensystem of the covariance operator. 
These new definitions are given through a proper metric related to the reproducing kernel Hilbert space of the covariance operator. We propose efficient algorithms to compute the proposed depths and establish estimation consistency. Through simulations and real data, we demonstrate that the proposed functional depths reveal important statistical information such as those captured by the median and quantiles, and detect outliers. 
\end{abstract} 
 
\vfill

\newpage
	\spacingset{1.5}

\section{Introduction}

The notion of statistical depth was first introduced \citep{tukey1975mathematics} as a tool to visualize bivariate data sets, and has later been extended to multivariate data over the last few decades. The depth is a measure of the centrality of a point with respect to certain data cloud, which helps to set up center-outward ordering rules of ranks. Alternatively, it can be treated as a multivariate extension of the notion of quantiles for univariate distributions. For instance, a deepest point in a given data cloud can be viewed as a ``multivariate median''. Based on different criteria on centrality, a large class of depths has been proposed, including the halfspace depth \citep{tukey1975mathematics}, convex hull peeling depth \citep{barnett1976ordering}, simplicial depth \citep{liu1990notion}, $L_1$-depth \citep{vardi2000multivariate}, and projection depth \citep{zuo2003projection}. The concept of statistical depth has been widely applied in outlier detection \citep{donoho1992breakdown}, multivariate density estimation \citep{fraiman1997multivariate}, non-parametric description of multivariate distributions \citep{liu1999multivariate}, and depth-based classification and clustering \citep{christmann2002classification}.

In many research areas such as medicine, biology, and engineering, it is natural to assume the observations being generated from infinite dimensional models, and analyze them using tools from functional data analysis (FDA).
Many efforts have attempted to extend the notion of depths from finite to infinite dimension in recent years. 
To name a few, \cite{fraiman2001trimmed} defined the integrated data depth for functional data based on integrals of univariate depths, and used it to construct an $\alpha$-trimmed functional mean to measure the centrality of given data. This method can reduce the effects of outlier bias in a sample set compared to the regular mean. 
In addition, \cite{cuesta2008random} extended the simple random Tukey depth (also called halfspace depth) to functional data analysis on a separable Hilbert space. A more comprehensive reviews on different notions of depths for functional data is provided in Section~\ref{sec:review_FD}.

%

Despite the broad variety and wide usage of statistical depths for both finite and infinite dimensional observations in exploratory data analysis, existing depth methods suffer from two apparent drawbacks: 1) They do not make use of any structural information from the generative model when defining or estimating the depths. Utilizing such information may enhance the power of the depth in tasks such as hypothesis testing, outlier detection, or classification. 2) For infinite-dimensional observations such as functional data, most depths are constructed via aggregating point-wise deviations, which fails to capture deviations of some more important global features such as phase variability and degree of smoothness.

In this paper, we propose a new model-based framework for defining and estimating statistical depths for both finite and infinite-dimensional data.
In particular, we propose to incorporate information from the data generative model in defining and estimating the statistical depth. When applied to functional data, our development leads to a new class of depths that captures global features such as shape and smoothness level. Our new model-based depth framework overcomes the aforementioned drawbacks and posses several attractive features: 
\begin{enumerate}
    \item It permits properly utilizing features in the generative model to define a data-dependent depth. Both computational efficiency and estimation accuracy of the depth can be benefited from the generative model via Monte Carlo sampling.
    \item The depth criterion is flexible, and can be chosen to better capture the underlying generative mechanism or meet specific application purposes.  Depending on the defining criterion, our framework can result in various forms and generalize commonly used depth functions.  
    \item The criterion may properly measure the metric distance between observations.  This naturally leads to the notions of centrality and variability in the given data.  In contrast, traditional depth methods are often procedure-based and do not provide such measurements.  
\end{enumerate}

\subsection{Related work on functional depth}\label{sec:review_FD}
Band depth \citep{lopez2009concept} is a very commonly used depth for functional data, which has been successfully used for tasks such as classification. 
Another important concept is half-region depth \citep{lopez2011half}, which is closely related to the band depth. It is considered to be applied to high-dimensional data with efficient computational cost. Based on the graph representation as in band depth, a number of extensions, modifications and generalizations have emerged.  For example,  
\cite{agostinelli2013ordering} proposed a so-called local band depth to deal with functional data which is considered to have multiple centers. It measures centrality conditional on a neighborhood of each point of the space and provide a tool that is sensitive to local features of the data, while retaining most features of regular depth functions. Set band depth \citep{whitaker2013contour} was proposed for the nonparametric analysis of random sets, and a generalization of the method of band depth. \cite{balzanella2015depth} introduced the spatial variability among the curves in the definition of band depth, and proposed a method -- spatially weighted band depth to incorporate the spatial information in the curves ordering.




More progress has been made in recent study of functional depth. 
\cite{chakraborty2014spatial} used the spatial distribution to define a so-called spatial depth, since the spatial distribution possesses an invariance property under a linear affine transformation. \cite{einmahl2015bridging} proposed to refine the empirical halfspace depth by setting extreme value to a so-called ``tail'' to avoid the problem of vanishing value outside the convex hull of the data, which benefits for inference on extremity.
\cite{narisetty2016extremal} introduced a notion called extremal depth, which satisfies the desirable properties of convexity and ``null at the boundary'', for which integrated data depth and band depth lack. These properties lead to a central region more resistant to outliers.
Based on an elastic-metric-based measure of centrality for functional data,  \cite{cleveland2018robust} adopted band depth and modified band depth to estimate the template for functional data with phase variability. They also showed their performance on outlier detection with new defined boxplots for time warping functions.

\vspace{0.5em}
The rest of this article is organized as follows: In Section 2, we first introduce our model-based framework for statistical depth. We then illustrate several forms of depth and their relations to commonly used depths. In Section 3, we elaborate on the application of our framework to functional data as generated from a second-order stochastic process. In Section 4, we investigate the statistical consistency of our depth estimation procedure. Simulations and real data analysis is provided in Section 5. Section 6 includes a summary and discusses some future directions. Other computational details and proofs are deferred to appendices in the supplementary material.

\section{Model-Based Statistical Depth}
\label{sec:new}

In this section, we introduce our model-based framework for statistical depth, where the model-based has two meanings: 1) the depth is defined based on a statistical model; and 2) the depth estimation procedure is two-stage, where we first estimate the model parameter, and then use a plug-in procedure for estimating the depth. The former view allows the depth definition itself to be data-dependent and automatically capture features underlying the data generating process, and the latter may lead to improved estimation accuracy of the depths due to the estimation efficiency of the model parameter.

\begin{figure}[htb]
\centering
\includegraphics[height=5cm]{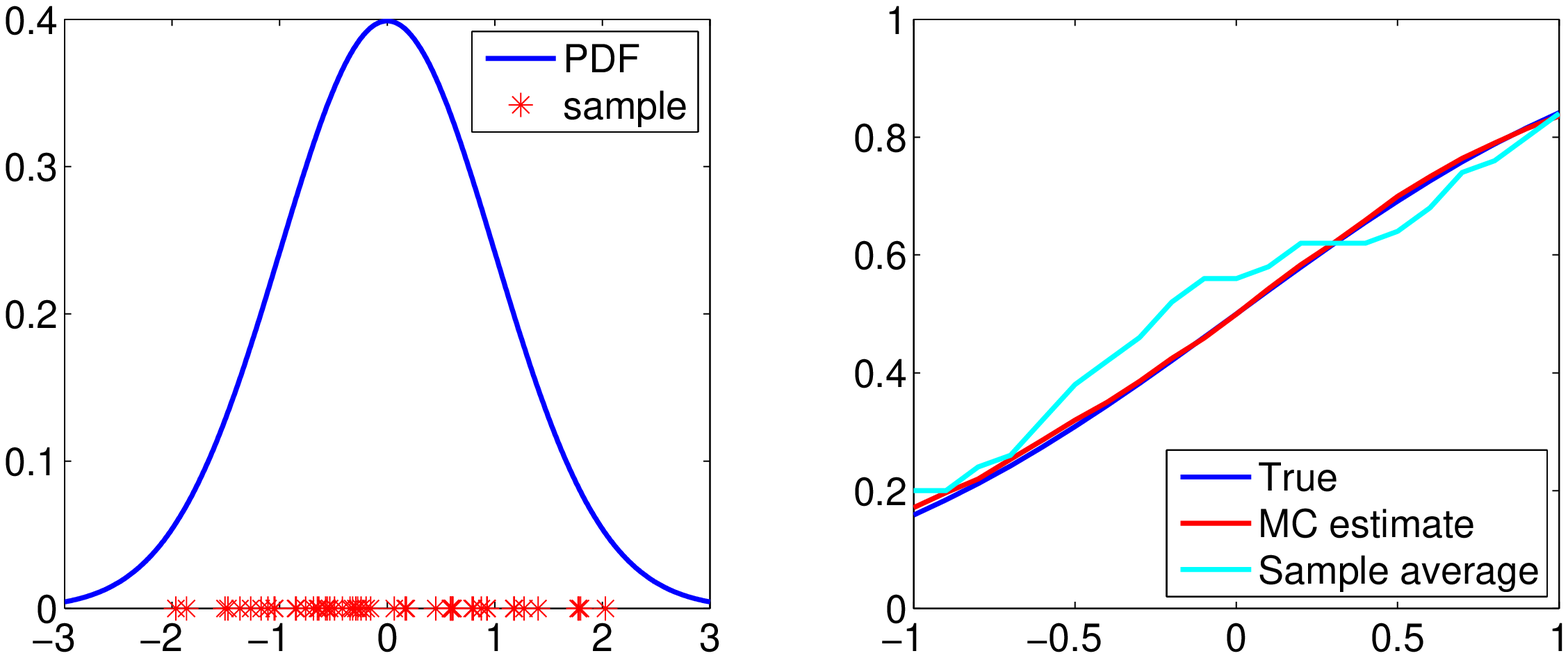}  \\
(a)  i.i.d. sample \hspace {1in} (b) CDF estiamte  \\
\vspace{12pt}
\begin{tabular}{|c|c|c|c|}
\hline
\bf{Method} & \bf{25\% quantile} &  \bf{median} & \bf{75\% quantile} \\
\hline
True & -0.67 & 0 & 0.67 \\
\hline
MC estimate & -0.71 &0.00 & 0.65 \\
\hline
Sample average & -0.76 & -0.26 & 0.78 \\
\hline 
\end{tabular}  \\
\vspace{3pt}
(c) quantile estimate

\caption{Toy example to compare the Monte Carlo method and sample average. {\bf (a)} 30 i.i.d. sample points from a standard normal distribution. {\bf (b)} Cumulative distribution functions of true model (blue), estimated using Monte Carlo method (red), and estimated using the sample average (cyan) in the range [-1, 1].  {\bf (c)} Quantile values at 25\%, 50\%, and 75\% of true and two estimate methods. 
}
\label{fig:toy}
\end{figure}

To illustrate the benefit in estimation accuracy via model-based procedures, we may compare the Monte Carlo (MC) method with the simple sample average approach.  A toy example is shown in Fig. \ref{fig:toy}, where we generate 30 i.i.d. sample points from a standard normal distribution (Fig. \ref{fig:toy}a).  In the MC method, we estimate mean and standard deviation from the sample, and then generate 2000 Monte Carlo sampling points to estimate the cumulative distribution within [-1,1].  In contrast, the sample average method estimates the cumulative distribution with the empirical distribution of the 30 points.  This comparison is shown in Fig. \ref{fig:toy}b.  Moreover, we compare the true and estimated quantiles at 25\%, 50\%, and 75\% in Fig. \ref{fig:toy}c.  It is apparent that the MC method provides more accurate and robust result.  


To begin with, we provide a general definition of depth by considering it as a functional of the underlying data generating model. Then, we provide a two-stage estimation procedure for the depth via Monte Carlo sampling. In the rest of the paper, we primarily focus on functional data for illustration, and the development naturally applies to finite-dimensional data.

\subsection{Depths within statistical models}
\label{sec:depth}
Let $\mathcal P = \{\mathbb P_\theta:\, \theta\in \Theta\}$ be a family of probability measures indexed by a parameter $\theta$ over a function (vector) space $\mathcal F \subset \mathcal L_2([0,1]):\,=\{f:\,[0,1]\to\mathbb R:\, \|f\|_2^2=\int_0^1f^2(x)\,dx<\infty\}$. For example, $\mathbb P_\theta$ can be the measure of a Gaussian Process GP$(m, C)$ with parameter $\theta=(m,C)$ collecting the mean function $m:\,[0,1]\to \mathbb R$ and the covariance function $C:\,[0,1]\times [0,1] \to \mathbb R$. Statistical depth should quantify how large a particular observed trajectory $f_{obs} \in \mathcal F$ deviates from certain notion of center $f_{c} \in \mathcal F$ under $\mathbb P_\theta$.  For example, in the case of the Gaussian Process (GP), a natural choice of the center would be its mean function.

\subsubsection{Definitions of Depths}
\label{subsec:def}
We will now provide the formal definition of a model-based functional depth, as well as the associated depth contour and central region. All these statistical terms 
 can be considered as infinite-dimensional generalization of the uni-variate survival function/$p$-value, quantiles, and highest-probability region.

Our proposed definition can be either norm-based or inner-product based.  We refer to the norm or inner-product as the criterion in the definition.  The norm-based depth is a generalization over various distance-based forms (see the discussion after the following definition).  In contrast, the inner-product depth is motivated with the classical halfspace depth by \cite{tukey1975mathematics}.  We at first define the norm-based depth in the following general form: 

\begin{definition}
{\bf (Norm-based statistical depth: general form):} The statistical depth $D_{ng}$ of $f_{obs} \in \mathcal F$ in the model $\mathbb P_\theta \in \mathcal P$ relative to the norm $\| \cdot \|$ and center $f_c \in \mathcal F$ is defined as
\begin{align*}
D_{ng}(f_{obs},\, \mathbb P_\theta, \| \cdot \|, f_c) \in [0,1],
\end{align*}
where $D_{ng}$ is strictly decreasing with respect to $\|f_{obs}- f_c\|$, and $D_{ng} \rightarrow 0$ when $\|f_{obs}- f_c\| \rightarrow \infty$.
\label{def:depthng}
\end{definition}

Norm-based depths are commonly used in statistics literature. For example, the $h$-depth \citep{nieto2011properties} and spatial depth \citep{sguera2014spatial}  are based on the $\mathbb L^2$ norm,  the $\mathbb L^p$-depth is based on the $\mathbb L^p$ norm \citep{zuo2000general, long2015study}, and the Mahalanobis depth is based on the Mahalanobis norm \citep{liu1999multivariate}. The depth in Definition \ref{def:depthng} generalizes these concepts and provides a broader framework for norm-based methods. In this paper, we study one specific form of this general definition. This specific form more resembles conventional depths and satisfies more desirable mathematical properties. The norm in the definition can be considered as a criterion function to compute the distance between any observation $f_{obs}$ and the center $f_c$ and we denote the criterion function as $\zeta(f_{obs},f_c)$ in the rest.

\begin{definition}
{\bf (Norm-based statistical depth: specific form):} The statistical depth $D_n$ of $f_{obs} \in \mathcal F$ in model $\mathbb P_\theta \in \mathcal P$ relative to norm $\| \cdot \|$ and center $f_c \in \mathcal F$ is defined as
\begin{align*}
D_n(f_{obs},\, \mathbb P_\theta, \| \cdot \|, f_c) :\,=\mathbb P_{\theta} \big [f\in \mathcal F: \|f-f_c\| \geq \|f_{obs}- f_c\| \big].
\end{align*}
\label{def:depthn}
\end{definition}
\vspace{-2em}
\noindent {\bf Remark 1:}  We point out that this specific form of depth is a representative of all norm-based depth in the general form as defined in Definition 1.  In fact, $D_n(f_{obs})$ measures the degree of extremeness of the observed function $f_{obs}\in\mathcal F$ under any normal-based depth $D_{ng}(f_{obs})$ in the following sense, 
\begin{eqnarray*}
\mathbb P_\theta \big [D_{ng}(f) \le D_{ng}(f_{obs})\big ] = \mathbb P_\theta \big [D_{n}(f) \le D_{n}(f_{obs})\big ] =  \mathbb P_\theta \big [\|f-f_c\| \ge \|f_{obs}-f_c\| \big ]   = D_n(f_{obs}).
\end{eqnarray*}

\noindent {\bf Remark 2:} 
One proper way to choose the center $f_c$ is to minimize $P(\|f-f_c\| \ge a)$ for any given $a>0$.  Note that 
$$P(\|f-f_c\| \ge a) \le \frac{E\|f-f_c\|^2}{a^2}.$$
When the norm $\|\cdot\|$ is inner-product induced (e.g. the classical $\mathbb L^2$ norm), it is easy to know that the optimal $f_c$ should be the expectation $Ef$. However, $f_c$ in general can take different form, dependent on different selection of the norm.  

Based on the definitions of the norm-based depth, we can naturally introduce the notions of depth contour and central region as follows. We adopt the specific form in Definition \ref{def:depthn} to simplify notation (same notion can be directly applied to the general form). 

\begin{definition} 
{\bf (Depth contour and central region for norm-based depth):} For any $\alpha\in [0,1]$, the $\alpha$-th depth contour in the model $\mathbb P_\theta \in \mathcal P$ relative to the norm $\| \cdot \|$ and center $f_c \in \mathcal F$ is defined as
\begin{align*}
C_n(\alpha,\, \mathbb P_{\theta}, \| \cdot \|, f_c) :\, = \big\{ f\in \mathcal F:\, D_n(f,\, \mathbb P_{\theta} , \| \cdot \|, f_c) = \alpha\big\}. 
\end{align*}
Also, the $\alpha$-th central region in the model $\mathbb P_\theta \in \mathcal P$ relative to the  norm $\| \cdot \|$  and center $f_c \in \mathcal F$ is defined as
\begin{align*}
R_n(\alpha,\, \mathbb P_{\theta}, \| \cdot \|, f_c) :\, = \big\{ f\in \mathcal F:\, D_n(f,\, \mathbb P_{\theta}, \| \cdot \|, f_c) \geq \alpha\big\}.
\end{align*}
\label{def:cr}
\end{definition}
\vspace{-2em}

Based on the multivariate halfspace depth, we now define the inner-product-based depth. In contrast to the general and specific forms in the norm-based case, the inner-product-based norm is defined only in a specific form as follows. 
\begin{definition}
{\bf (Inner-product-based statistical depth):} The statistical depth $D_{ip}$ of $f_{obs} \in \mathcal F$ in the model $\mathbb P_\theta \in \mathcal P$ relative to the inner-product $\left < \cdot, \cdot \right > $ and a subset $\mathcal G$ of $\mathcal F$ is defined as
\begin{align*}
D_{ip}(f_{obs},\, \mathbb P_\theta, \left < \cdot, \cdot \right > , \mathcal G) :\,=
\inf_{g \in \mathcal G, ||g|| = 1} \mathbb P_{\theta} \big [f \in \mathcal F: \left < f , g \right > \ge \left < f_{obs} , g \right > \big] 
\end{align*}
\label{def:depthip}
\end{definition}
\vspace{-2em}
\noindent {\bf Remark 3:}  There are two apparent differences between Definitions~\ref{def:depthn} and \ref{def:depthip}: 1) Definition~\ref{def:depthn} depends on the center $f_c$, whereas Definition~\ref{def:depthip} is independent of it.  However, we will point out in Section~\ref{subsec:mp} that when the distribution function has a center under certain form of symmetry, this center should be the deepest point under Definition~\ref{def:depthip}. 2) Definition~\ref{def:depthip} involves an infimum in order to match the half-region depth~\citep{lopez2011half} for finite-dimensional Euclidean data. Different from the usual half-region depth where $\mathcal G$ as the range of the infimum is taken as the entire function space $\mathcal F$, the following lemma shows that for infinite-dimensional functional data, $\mathcal G$ is necessary to be a proper (finite-dimensional) subset to avoid depth value degeneracy. A proof is provided in Appendix~\ref{sec:lemma:ip}.

\begin{lemma}
Let $\mathbb P_C$ be the probability measure of a zero-mean Gaussian process GP$(0, C)$, where the eigensystem $\{(\lambda_j,\phi_j)\}_{j=1}^\infty$ of the covariance operator $C$ has infinite number of positive eigenvalues $\{\lambda_j\}_{j=1}^\infty$. If $\langle\cdot,\cdot\rangle$ is an inner-product over $\mathcal F$ such that the $P\times P$ Gram matrix $[\langle \phi_j,\phi_k\rangle]_{j,k=1}^P$ of the first $P$ eigenfunctions $\{\phi_j\}_{j=1}^P$ is positive definite for any $P\in\mathbb N$, then
$$ D_{ip}(f, \mathbb P_C, \left < \cdot, \cdot \right >, \mathcal F) = 0 $$
almost surely for $f \in$ GP$(0, C)$.
\label{lem:ip}
\end{lemma}

\noindent {\bf Remark 4:}  This lemma indicates that special attention is needed for defining an inner-product-based depth for infinite-dimensional space $\mathcal F$. \cite{dutta2011some} also observed this anomalous behavior of halfspace depth in infinite-dimensional spaces. As a consequence, the halfspace depth (where $\mathcal G=\mathcal F$) is only meaningful for finite-dimensional space. In contrast, the norm-based depth can be effective for both finite- or infinite-dimensional space. To have a proper inner-product-based depth, either $\mathcal F$ itself is finite-dimensional, or we use a finite-dimensional subset $\mathcal G$ as shown in Definition \ref{def:depthip}.

Under this model-based framework, we can naturally estimate the proposed statistical depth $D(f_{obs},\, \mathbb P_{\theta}, \cdot, f_c)$ via the following two-stage procedure: 1. Find an estimate $\widehat{\theta}$ of the parameter $\theta$; 2. Compute the estimated depth $D(f_{obs},\, P_{\widehat{\theta}}, \cdot, f_c)$ by either using a closed-from expression of the depth or by a Monte Carlo method for an approximation.  For example, when $\mathbb P_{\theta}$ is a GP measure and the depth as a functional of parameter $\theta$ may not admit a closed-form expression, we may resort to Monte Carlo method for estimating the depth.
More details of the estimation will be provided in Appendix~\ref{sec:more_alg} in the supplementary material of the paper.


\subsubsection{Mathematical Properties}
\label{subsec:mp}

\cite{zuo2000general} introduced a list of favorable mathematical properties to be satisfied by good multivariate statistical depths. Based on this, \cite{nieto2016topologically} further explored the extensions of these properties for functional data. \cite{gijbels2017general} discussed these properties on commonly used methods such as the random Tukey depth, band depth, and spatial depth.  In this part, we discuss these properties on our norm-based and inner-product based depths. 

Before discussing basic properties of these two types depths, we need to clarify the concept of ``halfspace'' with the following definition:
\begin{definition}
A closed halfspace $H_{h,g}$ for $g,h \in \mathcal F$ is defined in the form 
$$ H_{h,g} = \big \{ f \in \mathcal F: \left < f - h, g \right > \ge 0 \big \}.$$ 
\end{definition}
To make the inner-product-based depth satisfy favorable properties, we need the following assumption on the ``center'' function $f_c$.  
\vspace{1em}

\noindent {\bf Assumption 1:} The distribution $\mathbb P_\theta$ of a random function $f \in \mathcal F$ is halfspace symmetric, or H-symmetric, about a unique function $f_c$. That is, $\mathbb P(f \in H) \ge 1/2$ for every closed halfspace $H$ containing $f_c$.  Moreover, we assume that $\mathbb P(f \in H) < 1/2$ for every closed halfspace $H$ that does not contain $f_c$.  
\vspace{1em}

Now we list four basic properties of the norm-based depth (Definition~\ref{def:depthn}) and the inner-product-based depth (Definition~\ref{def:depthip}), respectively, as following: 

\noindent {\bf Norm-based depth:}\vspace{-0.5em}
\begin{itemize}
\item[] P-1. ({\em Linear invariance}) Let $\mathbb P_{\theta, F}$ denote the distribution $P_\theta$ of a random variable $F \in \mathcal F$.  Then for any $a \in \mathbb R \setminus \{0\}$ and $h \in \mathcal F$,   
$$D(af_{obs}+h,\, \mathbb P_{\theta, aF+h}, \| \cdot \|, af_c+h) = D(f_{obs},\, \mathbb P_{\theta, F}, \| \cdot \|, f_c). $$
\item[]
P-2. ({\em Maximality at center}) $D(f_c,\, \mathbb P_{\theta}, \| \cdot \|, f_c) = \sup_{f_{obs} \in \mathcal{F}}D(f_{obs},\, \mathbb P_{\theta}, \| \cdot \|, f_c)$. \item[]
P-3. ({\em Monotonicity with respect to the deepest point}) Let the deepest function be $f_c \in \mathcal{F}$.  Then for any $f_{obs} \in \mathcal{F}$ and $\alpha \in (0,1)$, $D(f_{obs},\, \mathbb P_{\theta}, \| \cdot \|, f_c) \le D(f_c+\alpha(f_{obs}-f_c),\, \mathbb P_{\theta}, \| \cdot \|, f_c)$.
\item[]
P-4. ({\em Vanishing at infinity}) $D(f_{obs},\, \mathbb P_{\theta}, \| \cdot \|, f_c)\to 0$ as $\|f_{obs}\| \to \infty$.
\end{itemize}
\noindent {\bf Inner-product-based depth:}\vspace{-0.5em}
\begin{itemize}
\item[] P-1'. ({\em Linear invariance}) Let $\mathbb P_{\theta, F}$ denote the distribution $P_\theta$ of a random variable $F \in \mathcal F$.  Then for any $a \in \mathbb R \setminus \{0\}$ and $h \in \mathcal F$,   
$$D_{ip}(af_{obs}+h,\, \mathbb P_{\theta, aF+h}, \left < \cdot, \cdot \right > , \mathcal G) = D_{ip}(f_{obs},\, \mathbb P_{\theta, F}, \left < \cdot, \cdot \right > , \mathcal G). $$

\item[]
P-2'. ({\em Maximality at center}) $D_{ip}(f_c,\, \mathbb P_\theta, \left < \cdot, \cdot \right > , \mathcal G) = \sup_{f_{obs} \in \mathcal{F}}D_{ip}(f_{obs},\, \mathbb P_\theta, \left < \cdot, \cdot \right > , \mathcal G)$. \item[]
P-3'. ({\em Monotonicity with respect to the deepest point}) Let the deepest function be $f_c \in \mathcal{F}$.  Then for any $f_{obs} \in \mathcal{F}$ and $\alpha \in (0,1)$, $ D_{ip}(f_{obs},\, \mathbb P_\theta, \left < \cdot, \cdot \right > , \mathcal G) \le D_{ip}(f_c+\alpha(f_{obs}-f_c),\, \mathbb P_\theta, \left < \cdot, \cdot \right > , \mathcal G)$.
\item[]
P-4'. ({\em Vanishing at infinity}) $D_{ip}(f_{obs},\, \mathbb P_\theta, \left < \cdot, \cdot \right > , \mathcal G)\to 0$ as $\left < f_{obs}, f_{obs} \right > \to \infty$.
\end{itemize}

We examine these mathematical properties of the three defined depths in Sec. \ref{subsec:def}, as  summarized in Lemma \ref{lem:prop} below. The detailed proof is given in Appendix~\ref{sec:lem_prop}. 

\begin{lemma}
The three depths in Definitions \ref{def:depthng},  \ref{def:depthn},  and \ref{def:depthip} satisfy the mathematical properties given below: 
\begin{enumerate}
\item Norm-based depth in general form (Definition~\ref{def:depthng}): P-2, P-3, P-4. 
\item Norm-based depth in specific form (Definition~\ref{def:depthn}): P-1, P-2, P-3, P-4.
\item Inner-product-based depth (Definition~\ref{def:depthip}, given Assumption 1): P-1', P-2', P-3', P-4'.
\end{enumerate} 
\label{lem:prop}
\end{lemma}

\subsection{Illustration of the Depth Definitions}
\label{sec:cfdd}
We have defined two forms of model-based functional depth -- norm-based (as in Definitions \ref{def:depthng} and \ref{def:depthn}) and inner-product-based (as in Definition \ref{def:depthip}). In this section, we provide some examples, both finite-dimensional and infinite-dimensional, to illustrate these definitions.  We will at first adopt various norms in $D_n$, and then demonstrate the inner-product-based definition. Using these depths one can rank functional data based on their amplitude, continuity, smoothness, or phase variability. Moreover, we will show that some of the functional depths can also be directly applied to multivariate data. 

\subsubsection{Norm-based Depth}
\label{sec:nbc}

There are various norms on functional variables. One commonly used is the classical $\mathbb L^p$-norm, with $p \ge 1$.  That is, for $f$ in a proper space, its $\mathbb L^p$-norm is
\[ \| f \|_p = (\int_0^1 |f(t)|^p dt)^{1/p}. \]
In particular, $\mathbb L^2$-norm, the Euclidean distance from 0, is most often used in functional data analysis. Due to the nature of $\mathbb L^2$ norm, it is a great tool for data visualization and ranking based on their own amplitude information. 
Considering functions in a Sobolev Space \citep{hsing2015theoretical}, we can also use $\mathbb L^p$ norm on the derivatives functions to quantify continuity or smoothness feature. We may consider the norm-based depth in the following two forms:
 \begin{enumerate}
\item  $D_n(f_{obs},\, \mathbb P_\theta, \| \cdot \|, f_c) :\,=\mathbb P_{\theta} \big [f\in \mathcal F: \|f-f_c\|_p \geq \|f_{obs}- f_c\|_p \big]$
\item  $D_n(f_{obs},\, \mathbb P_\theta, \| \cdot \|, f_c) :\,=\mathbb P_{\theta} \big [f\in \mathcal F: \|D^r f - D^r f_c\|_p \geq \|D^r f_{obs} - D^r f_c\|_p \big]$, where $D^r$ indicates $r$-th order differentiation.  
\end{enumerate}
When we adopt the $\mathbb L^p$ norm,  the resulting depth can approxiamte band depth \citep{lopez2009concept} for functional observations from a distribution with mean 0.

In addition to having variability in amplitude (characterized by $\mathbb L^p$ norms), functional observations often exhibit variability in phase.  Such variability has been extensively studied over the past two decades and various methods were proposed to separate phase and amplitude, and quantify each variability in the given data \citep{ramsay1998curve,liu2004functional,tang2008pairwise,cleveland2018robust}.  In particular, phase is represented with time warping functions --
Let $\Gamma$ be the set of orientation-preserving diffeomorphisms of the unit interval $[0,1]$: $\Gamma = \{\gamma : [0,1] \to [0,1] | \gamma(0)=0,\gamma(1)=1,  \dot \gamma  > 0\}$ (the dot indicates derivative operation), and $\gamma$ is called a \emph{warping function}. Given two functions $u, v$, we denote $\gamma_{uv}$ as the optimal warping from $u$ to $v$. There are various forms to define the ``optimal'' warping, and here we adopt the well-known Fisher-Rao framework \citep{srivastava2011registration} and 
$$\gamma_{uv} = \mbox{arginf}_{\gamma \in \Gamma} \| (q(u) \circ \gamma) \sqrt{\dot \gamma} - q(v)\|_2$$ 
where $\circ$ denotes function composition and $q(\cdot)$ is a transformation on the given function defined as $q(x) = sign(\dot x)\sqrt{|\dot x|}$.  The degree of warpingness from the identity $\gamma_{id}(t) = t$ can be properly measured by two distances, namely, the $\mathbb L^2$ distance and the Fisher-Rao distance.  We may consider the norm criterion based on each of these distances:
\begin{enumerate}
\item  $D_n(f_{obs},\, \mathbb P_\theta, \| \cdot \|, f_c) :\,= \mathbb P_{\theta} \big [f\in \mathcal F: \|\gamma_{ff_c} - \gamma_{id}\|_2 \geq \|\gamma_{f_{obs}f_c} - \gamma_{id}\|_2 \big]$
\item  $D_n(f_{obs},\, \mathbb P_\theta, \| \cdot \|, f_c) :\,=\mathbb P_{\theta} \big [f\in \mathcal F: d_{FR}(\gamma_{ff_c}, \gamma_{id}) \geq d_{FR}(\gamma_{f_{obs}f_c}, \gamma_{id}) \big]$, where $d_{FR}(\gamma_{uv}, \gamma_{id}) = cos^{-1}(\int_0^1 \sqrt{\dot{\gamma}_{uv}(t)} \sqrt{\dot{\gamma}_{id}(t)} dt) = cos^{-1}(\int_0^1 \sqrt{\dot{\gamma}_{uv}(t)} dt),$
\end{enumerate}
Due to the nature of the Fisher-Rao distance, depth based on this criteria in our framework is sensitive to smoothness in the warping function.

\subsubsection{Inner-product-based Depth}
\label{sec:ipbc}


For multivariate data, Tukey's halfspace depth \citep{tukey1975mathematics} is one of the most popular depth functions available in literature. \cite{dutta2011some} investigated an extension on any Banach space, and proposed a specialization on a Hilbert space $\mathcal{H}$. If $\textbf{X}$ is a random element in $\mathcal{H}$ having the distribution $F$, then the {\it halfspace depth} of an observation $x \in \mathcal{H}$ is defined as
$$ HD (x,F) = \inf_{h \in \mathcal{H}} P\{ \left < h, \textbf{X}-x  \right >  \ge 0\},$$
where $\left < \cdot, \cdot \right > $ stands for the inner product defined on $\mathcal{H}$.  
Note that the inner-product based depth in Definition \ref{def:depthip} can be rewritten as
$$
D_{ip}(f_{obs},\, \mathbb P_\theta, \left < \cdot, \cdot \right > , \mathcal G) :\,=
\inf_{g \in \mathcal G, ||g|| = 1} \mathbb P_{\theta} \big [f \in \mathcal F: \left < f , g \right > \ge \left < f_{obs} , g \right > \big] = \inf_{g \in \mathcal G, ||g|| = 1} \mathbb P_{\theta} \big [ \left < f - f_{obs}, g \right > \ge 0\big].$$
Therefore, the halfspace depth can be treated as one special case in the proposed framework. However, Lemma~\ref{lem:ip} illustrates that the halfspace depth may collapse to zero for infinite-dimensional functional data unless the underlying data generating model is intrinsically finite-dimensional. As a consequence, the choice of the range $\mathcal G$ of the infimum in the preceding display becomes important.

In general, there is no simple solution to the above minimization process \citep{tukey1975mathematics,rousseeuw1996algorithm,dutta2011some}. However, if the functions are samples from a finite-dimensional stochastic process, an optimal $g$ can be found in closed forms. For illustration purpose, let us assume that the data generating process is a finite-dimensional Gaussian process.  Then the minimization takes the the following closed-form (see detailed derivation in Appendix~\ref{sec:opt_col})
$$ D_{ip}(f_{obs}) = 1 - \Phi (\|f_{obs}\|_{\mathbb H_K}),$$
where $\Phi$ is the c.d.f. of a standard normal random variable and the norm $\| \cdot \|_{\mathbb H_K}$ is the induced RKHS norm (formal definitions are provided in Section \ref{sec:MBD}).
Note that as $\Phi$ is a c.d.f. function, the depth value of $f_{obs}$ is in the range $[0, 1/2]$, which is consistent to the notion of halfspace depth in function space \citep{dutta2011some}.  
It is well known that, for the halfspace depth, if we have a symmetric distribution in a Hilbert space, then the maximum depth is $0.5$, and the point of symmetry will achieve at the halfspace median. In this case, it is easy to see that $D_{ip}(f_{obs}) = 1/2 \Leftrightarrow f_{obs} = 0$. Therefore, the median (i.e. function with largest depth value) is our center function $f_c= 0$. 


\noindent {\bf Remark 5:} The above result is based on the assumption that the stochastic process is a Gaussian process.  However, the Gaussianity is only used in the step that the c.d.f. $\Phi$ is independent of $g$ after the standardization (i.e., $X \rightarrow \frac{X-\mu_g}{\sigma_g}$), and the results can be generalized to any second-order stochastic process. 
\vspace{0.5em}

\noindent {\bf Simplifications in Multivariate Data:}
The above inner-product-based depth can also be applied to multivariate data where the Gaussian process reduces to a multivariate Gaussian distribution, denoted as $\mathcal N(\mu,\Sigma)$. In particular, the corresponding inner-product criterion function reduces to a variant of the well-known Tukey's \textit{halfspace} depth, or location depth \citep{tukey1975mathematics}, 
$$D_{ip}(x) = \inf_{u\in\mathbb R^d,\,\|u\|=1}P\{X: \langle u,X-x \rangle \geq 0\},$$ 
where the new halfspace depth incorporates the second moment information $\Sigma$ through
the (zero-mean) inner product $ \left < x,y \right >=x^T\Sigma^{-1} y$ and the norm $\|x\|^2=x^T\Sigma^{-1} x$ induced from the covariance matrix of the multivariate data generating distribution, and $\mathcal G$ becomes the unit ball of $\mathbb R^d$ relative to this inner-product.
In the special case when $X$ is a random realization from a zero-mean multivariate normal distribution, or $X \sim \mathcal{N}(0,\Sigma)$, then the half-space depth $D_{ip}$ admits a closed form. More concretely, using the singular value decomposition on the covariance matrix $\Sigma = U \Lambda U^T$, where $\Lambda$ is a diagonal matrix with elements of eigenvalues $\{\lambda_p\}_{p=1}^d$, we can express $X$ through a finite-dimensional version of the Karhunen Lo\`{e}ve expansion $X = \sum_{p=1}^d \xi_p U_p$,
where $U_p$ is the $p$-th column of $U$ (i.e. the eigenvector corresponding to $\lambda_p$), and $\xi_p\sim \mathcal N(0,\lambda_p)$, $p=1,2,\ldots,d$ are independent random variables. Correspondingly, the depth of any $x\in\mathbb R^d$ is given as 
$$D_{ip}(x) = 1 - \Phi(\sqrt{\sum_{p=1}^d \frac{\xi_p^2}{\lambda_p}}).$$
Note that the maximum depth value computed by this way is the same as maximum via Tukey's half space depth, which is $1/2$.

\section{A New Model-Based Depth for Functional Data}
\label{sec:MBD}
In this section, we apply our proposed depth framework to functional data and propose a new data-dependent functional depth. As we will illustrate, our new model-based functional depths can capture and adapt to global features such as smoothness and shapes in the underlying data generating processes. Our proposed methods incorporate information from the reproducing kernel Hilbert space (RKHS) associated with the covariance operator of the underlying stochastic process. 

\subsection{Depths induced by reproducing kernels}
\label{sec:inorm}

We will provide a construction of norm-based depth for zero-mean second-order stochastic processes $\mathcal F$, where the norm itself is model-dependent and learned from the data. Recall that a stochastic process $\{f(t):\,t\in[0,1]\}$ is a second-order process if $\mathbb E[f^2(t)]<\infty$ for all $t\in[0,1]$, so that its covariance function $\mathbb E[f(s) f(t)]$ is well-defined. If the process has a nonzero mean function $m$, then we can always subtract the mean by choosing the center $f_c$ as $m$.  

\subsubsection{Background on covariance kernels}
Since $\{f(t):\,t\in[0,1]\}$ is a second-order process, its covariance kernel $K \in [0,1] \times [0,1] \mapsto \mathbb R$, $K(s,t):=\mathbb E [f(s)f(t)]$ is a well-defined function for all $(s,t)\in[0,1]^2$. In addition, $K(\cdot,\cdot)$ is a symmetric, positive semi-definite real-value function, that is, 
\begin{align*}
   i)& \ \ \ \  K(s,t) = K(t,s),  \\
   ii)& \ \ \  \int_0^1\int_0^1 K(s,t)\, h(s)\, h(t)\, ds\, dt \geq 0\ \text{for any $h \in \mathcal F$}.
\end{align*}

According to Mercer's Theorem \citep{j1909xvi}, there exists a sequence of orthonormal  eigenfunctions $\{ \phi_1(t), \phi_2(t), \cdots \}$ over $[0,1]$ and a sequence of corresponding non-negative eigenvalues $\lambda_1\ge\lambda_2\ge \cdots\ge 0$  \citep{riesz1990functional} satisfying
\begin{align}
    \int_0^1 K(s,t)\phi_p(s)ds = \lambda_p\phi_p(t),\ \text{for any $p\geq 1$, \ and \ }
    K(s,t) = \sum_{p = 1}^{\infty}\lambda_p\phi_p(s)\phi_p(t),\label{eq:mercer}
\end{align}
which implies $\int_0^1 \int_0^1 K^2(s,t) \,ds \, dt = \sum_{p = 1}^{\infty} \lambda_p^2$.
The convergence in Equation~\eqref{eq:mercer} is absolute and uniform on $[0,1] \times [0,1]$ \citep{cucker2007learning}. 

By the Karhunen Lo\`eve theorem \citep{ash1990information}, a random observation $f$ has the following representation 
\begin{align}
    f(t) = \sum_{p=1}^{\infty} f_p\phi_p(t)
\label{eq:KL}
\end{align}
where $f_1,f_2,\cdots$ are uncorrelated random variables with mean
$\mathbb{E} f_p = 0$, and variance $\mathbb{E} f_p^2 = \lambda_p$.  Each coefficient $f_p$ is unique and can be obtained by
$ f_p = \int_0^1 f(s)\phi_p(s)ds.$   In particular, if the stochastic process is a GP,  $f_1,f_2,\cdots$ will be independent Gaussian random variables.

\subsubsection{Reproducing kernel Hilbert space and its induced norm}
\label{sec:inRKHS}

Any symmetric, positive semi-definite function $K$ on $[0,1] \times [0,1]$ corresponds to a unique RKHS with $K$ as its reproducing kernel \citep{wahba1990spline}. We denote this RKHS by $\mathbb H_K$ with inner-product
\begin{align*}
    \langle K(s,\cdot),K(t,\cdot)\rangle_{\mathbb H_K} = \langle K(t,\cdot),K(s,\cdot)\rangle_{\mathbb H_K} = K(s,t).
\end{align*}
Moreover, the reproducing property ensures that for any $f \in \mathbb H_K$, $\langle f,K(t,\cdot)\rangle_{\mathbb H_K} = f(t)$.
The inner product induces the RKHS norm $\|f\|_{\mathbb H_K}=\sqrt{ \langle f,\,f\rangle_{\mathbb H_K}}$. This leads to an equivalent definition of the RKHS as $\mathbb H_K=\{f:\,[0,1]\to\mathbb R,\, \|f\|_{\mathbb H_K}<\infty\}$.
Therefore, under the representations in Equations ~\eqref{eq:mercer} and \eqref{eq:KL}, we have $f \in \mathbb H_K$ if and only if $\| f \|_{\mathbb H_K}^2 = \sum_{p: \lambda_p > 0} \frac{f_p^2}{\lambda_p}< \infty$.

For a random trajectory $f \in \mathcal F$ from a second-order stochastic process with covariance kernel $K$, it is important to examine if the norm $ \| f \|_{\mathbb H_K}^2$ is finite.  If $K$ has only finite number of positive eigenvalues, this conclusion certainly holds.  However, if $K$ has infinite number of positive eigenvalues, in general $ \sum_{p: \lambda_p > 0} \frac{f_p^2}{\lambda_p} = \infty (a.s.)$ since by the SLLN,  
$$ \frac{1}{n}  \sum_{p=1}^n  \frac{f_p^2}{\lambda_p} \overset{a.s.}{\longrightarrow}  E( \frac{f_p^2}{\lambda_p} ) = 1$$ (the case for GP is discussed in  \citep{wahba1990spline}). 

Consequently, although the RKHS norm $\|\cdot\|_{\mathbb H_K}$ contains important global features of the underlying data generating process, we cannot use the RKHS norm to define the depth since the RKHS norm of the observations are infinite almost surely. For example, for one-dimensional integrated Bownian motions \citep{vaart2011information}, it is known that smoothness level of the sample trajectories is $0.5$ smaller than that of its associated RKHS (for Brownian motion, see \citep{karatzas2012brownian}), where the corresponding RKHS norm coincides with the Sobolev norm.
 In this paper, we aim to combine these global features reflected in the RKHS norm into the construction of model-based functional depth.
In particular, to solve this divergent issue of the RKHS norm,  we propose a modified RKHS norm in the construction of our norm induced depth for functional data by weakening the impact of high-frequency signals, which are usually hard to estimate, on the modified norm.

\subsection{Depth induced by modified RKHS norm}
\label{subsec:modified}
In this section, we propose a modified inner product structure for functions in $\mathcal F$. This new inner product will induce a modified RKHS norm that is almost surely finite for the sample trajectories from the second-order stochastic process. 

\subsubsection{Modified Inner Product and Norm}

Suppose $f,g$ are two random realizations over $\mathcal F$ from a second-order stochastic process with covariance kernel $K$. Recall the eigen-decomposition $K(s,t) = \sum_{p = 1}^{\infty}\lambda_p\phi_p(s)\phi_p(t)$ for any $s,t \in [0,1]$. Without loss of generality, we assume all eigenvalues $\{\lambda_p\}$ are positive to avoid zero appearing in the denominator.

Recall the Karhunen Lo\`eve expansion, $f(t) = \sum_{p=1}^{\infty} f_p\phi_p(t)$  and $g(t) = \sum_{p=1}^{\infty} g_p\phi_p(t)$, 
with $ f_p = \int_0^1 f(s)\phi_p(s)ds$ and $g_p = \int_0^1 g(s)\phi_p(s)ds$. In addition, the RKHS induced inner-product and norm are given in the following forms, respectively. 
\begin{align*}
    \langle f,g \rangle_{\mathbb H_K} = \sum_{p=1}^{\infty} \frac{f_p g_p}{\lambda_p}  \ \ \mbox{and} \ \  \|f\| =  \langle f,f \rangle_{\mathbb H_K}^{1/2}
\end{align*}
As we discussed earlier in Sec. \ref{sec:inorm}, the RKHS norm diverges almost surely.  This divergence motivates us to a modified inner-product as follows: 
\begin{align*}
    \langle f,g \rangle_{mod} := \sum_{p=1}^{\infty} \frac{f_p g_p}{\lambda_p} a_p^2 
\end{align*}
where $\{a_p\}_{p=1}^\infty$ is any real sequence satisfying $\sum_{p=1}^{\infty} a_p^{2} < \infty$.  In practice, we may adopt commonly used convergent sequence $\{a_p = \frac{1}{p^s}\}_{p=1}^\infty$ or $\{a_p = \frac{1}{\sqrt{p} (\log p)^s}\}_{p=1}^\infty$with $s > 1/2$. 
Our idea is to assign a decaying weight to each positive eigenvalue, so that the overall sum converges after the adjustment. This modified inner product yields a squared \emph{modified RKHS norm} as
\begin{align*}
    \| f \|_{mod}^2 =  \langle f,f \rangle_{mod} = \sum_{p=1}^\infty \frac{f_p^2}{\lambda_p} a_p^2.
\end{align*}
Straightforward calculations yield $\mathbb E (\| f \|_{mod}^2) =  \sum_{p=1}^\infty \frac{E(f_p^2)}{\lambda_p} a_p^2 = \sum_{p=1}^{\infty} a_p^2 < \infty$. As a consequence,  $\| f \|_{mod} < \infty$ almost surely, and the above modified inner product and norm are well-defined for the observed trajectories.  We can use this modified RKHS norm to define  a model-based norm-induced depth as described in Section~\ref{sec:examples}.

Recall of Definition 2 of depth in Sec. \ref{sec:depth}: $ D(f_{obs},\, \mathbb P_{\theta}, \|\cdot\|, f_c) = \mathbb P_{\theta} \big [ \|f - f_c\| \geq \| f_{obs} - f_c\| \big]$. In this case, the central function $f_c = 0$ is the mean function in our model; the norm function is the modified RKHS norm $\| \cdot \| = \| \cdot \|_{mod}$; $\mathbb P_{\theta}$ is a probability measure defined by the probability density on $\|f\|_{mod}$ or $\|f\|_{mod}^2$. Apparently, with different settings of the decaying sequence $\{a_p\}_{p=1}^{\infty}$, we will have different probability density for $\|f\|_{mod}$ or $\|f\|_{mod}^2$.  It is often intractable to derive a closed-from expression on the density.  Fortunately, our model-based depth framework provides a natural way of estimating the depth through Monte Carlo sampling, where the coefficients ($\{f_p\}$ in the Karhunen-Lo\`eve expansion) can be simulated with re-sampling techniques such as the Bootstrap.

\subsubsection{Depth estimation procedure and algorithm}

Suppose we have $n$ zero-mean independent sample functions $f_1, \cdots, f_n \in \mathcal F$ on $t \in [0,1]$, and our goal is to compute the model-based depth of any observed sample $f_{obs} \in \mathcal{F}$. We propose an estimation algorithm as follows.

\textbf{Algorithm I.}  (Input: functional data $\{f_1, \cdots, f_n\}$, any observation $f_{obs}$, a small threshold $\delta_n >0$, and a sequence $a_1, \cdots, a_n$.)  
    \begin{enumerate}
        \item Compute the sample mean function $\hat{f}(t) = \frac{1}{n}\sum_{i=1}^n f_i(t)$, and empirical covariance kernel $\hat{K}(s,t) = \frac{1}{n}\sum_{i=1}^n [f_i(s)-\hat{f}(s)][f_i(t)-\hat{f}(t)]$;
        \item Eigen-decompose $\hat{K} = \sum_{p=1}^n \hat{\lambda}_{p,n} \hat{\phi}_{p,n}(s) \hat{\phi}_{p,n}(t)$;
        \item Set $\hat{\lambda}_{p,n} = 0$ if $\hat{\lambda}_{p,n} < \delta_n$;
        \item Set $M_n = \operatorname*{arg\ max}_m \{ \hat \lambda_{m,n} \geq \delta_n\}$, and $C_n = M_n \land n$ (minimum of $M_n$ and $n$);
        \item Compute $\hat{f}_{i,p} = \int_0^1 f_i(t) \hat{\phi}_{p,n}(t) dt$ for all $i = 1,\cdots,n$ and $p = 1,\cdots,C_n$, and compute $\hat{f}_{p} = \int_0^1 f_{obs}(t) \hat{\phi}_{p,n}(t) dt$;
        \item For each $p \in {1,\cdots,C_n}$, re-sample (with replacement) a large number $N$ of coefficients $\{\hat{g}_{j,p}\}_{j=1}^N$ based on $\{ \hat{f}_{1,p}, \cdots, \hat{f}_{n,p} \}$;
        \item Construct $g_j(t) = \sum_{p=1}^{C_n} \hat{g}_{j,p} \hat{\phi}_{p,n}(t)$;
        \item Compute $||f_{obs}||_{\hat{mod}}^2 = \sum_{p = 1}^{C_n} \frac{ \hat{f}_{p}^2}{\hat{\lambda}_{p,n}} a_p^2$, and $||g_j||_{\hat{mod}}^2 = \sum_{p = 1}^{C_n} \frac{ \hat{g}_{j,p}^{2}}{\hat{\lambda}_{p,n}} a_p^2$;
        \item Estimate the depth of $f_{obs}$ using $\{g_j\}$:
            \begin{align*}
                D_n(f_{obs};\{g_j\}_{j=1}^N) = \frac{1}{N}  \sum_{j=1}^N 1_{\|f_{obs} \|_{\hat{mod}}^2 \leq \|g_j \|_{\hat{mod}}^2}.
            \end{align*} 
    \end{enumerate}
    

The first 4 steps aim to estimate the eigen-system of the covariance kernel via given observations. In particular,  the Karhunen Lo\`eve expansion \citep{ash1990information} is used in Step 2 to decompose the covariance kernel, and offer a method to reconstruct samples. Using a functional principal component analysis \citep{ramsay2005functional}, we retain the eigen-functions which explain meaningful variance in our system by truncating the empirical eigenvalues in Step 3 \citep{nicol2013functional}.

Steps 5-8 are the second part of the algorithm. They estimate the depth value with  the modified RKHS norm, where we need re-sampling techniques and Monte Carlo approximations. This algorithm can be easily adapted to the multivariate data. In such case, the dimension of the data is already given and the principal component analysis and the multivariate metric can be directly applied. 
Step 9 estimates the probability in the depth definition by resampling from the empirical distribution of the sample basis expansion coefficients $\{\hat f_{i,p}\}_{i=1}^p$ for each coordinate $p=1,\ldots,C_n$.

In Appendix~\ref{sec:FDGP} in the supplementary material of the paper, we specialize these developments to finite-dimensional processes (or multivariate data).

\section{Asymptotic Consistency}
In this section, we will prove the consistency for the new model-based depths in Sec. \ref{sec:MBD}. We assume the functional data are fully observed over its domain $ [0,1]$.  This assumption is commonly used in asymptotic theory for various depths in functional data such as the integrated data depth \citep{fraiman2001trimmed}, the band depth \citep{lopez2009concept}, the half-region depth \citep{lopez2011half}, and the extremal depth \citep{narisetty2016extremal}. 

As our framework is model-based, there will be a main difference in the proofs between our framework and the traditional functional depth methods.  In particular, since previous depths are independent of the generative model, usually an LLN suffices to show the consistency.  In contrast, our method is considerably more involved since the depth itself is data dependent --- it depends on the estimated model or parameters from the observations. Despite this extra difficulty in the theory, our new model-based depth can better utilize the generative patterns in the data, and therefore yields better (discriminative) power and efficiency in a variety of applications. 


We start by introducing the notation used throughout in our proofs. Recall that $\mathcal{F} \subseteq \mathbb L^2([0,1])$ is the function space supporting the observations, which are generated from a second-order stochastic process with covariance function $K(s,t)=\mathbb{E}[(f(s)-\mathbb{E}(f(s)))(f(t)-\mathbb{E}(f(t)))]$. Suppose we have $n$ functional replicates $f_1,\cdots,\ f_n \ \in \mathcal{F}$. Note that the empirical approximation of $K(s,t)$ is $\hat{K}(s,t) = \frac{1}{n} \sum_{i=1}^n [(f_i(s)-\frac{1}{n}\sum_{p=1}^n f_p(s))(f_i(t)-\frac{1}{n}\sum_{p=1}^n f_p(t))]$. It is clear that $\hat{K}$ is also a symmetric positive semi-definite kernel. By Mercer's theorem, we have 
\begin{align*}
    K(s,t) = \sum_{p=1}^\infty \lambda_p \phi_p(s)\phi_p(t)\quad\mbox{and}\quad \hat{K}(s,t) = \sum_{p=1}^n \hat{\lambda}_{p,n} \hat{\phi}_{p,n}(s) \hat{\phi}_{p,n}(t),
\end{align*}
where eigenvalues $\lambda_1 \geq \lambda_2 \geq \cdots$ and $\hat{\lambda}_{1,n} \geq \hat{\lambda}_{2,n} \geq \cdots \geq \hat{\lambda}_{n,n}$ are non-negative, and their corresponding eigenfunctions $\{ \phi_p\}_{p=1}^\infty$ and $\{ \hat{\phi}_{p,n}\}_{p=1}^n$ are continuous on [0,1]. In this section, we primarily study the consistency of the proposed depth in the infinite-dimensional case where $\lambda_p > 0$ for any $p \in \mathbb N$. Due to space constraint, a counterpart result in the
finite-dimensional case where $\lambda_p = 0$ for all $p>P$, where $P\in\mathbb N$, is deferred to Appendix~\ref{sec:consistency} in the supplementary material.


\subsection{Depth estimation consistency}

We study the general case when all eigenvalues $\{ \lambda_p\}_{p=1}^\infty$ are positive. For any $f_{obs} \in \mathcal{F}$, we have shown in Sec. \ref{subsec:modified} that the squared modified norm
\begin{equation}
     \|f_{obs} \|_{mod}^2 = \sum_{p=1}^\infty \frac{\langle f_{obs},\phi_p\rangle^2}{\lambda_p} a_p^2 
\label{eq:pnorm}
\end{equation}
where ${\langle \cdot,\cdot\rangle}$ is the classical $\mathbb L^2$ inner-product and $\{ a_p\}_{p=1}^\infty$ is a real-valued sequence satisfying $\sum_{p=1}^\infty a_p^2 < \infty$.  Based on the modified norm, the depth of $f_{obs}$ is given as follows: 
\begin{align}
   d_{mod}(f_{obs}) &= D_n(f_{obs},\, \mathbb P, \|\cdot\|_{mod}, 0) = \mathbb P \big [f: \|f\|_{mod} \geq \|f_{obs}\|_{mod} \big] 
   \nonumber
   \\ &= 1 - \mathbb P \big [f: \|f\|_{mod}^2 \leq \|f_{obs}\|_{mod}^2 \big] = 1 - F(\|f_{obs}\|_{mod}^2),
   \label{eq:pd}
\end{align}
where $F(x)$ denotes the cumulative distribution function of $\|f\|_{mod}^2$ for the random function $f$.

As given in Algorithm I, the sample version of the squared modified norm is given as 
\begin{equation}
\|f_{obs}\|_{\hat{mod}}^2 = \sum_{p=1}^{C_n} \frac{\langle f_{obs}, \hat{\phi}_{p,n} \rangle^2}{\hat{\lambda}_{p,n}}a_p^2
\label{eq:snorm}
\end{equation}
where $C_n = M_n \land n$ (minimum of $M_n$ and $n$) and $M_n = \operatorname*{arg\ max}_m \{ \lambda_m \geq \delta_n\}$ for a given small threshold $\delta_n >0$.  In our framework, we adopt the sample version of the depth of $f_{obs}$ given as 
\begin{equation}
  d_{mod,n}(f_{obs}) =  \mathbb P \big [f: \|f\|_{mod} \geq \|f_{obs}\|_{\hat {mod}} \big] = 1 - F(\|f_{obs}\|_{\hat {mod}}^2).
  \label{eq:sd}
\end{equation}

In this section, we focus on proving $d_{mod,n}(f_{obs})$ converges to $d_{mod}(f_{obs})$ when $n$ is large.       
Before we proceed to find consistency of the modified norm, we make the following two assumptions:
\begin{assumption}
$\exists \beta > 1,\ C,\ C_1,\ C_2 >0,\ s.t.$
\begin{align*}
    C_1 p^{-\beta} \geq \lambda_p \geq C_2 p^{-\beta} \quad and \quad \lambda_p - \lambda_{p+1} \geq C p^{-(\beta+1)} \quad \forall p \in \mathbb{N}.
\end{align*}
\vspace{-24pt}
\label{assump:con}
\end{assumption}
 
\begin{assumption}
There exists a real sequence $\{b_p\}_{p=1}^\infty$ and some constant $\alpha>0$, such that  $ \sum_p b_p^2 < \infty,$
and $a_p\leq b_p \,p^{-\alpha}$ as $p$ goes to $\infty$.
\label{assump:ub}
\end{assumption}

For convenience, we abuse the notation $``C"$ to denote any constant coefficient.  Followed by Assumption \ref{assump:con}, it is apparent that the multiplicity of each $\lambda_p$ is strictly $1$.   We point out that Assumption \ref{assump:ub} can be easily satisfied in commonly used sequences of $\{a_p\}$.   For example, if we choose $a_p = p^{-(0.5 + \gamma)}$ for $\gamma>0$, then we can choose $b_p = p^{-(0.5 + \gamma / 2)}$ (with $\alpha = \gamma/2$).   Using the sequence $\{b_p\}$, we can define another type of modified form for any $f_{obs} \in \mathcal F$.  As compared to modified norm in Equation \eqref{eq:pnorm}, we only change the sequence $\{a_p\}$ to $\{b_p\}$.  That is, 
\begin{equation}
\|f_{obs} \|_{b}^2 = \sum_{p=1}^\infty \frac{\langle f_{obs},\phi_p\rangle^2}{\lambda_p} b_p^2. 
\label{eq:bnorm}
\end{equation}

Our main convergence result is given in Theorem \ref{thm:infinite} as follows, where the proof is given in Appendix~\ref{sec:thm_infinite}. 

\begin{theorem}
Under Assumptions \ref{assump:con} and  \ref{assump:ub}, if the covariance kernel $K$ has infinite number of positive eigenvalues $\{\lambda_p\}$, then the following holds with probability tending to one as $n\to\infty$,
\begin{equation}
\sup_{f_{obs} \in \mathcal{F}, ||f_{obs}|| \le 1,  ||f_{obs}||_b \le 1} | \| f_{obs} \|_{\hat{mod}}^2 - \| f_{obs} \|_{mod}^2 | \leq C\,n^{-\kappa}\to 0,
\label{eqn:conv1}
\end{equation}
where $(C,\kappa)$ are some positive constants, $\| \cdot \|$ is the classical $\mathbb L^2$ norm and $\| \cdot \|_{mod}$, $\| \cdot \|_{\hat{mod}}$, $\| \cdot \|_{b}$ are the norms defined in Equations \eqref{eq:pnorm},  \eqref{eq:snorm}, and \eqref{eq:bnorm}, respectively.   Moreover, for any $f_{obs} \in \mathcal{F}$ 
\begin{equation}
\lim_{n \rightarrow \infty} d_{mod,n}(f_{obs}) = d_{mod}(f_{obs}),
\label{eqn:conv2}
\end{equation}
where the two depths $d_{mod,n}(f_{obs})$ and $d_{mod}(f_{obs})$ are given in Equations \eqref{eq:pd} and \eqref{eq:sd}, respectively.  

\label{thm:infinite}
\end{theorem}


\subsection{Monte-Carlo method and sample average}

We have proven the convergence of the sample depth to the population depth.  In practical computation such as Algorithm I, the sample depth is obtained using samples.  In the proposed model-based framework, the depth is computed using Monte-Carlo samples.  Alternatively, we can simply use the given sample and the estimate will be the sample average.  In this subsection, we will prove that either of the methods can lead to accurate estimate asymptotically.  

The main result on the Monte-Carlo approximation and sample average can be summarized in the following two theorems, where the detailed proofs are given in Appendix~\ref{sec:thm_samc}.   The main result will be based on the following assumption. 

\begin{assumption}
Let $f$ denote an observed sample from the true model. Then $\|f\|_b$ is sub-Gaussian, that is, there exists some constant $\sigma>0$, such that $\mathbb E[\exp( t\,\|f_p\|_b)]\leq\exp(\sigma^2t^2/2)$ for all $t\in\mathbb R$.
\label{assump:tail}
\end{assumption}
This assumption essentially controls the tail probability bound for $\|f\|_b$ as a random variable. In particular, it controls the maximal norm $\max_{i=1,\ldots,n}\|f_i\|_b$ as of order $O_p(\sqrt{\log n})$ for an i.i.d.~sample $\{f_i\}_{i=1}^n$ of size $n$ from the true model, so that we can apply Theorem~\ref{thm:infinite} to control the approximation errors $\big|\|f_i\|^2_{\hat{mod}} - \|f_i\|^2_{{mod}}\big|$ uniformly over all $i=1,2,\ldots,n$.

\begin{theorem}
Let the sample depth  $d_{mod,n}(f_{obs}) =  \mathbb P \big [f: \|f\|_{mod} \geq \|f_{obs}\|_{\hat {mod}} \big]$ be estimated as:
$ \frac{1}{n} \sum_{p=1}^n 1_{\|f_p\|_{\hat {mod}} \geq \|f_{obs}\|_{\hat mod}},$
where $\{f_p\}$ are observed i.i.d. sample from the true model and the model paramenters are estimated from this sample.  Then under Assumptions 1, 2, and 3, we have 
\begin{align*}
\frac{1}{n} \sum_{p=1}^n 1_{\|f_p\|_{\hat {mod}} \geq \|f_{obs}\|_{\hat mod}} \to  1 - F(\|f_{obs}\|_{ {mod}}^2),
\end{align*}
in probability as $n\to\infty$.
\label{thm:sa}
\end{theorem}

For the Monte Carlo approximation, we consider the simpler case where the true model is a zero mean Gaussian process with covariance function given by $K$ for technical simplicity, and the Monte Carlo samples are also from a zero mean Gaussian process, but with the estimated covariance function $\hat K$.
\begin{theorem}
Assume the true model is a zero-mean Gaussian process and 
let the sample depth  $d_{mod,n}(f_{obs}) =  \mathbb P \big [f: \|f\|_{mod} \geq \|f_{obs}\|_{\hat {mod}} \big]$ be estimated as:
$ \frac{1}{N} \sum_{p=1}^N 1_{\|g_p\|_{\hat {mod}} \geq \|f_{obs}\|_{\hat mod}},$
where $\{g_p\}$ are an i.i.d. sample from the estimated distribution.  
Then under Assumptions 1 and 2 we have 
\begin{align*}
\frac{1}{N} \sum_{p=1}^N 1_{\|g_p\|_{\hat {mod}} \geq \|f_{obs}\|_{\hat {mod}}} \to  1 - F(\|f_{obs}\|_{ {mod}}^2)
\end{align*}
almost surely as $N, n\to\infty$.
\label{thm:mc}
\end{theorem}

\section{Simulation and real data analysis}
In this section, we illustrate applications of our proposed model-based depths to synthetic data and real data.

\subsection{Simulation Examples}
\label{sec:examples}
We will at first use several simulations to illustrate the uses of the norm-based and inner-product-based forms in Section~\ref{sec:nbc} and Section~\ref{sec:ipbc} for exploratory data analysis of both multivariate and functional data. In particular, Simulations 1-2 focus on several commonly used norms (inner-products) for model-based depth developed in Section~\ref{sec:new}, and Simulations 3-4 consider the new model-based functional depth introduced in Section~\ref{sec:MBD}. More simulation examples, including multivariate depth, are provided in Appendix~\ref{sec:more_sims}.

\noindent \textbf{Simulation 1.}
In this example, we illustrate the $\mathbb L^p$ induced norms as criteria functions which are discussed in the first part of Section~\ref{sec:nbc}. We demonstrate our framework by observations from zero-mean Gaussian Process with Mat\'ern class kernel on $[0,1]$. The generative formula for Mat\'ern kernel is 
\begin{align*}
    K_M(x_{i},\, x_{j}) = \frac{2^{1-\nu}}{\Gamma(\nu)}(\frac{\sqrt{2\nu}|x_{i}-x_{j}|}{l})^{\nu}K_{\nu}(\frac{\sqrt{2\nu}|x_{i}-x_{j}|}{l}), 
    \ \ \ \ \  x_i, x_j \in [0, 1]
\end{align*}
where $K_{\nu}$ is the modified Bessel function of order $\nu$, and the parameter $l$ is the characteristic length-scale of the process. For instance, if $\nu = \frac{1}{2}$ and $l=1$, then the Mat\'ern kernel $K_1(s,t) = \exp(-|s-t|)$, and if  $\nu = \frac{3}{2}$ and $l=1$, $K_2(s,t) = (1+\sqrt{3}|s-t|)\exp(-\sqrt{3}|s-t|)$, for $s, t \in [0, 1].$

\begin{figure}[ht!]
\captionsetup[subfigure]{}

\begin{subfigure}{.5\textwidth}
\centering
\includegraphics[height=5cm]{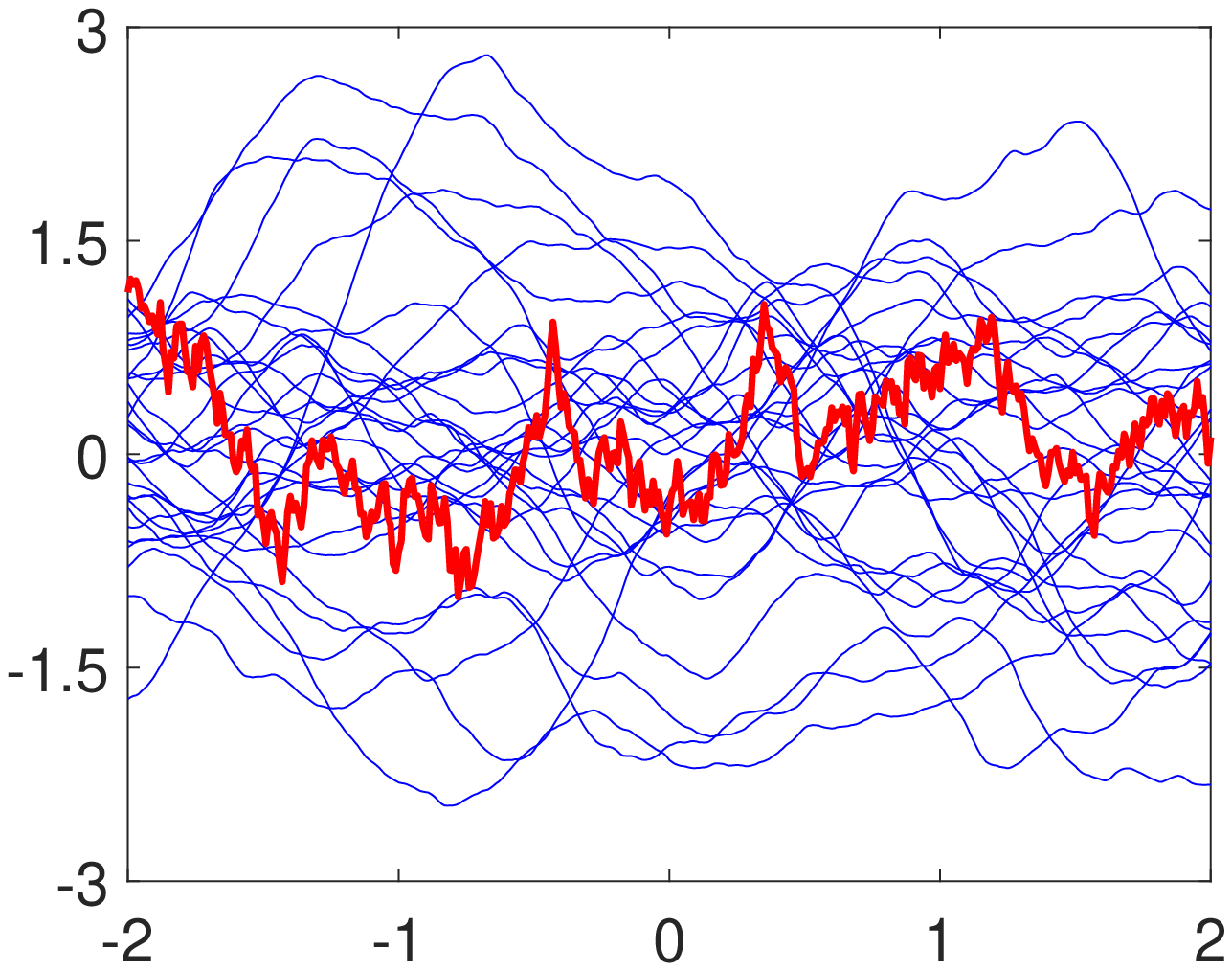}
\caption{Given Observations}
\end{subfigure}%
\begin{subfigure}{.5\textwidth}
\centering
\includegraphics[height=5cm]{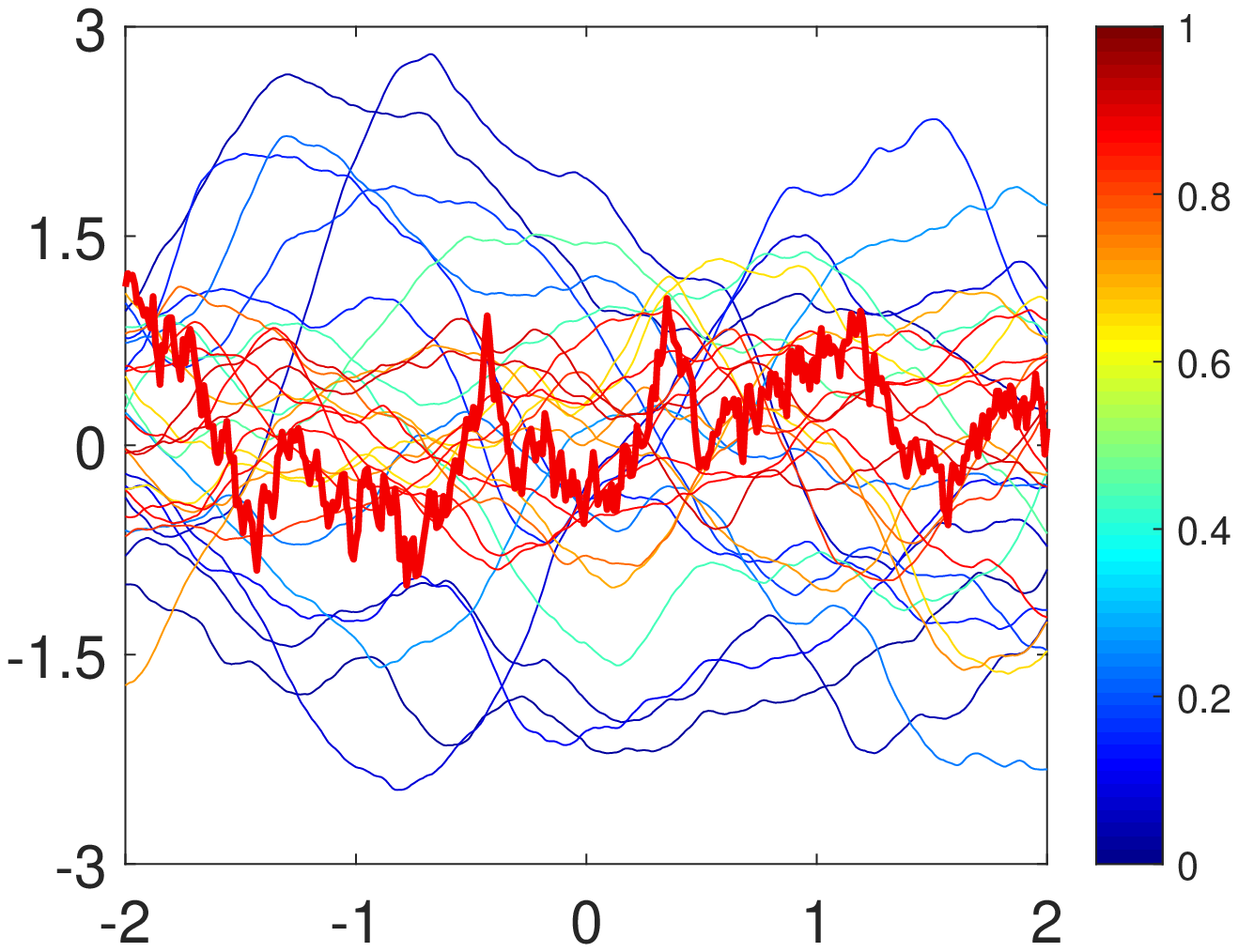}
\caption{$\zeta(f,0) = \|f \|_2$}
\end{subfigure}%

\begin{subfigure}{.5\textwidth}
\centering
\includegraphics[height=5cm]{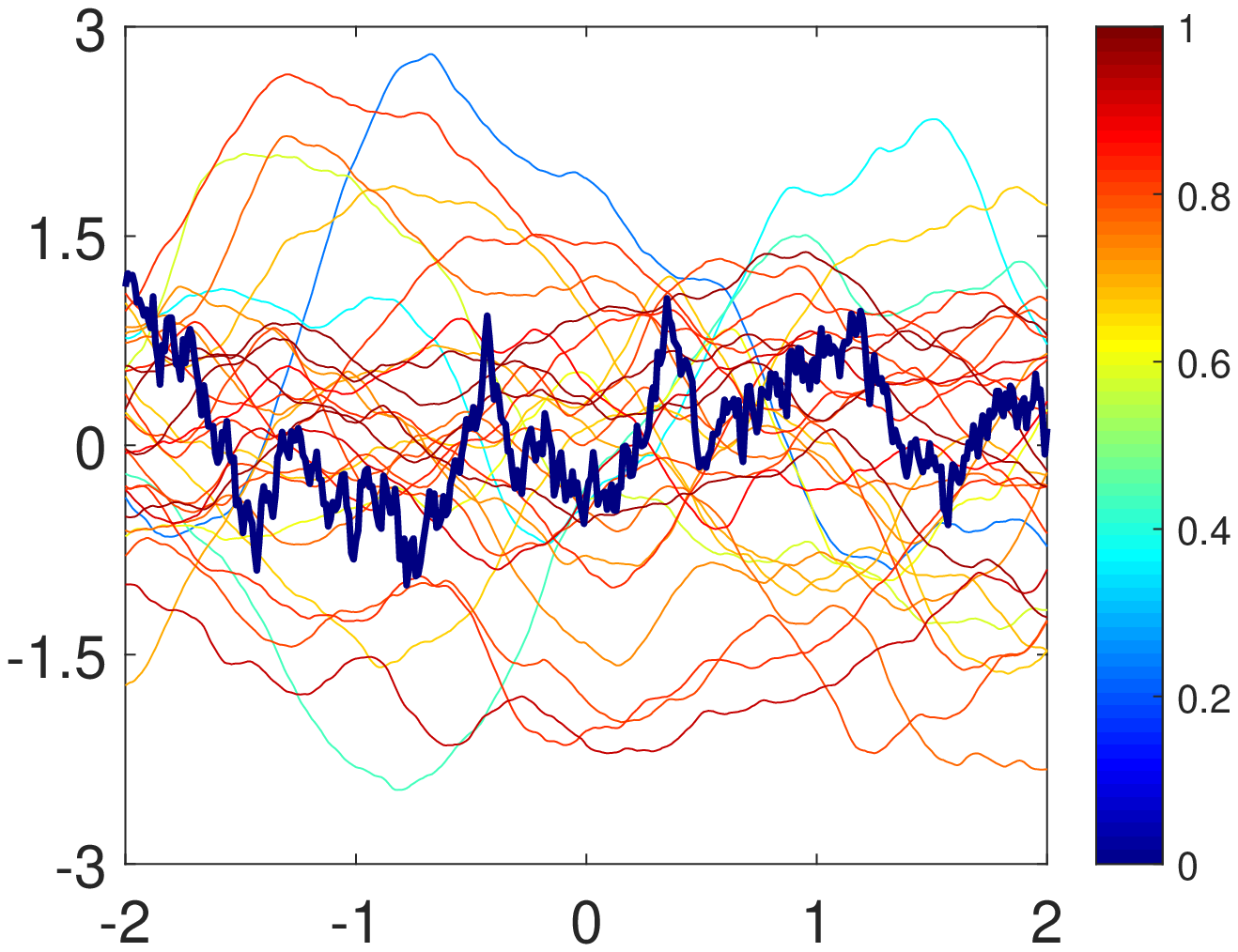}
\caption{$\zeta(f,0) = \|f' \|_2$}
\end{subfigure}%
\begin{subfigure}{.5\textwidth}
\centering
\includegraphics[height=5cm]{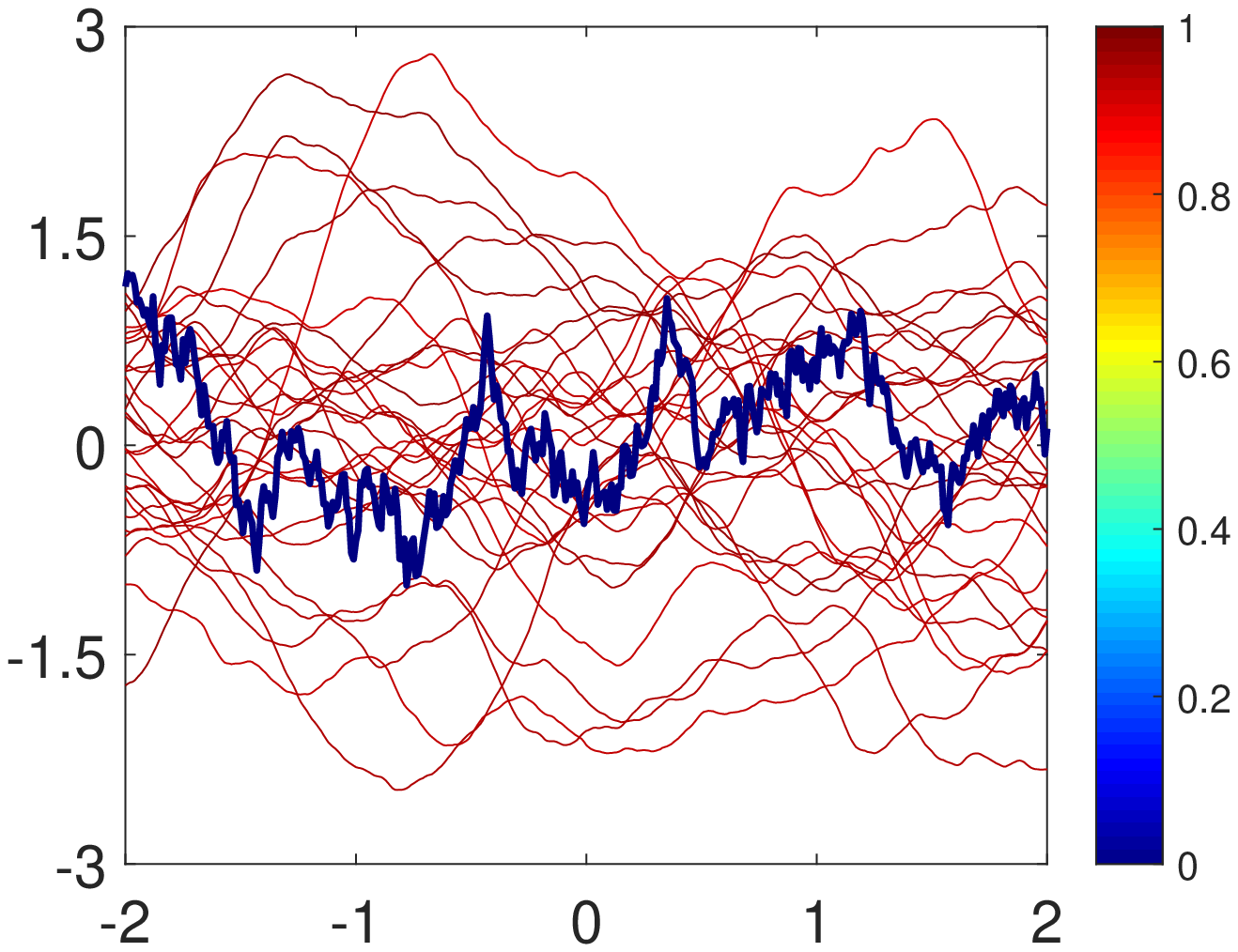}
\caption{$\zeta(f,0) = \|f{''} \|_2$}
\end{subfigure}%
\caption{Simulation 1: {\bf (a)} 30 observed functions, where the red one is generated from $GP(0,K_1)$ and 29 blues ones are generated from $GP(0,K_2)$. {\bf (b)} The 30 functions with color-labeled depth using $\mathbb L^2$ norm. Observations assigned with color closer to red are considered to be deeper than those assigned with color closer to blue.  {\bf (c),(d)} Same as (b) except for $\mathbb L^2$ norm on the first and second-order derivative functions, respectively.  
}
\label{fig:DemoMK1}
\end{figure}

For better visualization, we sample only one function from $GP(0,K_1)$ on $[0,1]$, and then mix it with another $n=29$ simulated samples from $GP(0,K_2)$ on $[0,1]$.  All these 30 functions are shown in Figure \ref{fig:DemoMK1}(a).  It is apparent that the one function from $K_1$ is near the zero-line, but somewhat ``noisy''.  In contrast, the 29 functions from $K_2$ have high variability in the amplitude, but are very smooth. 
We then color-labeled them differently in Panels (b)-(d) using their depth values with respect to different criterion functions, namely, $\mathbb L^2$ norm on each function, $\mathbb L^2$ norm on the first-order derivative function, and  $\mathbb L^2$ norm on the second-order derivative function.  The results clearly illustrate that criteria properly characterize the desirable features in the data. In Panel (b), we rank the function with respect to their $\mathbb L^2$ norm. The one function from $K_1$ is near the zero-line and has the highest depth value. In contrast, since this function is not smooth, it has the least depth values with derivative-based norms in Panels (c) and (d).  

\noindent \textbf{Simulation 2.} 
In this example, we illustrate the time warping distance in the depth computation. We study a set of simulated functions $\{ f_1, \cdots, f_{21}\}$ on $[-3,3]$. For $i = 1,\cdots,21$, we first simulate a set of functions by $h_i(t) = \phi_{i,1}e^{-(t-1.5)^2/2} + \phi_{i,2}e^{-(t+1.5)^2/2}$, where $\phi_{i,1}$ and $\phi_{i,2}$ are i.i.d. normal with mean one and variance $1/16$.  
Let the warping function $\gamma_i(t) = 6(\frac{e^{a_i(t+3)/6}-1}{e^{a_i}-1})-3$ if $a_i \neq 0$ otherwise $\gamma_i = \gamma_{id}$, where $a_i$ are equally spaced between $-1$ and $1$.  The observations are $f_i(t) = h_i(\gamma_i(t))$ on $[-3, 3], i = 1, \cdots, 21$. At the final step, we add some noise to the original $f_{11}$ by $\tilde{f}_{11}(t) = f_{11}(t) + \epsilon(t)$, where $\epsilon(t)$ is a Gaussian process with mean 0 and covariance function $C(s,t) = 0.01\delta_{s,t}$. To simplify the notation, we abuse $f_{11}$ to denote the noise contaminated $\tilde{f}_{11}$. 

\begin{figure}[ht]
\captionsetup[subfigure]{}
\begin{subfigure}{.33\textwidth}
\centering
\includegraphics[height=4.2cm]{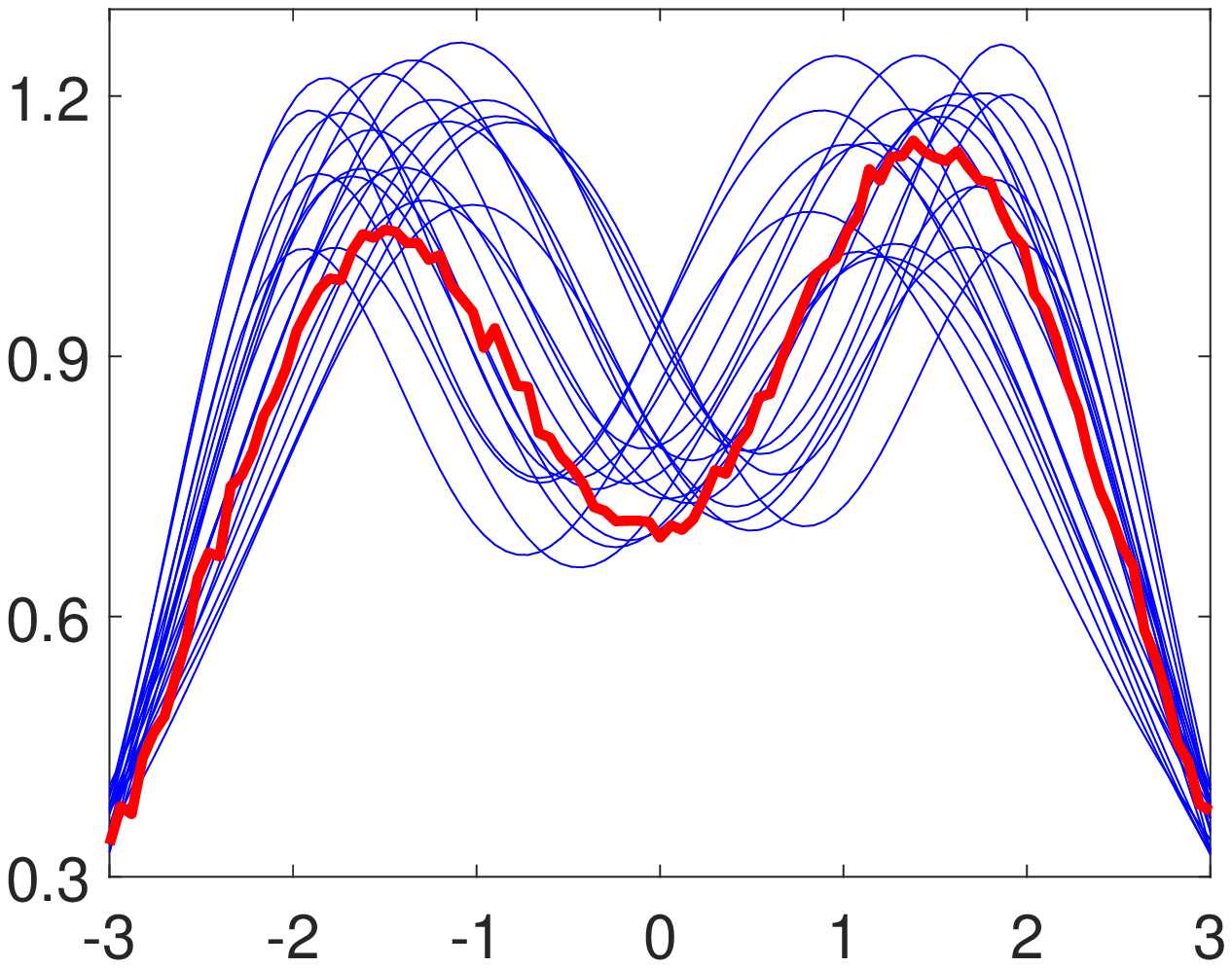}
\caption{Observed functions}
\end{subfigure}%
\begin{subfigure}{.33\textwidth}
\centering
\includegraphics[height=4.2cm]{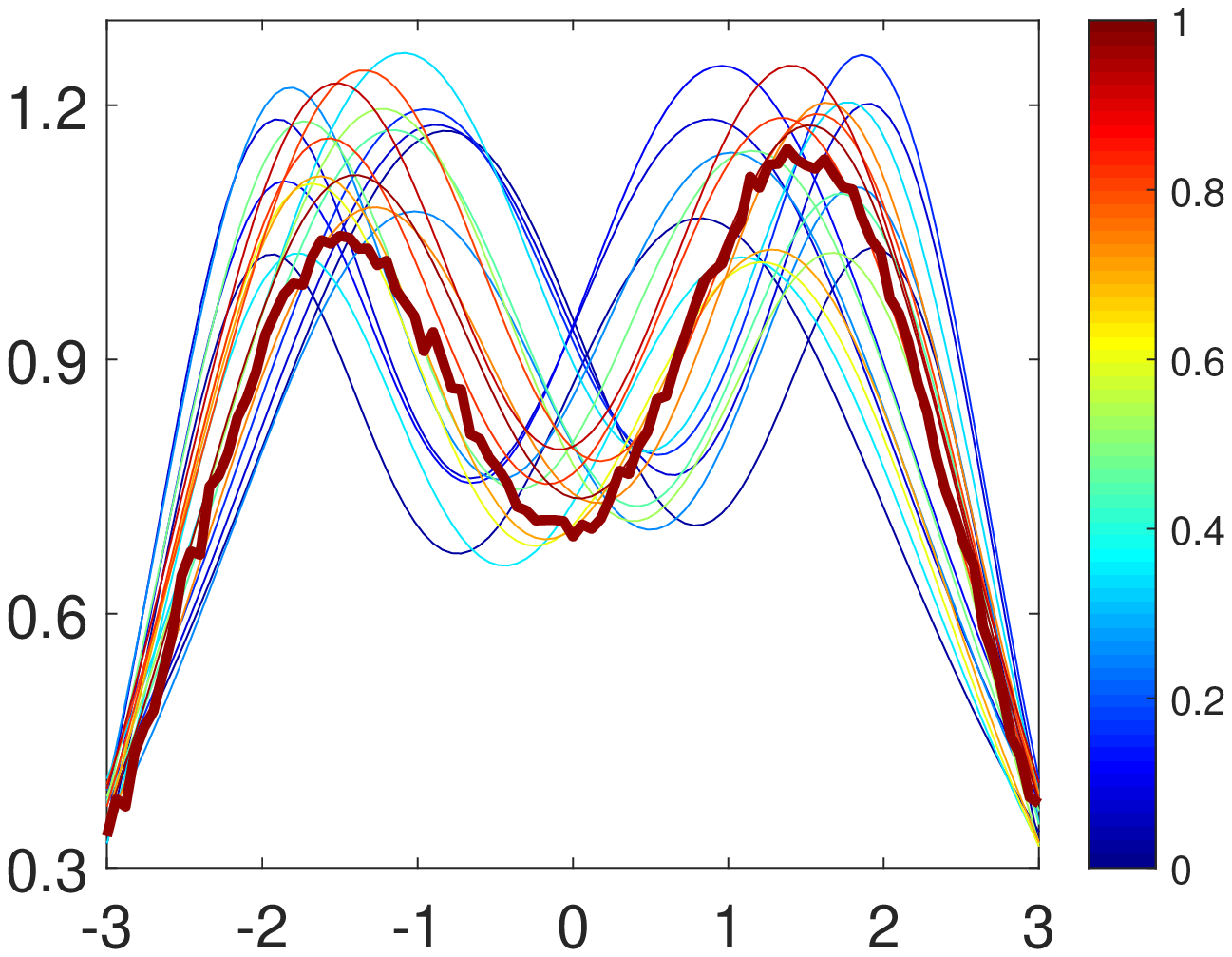}
\caption{$\zeta(f_i,f_c) = \|\hat{\gamma}_i - \gamma_{id}\|_2$}
\end{subfigure}%
\begin{subfigure}{.33\textwidth}
\centering
\includegraphics[height=4.2cm]{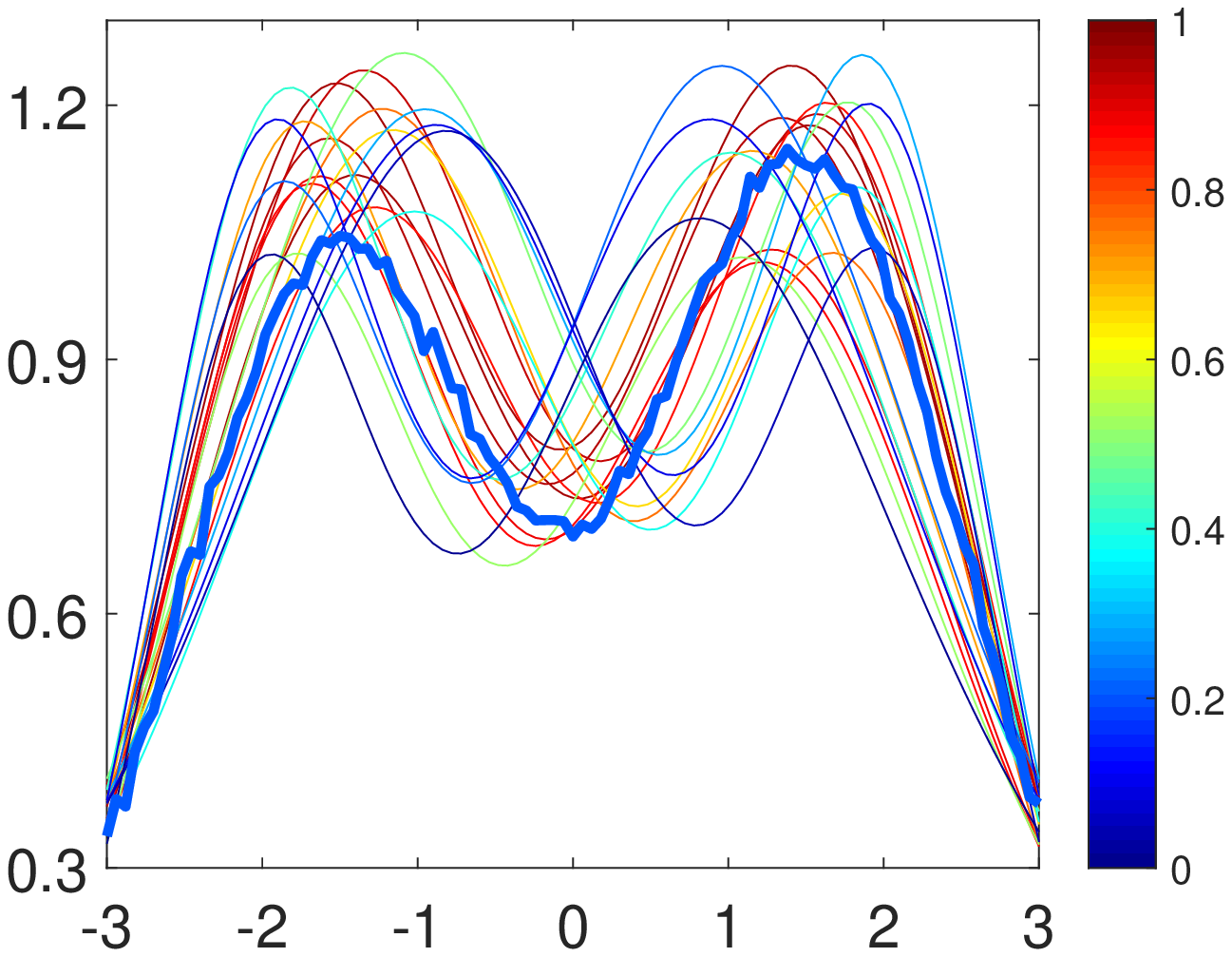}
\caption{$\zeta(f_i,f_c) = d_{FR}(\hat{\gamma}_i,\gamma_{id})$}
\end{subfigure}%
\caption{Simulation 2: {\bf (a)} 21 observed functions, where $f_{11}$ is emphasized in red color. {\bf (b)} The 21 functions with color-labeled depth via the $\mathbb L^2$ warping distance $\|\hat{\gamma}_i - \gamma_{id}\|_2$. {\bf (c)} Same as (b) except for the Fisher-Rao distance $d_{FR}(\hat{\gamma}_i,\gamma_{id})$.}
\label{fig:DemoTW}
\end{figure}

All these 21 functions are shown in Figure \ref{fig:DemoTW}(a), where we use red line to represent $f_{11}$ and blue lines to represent the others. Before  computing depth values, we conduct the Fisher-Rao alignment procedure to align the observed functions and obtain the corresponding time warping functions $\{\hat{\gamma}_1(t),\cdots, \hat{\gamma}_{21}(t)\}$ \citep{srivastava2011registration}. Let $f_c$ denote the Karcher mean of $\{f_1, \cdots, f_{21}\}$ (in the sense of SRVF space). Then the optimal time warping function from $f_i$ to $f_c$ is $\hat{\gamma}_i, i = 1, \cdots, 21$.  
 
We take the criterion function $\zeta(f_i, f_c) = \| \hat{\gamma}_i - \gamma_{id} \|_2$ for the depth computation. The 21 color-labeled functions using depth values are shown in Figure \ref{fig:DemoTW}(b). In general, functions in the middle along x-axis have large depth values, whereas those at each side have low values. In particular, because $f_{11}$ stays in the middle of the observations, it has the least $\mathbb L^2$ warping distance from $f_c$ and largest depth values. As comparison we also use the well-known Fisher-Rao distance function $\zeta(f_i, f_c) = d_{FR}(\hat \gamma_i, \gamma_{id})$ for the depth computation. The 21 color-labeled functions using depth values are shown in Figure \ref{fig:DemoTW}(c). As the Fisher-Rao distance is derivative-based, small perturbation on time warping results in large difference. The small noise on $f_{11}$ makes it have smallest depth value in the 21 functions. For other 20 smooth functions, their depth values are consistent to those in Panel (b).

\noindent{\bf Simulation 3.} In this illustration, we demonstrate Algorithm I in Section~\ref{sec:MBD} for modified norm-based depth estimation on a variant of the continuous-time Brownian Bridge on $[0,1]$, with different choices of decaying weight sequences $\{a_p\}$, where the covariance kernel function is $K(s,t) = min(s,t)-st$ for any $s,t \in [0,1]$. According to the notation in Equations \eqref{eq:mercer} and \eqref{eq:KL}, we have $\lambda_p = \frac{1}{p^2\pi^2}, \phi_p(t) = \sqrt{2}sin(\pi pt)$, and can simulate $f_p$ from {\it independent Laplace distribution} with mean 0 and variance $\lambda_p$ for $p = 1,2,\cdots$ (note that this is different from the normal distribution $\mathcal{N}(0,\lambda_p)$ in a Brownian bridge). 

\begin{figure}[ht]
\captionsetup[subfigure]{}
\begin{subfigure}{.33\textwidth}
\centering
\includegraphics[height=4.2cm]{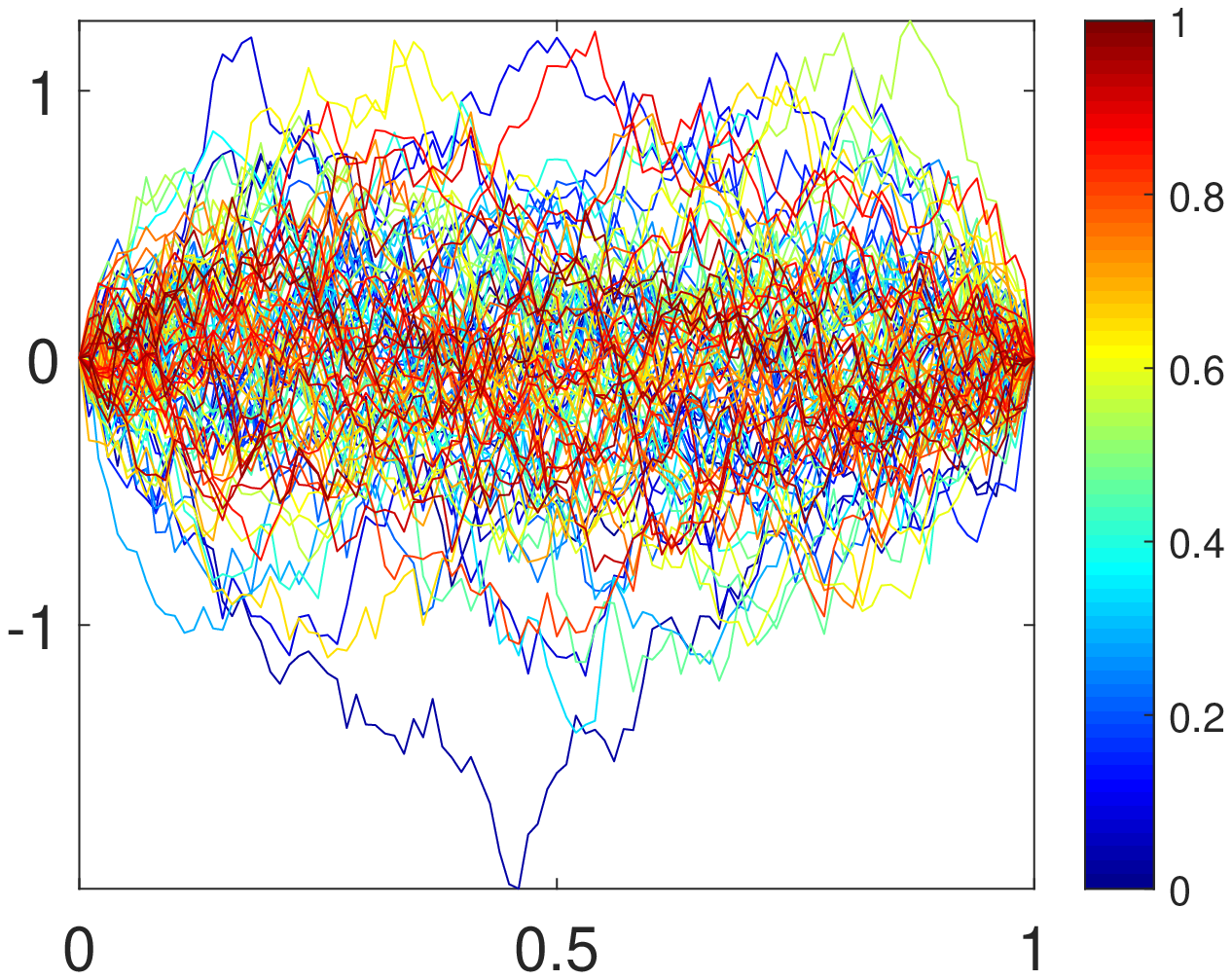}
\caption{$a_p = 1$}
\end{subfigure}%
\begin{subfigure}{.33\textwidth}
\centering
\includegraphics[height=4.2cm]{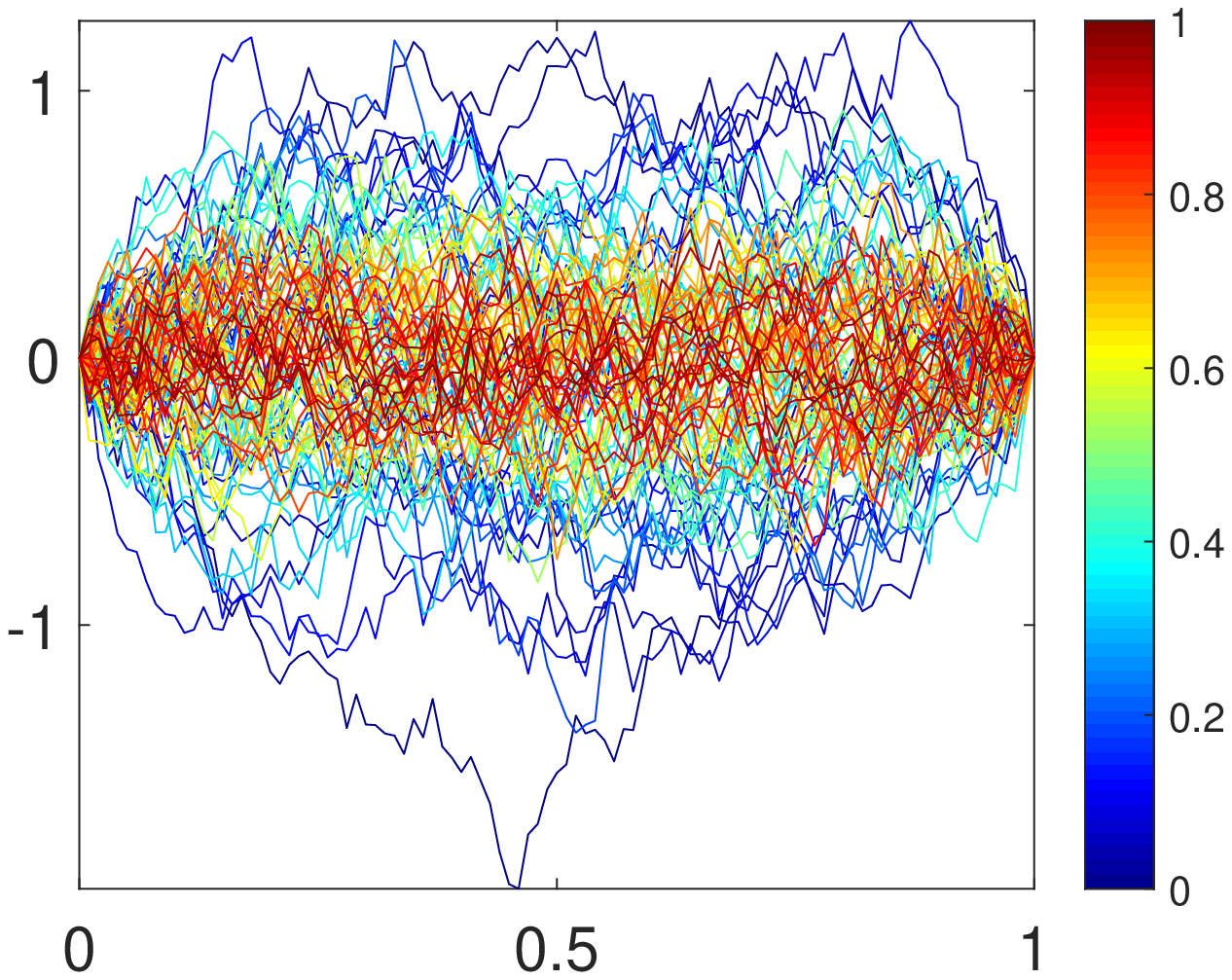}
\caption{$a_p = 1/p$}
\end{subfigure}%
\begin{subfigure}{.33\textwidth}
\centering
\includegraphics[height=4.2cm]{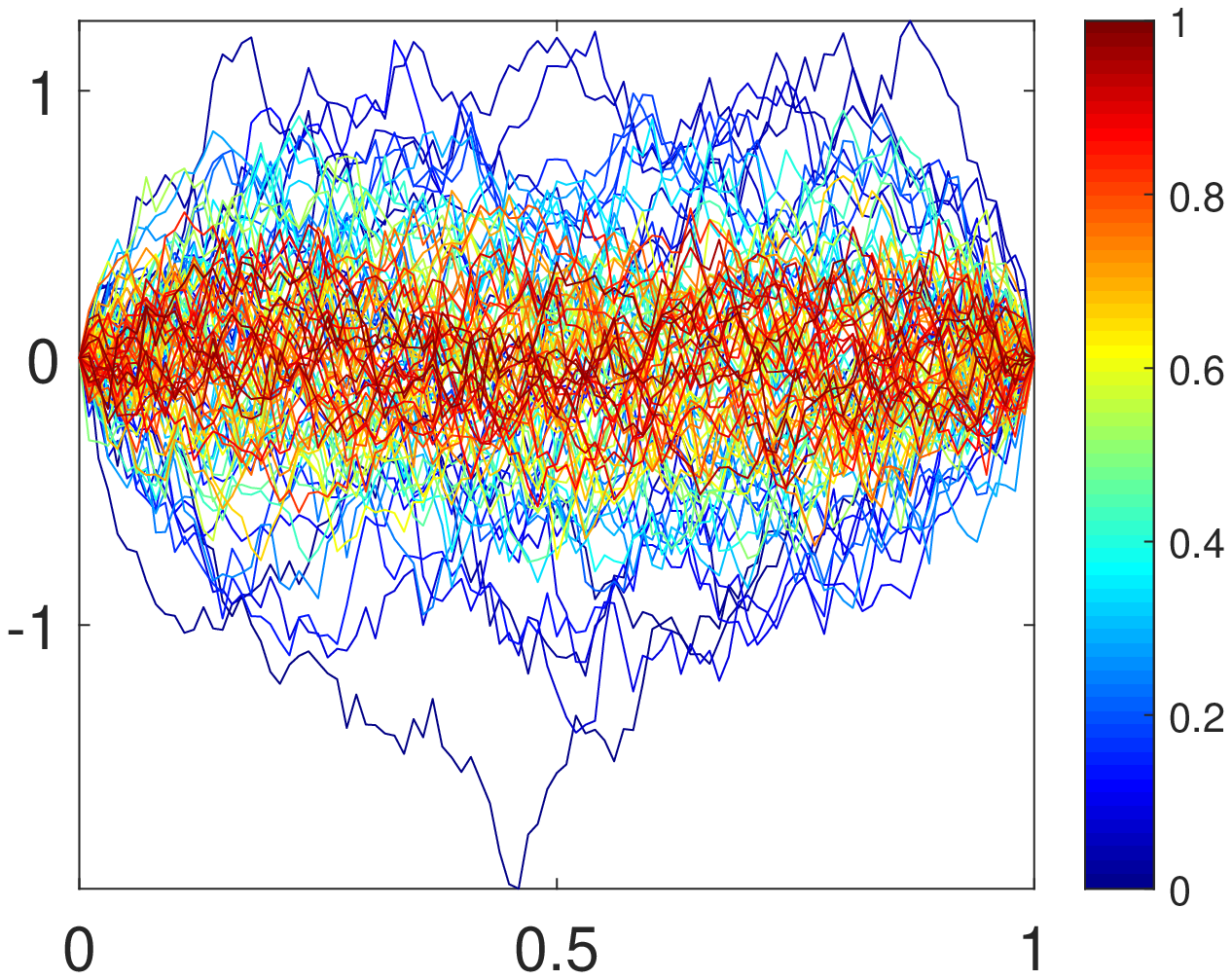}
\caption{$a_p = 1/[\sqrt{p}\log (p+1)]$}
\end{subfigure}%
\caption{Simulated functions with color-labeled depth.  {\bf (a)} Each function is color-labeled using its depth value, where the sequence $\{a_p\}$ is constant 1. Observations assigned with color closer to red are considered to be deeper than those assigned with color closer to blue.  {\bf (b)} and {\bf (c)}, Same as (a) except that the coefficients $a_p = 1/p$ and $a_p = 1/1/[\sqrt{p}\log (p+1)], p = 1, \cdots, N$, respectively.}
\label{fig:DemoAI}
\end{figure}

More specifically, we sample $\{f_i(t)\}$ by the linear combination $f_i(t) = \sum_{p=1}^{1000} f_{i,p}\phi_p(t), i = 1, \cdots, n (=100)$ to approximate the infinite-dimensional stochastic process. We set $N=1000$ in the Monte Carlo sampling.
We have three different settings to choose the weight coefficients $\{a_p\}$: $ a) \ a_p \equiv 1, \ b) \ a_p = 1/p,$ and $ c) \ a_p = 1/[\sqrt{p}\log (p+1)], p = 1, \cdots, N$.   In Case a), there is actually no weight terms, and the modified norm is equal to the RKHS induced norm.  In Figure \ref{fig:DemoAI}(a), we show the 100 functions with color-labeled using its depth value from this norm. 
Note that we compute this norm in a finite-dimensional setup and it will diverge to $\infty$ when $N$ is large.  It is straightforward to find that 
\begin{equation}
\|f_i^{'}\|_{\mathbb L^2}^2 = \sum_{p=1}^{N} \int_0^1  2\pi^2 p^2 \cdot f_{i,p}^2 cos(\pi p t)^2 dt = \sum_{p=1}^{N} \pi^2p^2 f_{i,p}^2 = \sum_{p=1}^{N} \frac{f_{i,p}^2}{\lambda_p} = \|f_i\|_{\mathbb H_K}^2.
\label{eq:equ1}
\end{equation}
That is, the RKHS induced norm is the same as $\mathbb L^2$ norm on the derivative function, a common measure of smoothness of a function. 

In Cases b) and c), the series satisfies the convergent requirement $\lim_{N\rightarrow \infty} \sum_{i=1}^N a_P^2 < \infty$, and therefore the modified norms are well-defined.  In particular, we find that when $a_p = 1/p$, the classical $\mathbb L^2$ norm  
\begin{equation}
 \| f_i\|_{\mathbb L^2}^2 = \sum_{p=1}^{N} \int_0^1 f_{i,p}^2 \phi_p^2(t)dt = \sum_{p=1}^{N} f_{i,p}^2 = \frac{1}{\pi^2}\sum_{p=1}^{N} f_{i,p}^2 \cdot \frac{1}{1/\pi^2 p^2} \cdot \frac{1}{p^2} \propto \|f_i\|_{mod}^2. 
\label{eq:equ2}
\end{equation}
That is, the modified norm in Case b) is proportional to the $\mathbb L^2$ norm.  This explains the result in Figure \ref{fig:DemoAI}(b) where the 100 functions are color-labeled using its depth value from this norm.  We can see that high-depth functions are near the zero-line and low-depth functions are near boundary lines.  This depth reflects the traditional functional depths such as band depth and half-region depth \citep{lopez2009concept,lopez2011half}.   
In Case c), we use another type of weight coefficient and the depth result is shown in  Figure \ref{fig:DemoAI}(c), which is very similar to the result in Case b).  In summary, we have found that 1) the modified norms can provide different forms of measurement on the center-out rank on the given functional observation and some of the special forms are consistent to the classical norms; and 2) the rank may be robust with respect to different choices of norm.   
    
\begin{figure}[h!]
\captionsetup[subfigure]{}
\begin{subfigure}{.5\textwidth}
\centering
\includegraphics[height=5cm]{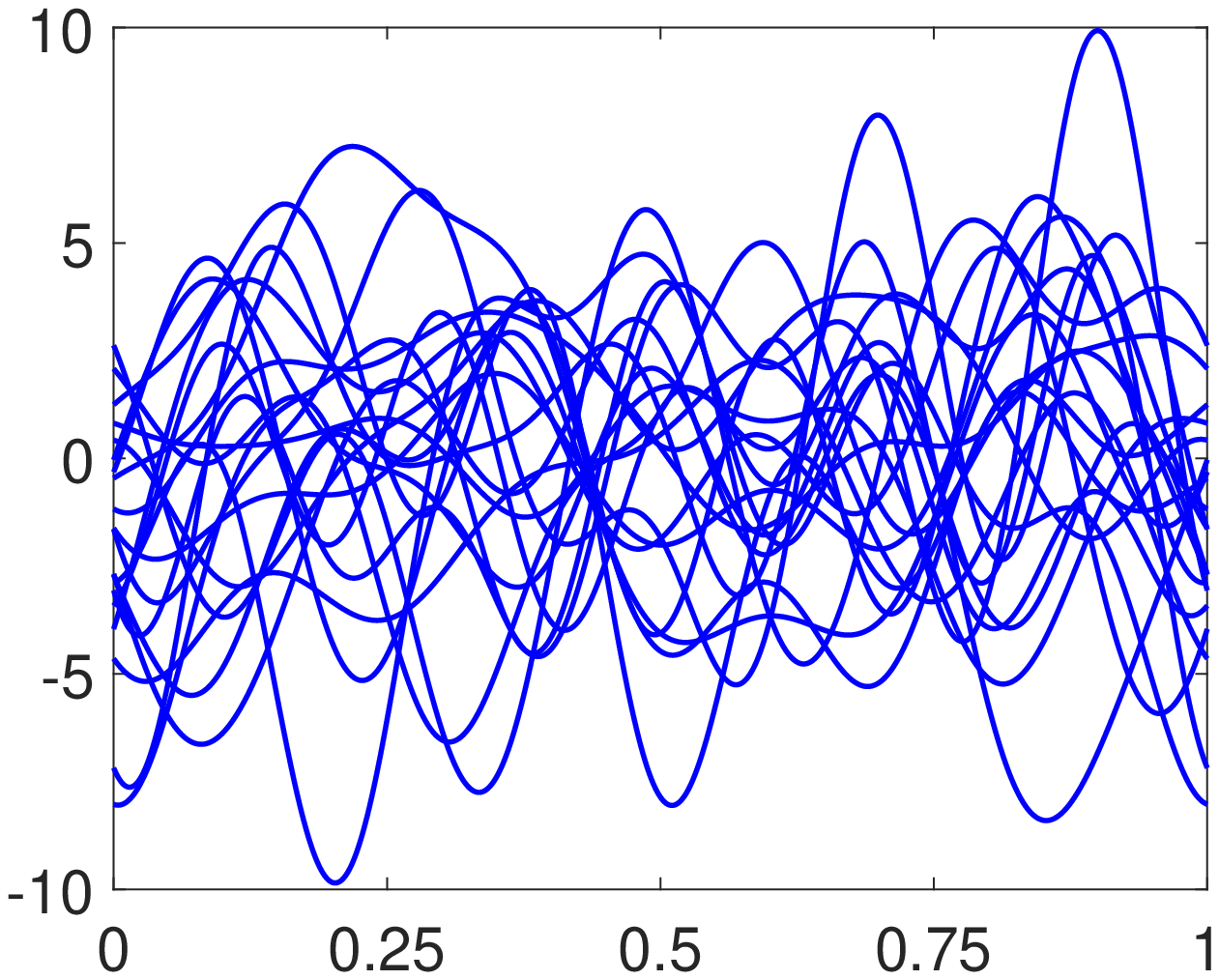}
\caption{21 functions from the original sample}
\end{subfigure}%
\begin{subfigure}{.5\textwidth}
\centering
\includegraphics[height=5cm]{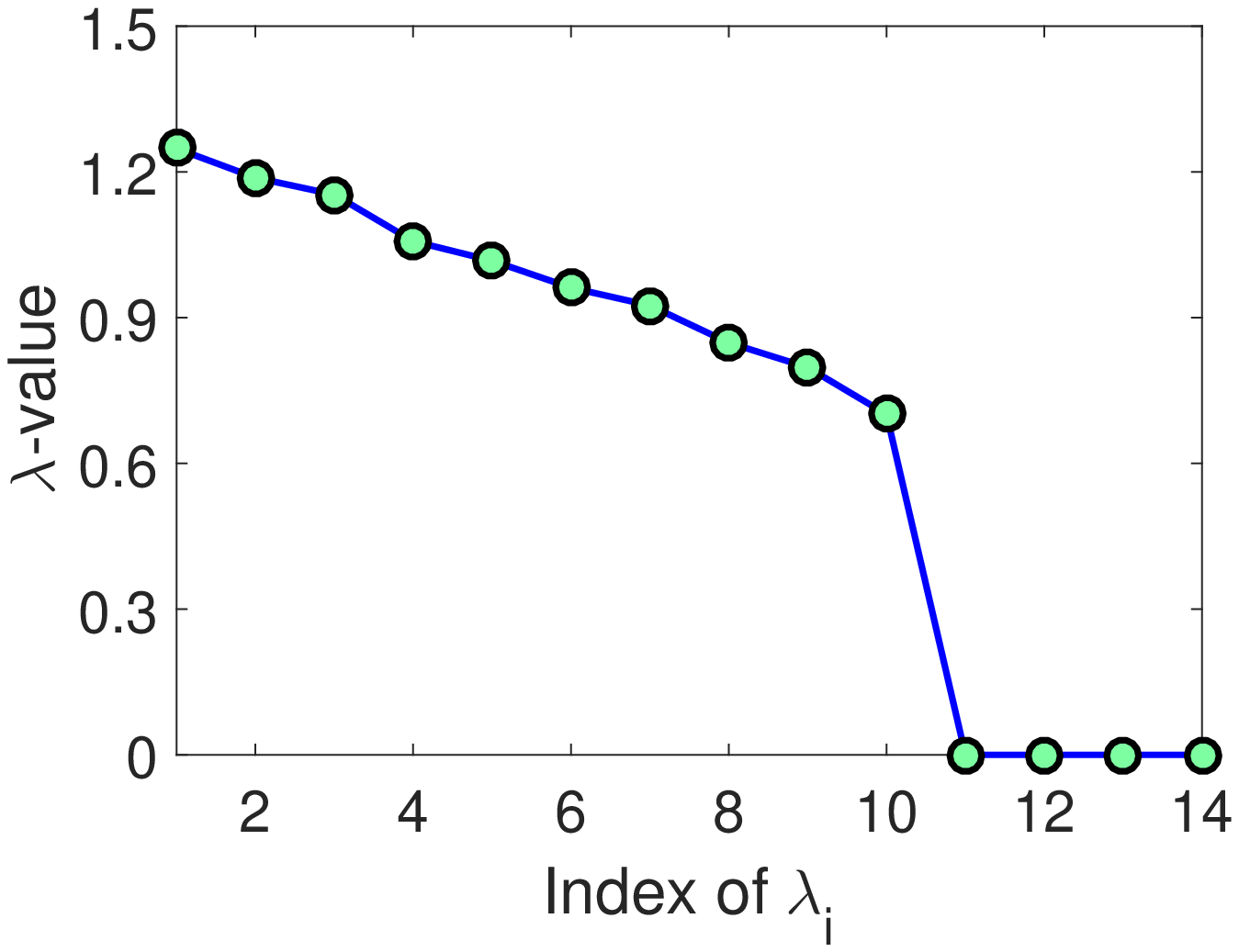}
\caption{Estimated eigenvalues}
\end{subfigure}%

\begin{subfigure}{.5\textwidth}
\centering
\includegraphics[height=5cm]{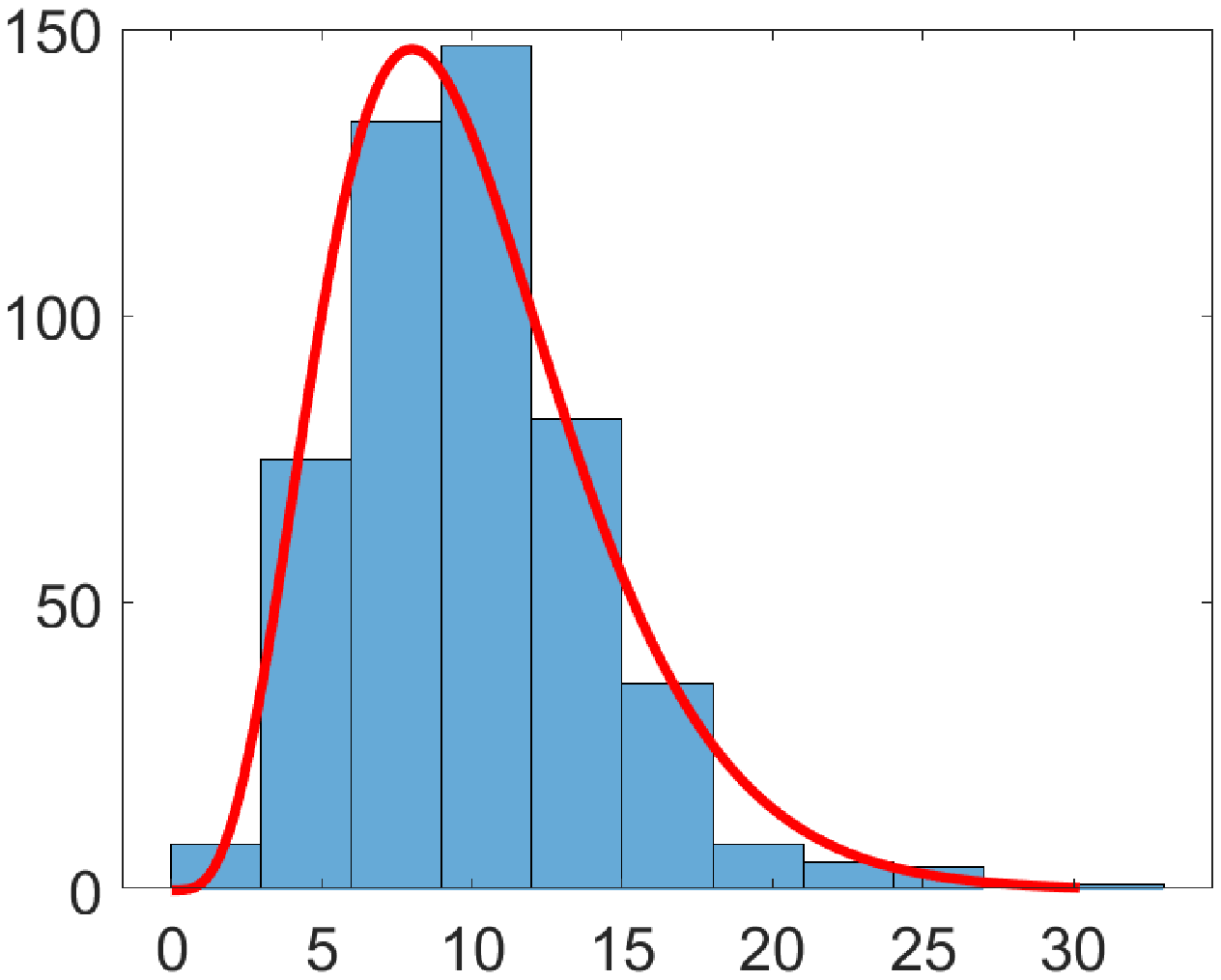}
\caption{Histogram of squared RKHS norm}
\end{subfigure}%
\begin{subfigure}{.5\textwidth}
\centering
\includegraphics[height=5cm]{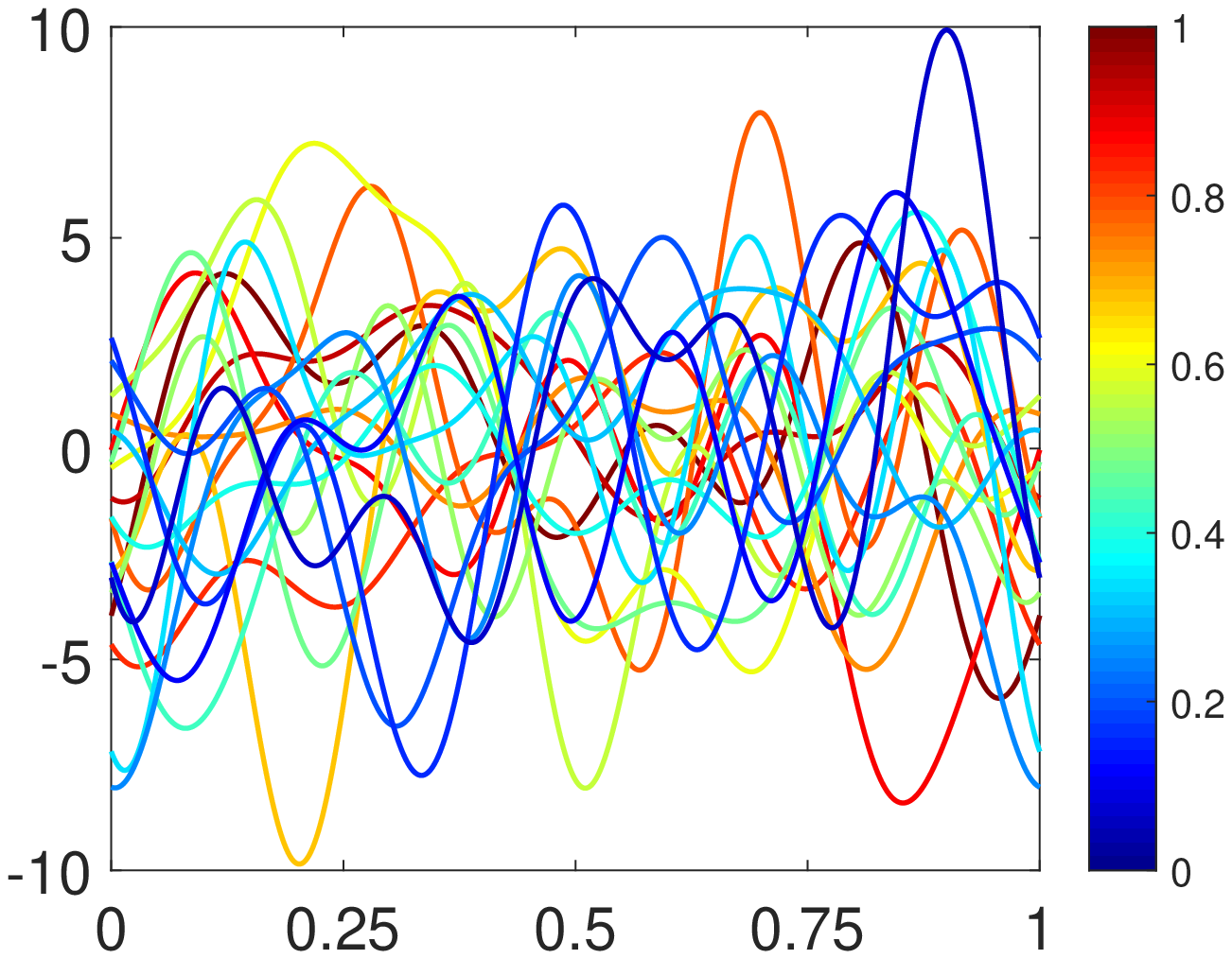}
\caption{Samples with color-labeled depth}
\end{subfigure}%
\caption{Finite Gaussian Process Illustration: {\bf (a)} 21 randomly selected samples; {\bf (b)} Estimated eigenvalues $\hat{\lambda}_p$ from the covariance $\hat{K}$; {\bf (c)} Histogram of squared RKHS norm $\| f_{i} \|_{\mathbb H_{\hat{K}}}^2 = \sum_{p=1}^{10} \frac{\hat{a}_{i,p}^2}{\hat{\lambda}_{p,n}}$, where the red line indicates a fit to chi-square distribution $\chi^2(10)$; {\bf (d)} Estimated depth of the 21 samples with color-label, where blue to red indicates the depth value range of $[0,1]$. }
\label{fig:FG}
\end{figure}

\noindent {\bf Simulation 4.} We consider a finite-dimensional Gaussian process by selecting a sequence of orthonormal Fourier basis functions up to order $P=10$ on $[0,1]$ such that
\begin{align*}
\phi_p(t)
= \left\{
\begin{array}{ccl}
1  &  & p=1\\
\sqrt{2}\cos(\pi pt)  &  & p=2,4,6,8,10\\
\sqrt{2}\sin(\pi (p-1)t)  &  & p=3,5,7,9
\end{array} \right. ,
\end{align*}
and a set of coefficients $\{ a_1, \cdots,a_P\} \sim N(0, I_{10})$. Then we generate $N=500$ functions via linear combination $f_i = \sum_{p=1}^P a_{i,p} \phi_p$. Panel (a) in Figure \ref{fig:FG} shows $n=21$ randomly selected samples from $\{f_i(t), t \in [0,1] \}_{i=1}^N$.

From Panel (b) in Figure \ref{fig:FG}, it is clear that there exists a significant gap in the decreasing sequence of estimated eigenvalues, and the gap locates just after the order of the dimension $P =10$. This indicates the correct dimension can be easily estimated.  From Panel (c), we can tell that the squared RKHS norm fits $\chi^2(P)$ well. This is also consistent to the above theoretical derivation. The estimated depth values are color-labeled in Panel (d). We note that the RKHS induced norm does not have a conventional $\mathbb L^2$ type of norm, so there is no direct visualization to evaluate the depth in this example. However, we point out that if the data are generated from two random processes, these depth values can help differentiate the observations, as illustrated in the following.

\begin{figure}[hbt!]
\captionsetup[subfigure]{}

\begin{subfigure}{.33\textwidth}
\centering
\includegraphics[height=4.2cm]{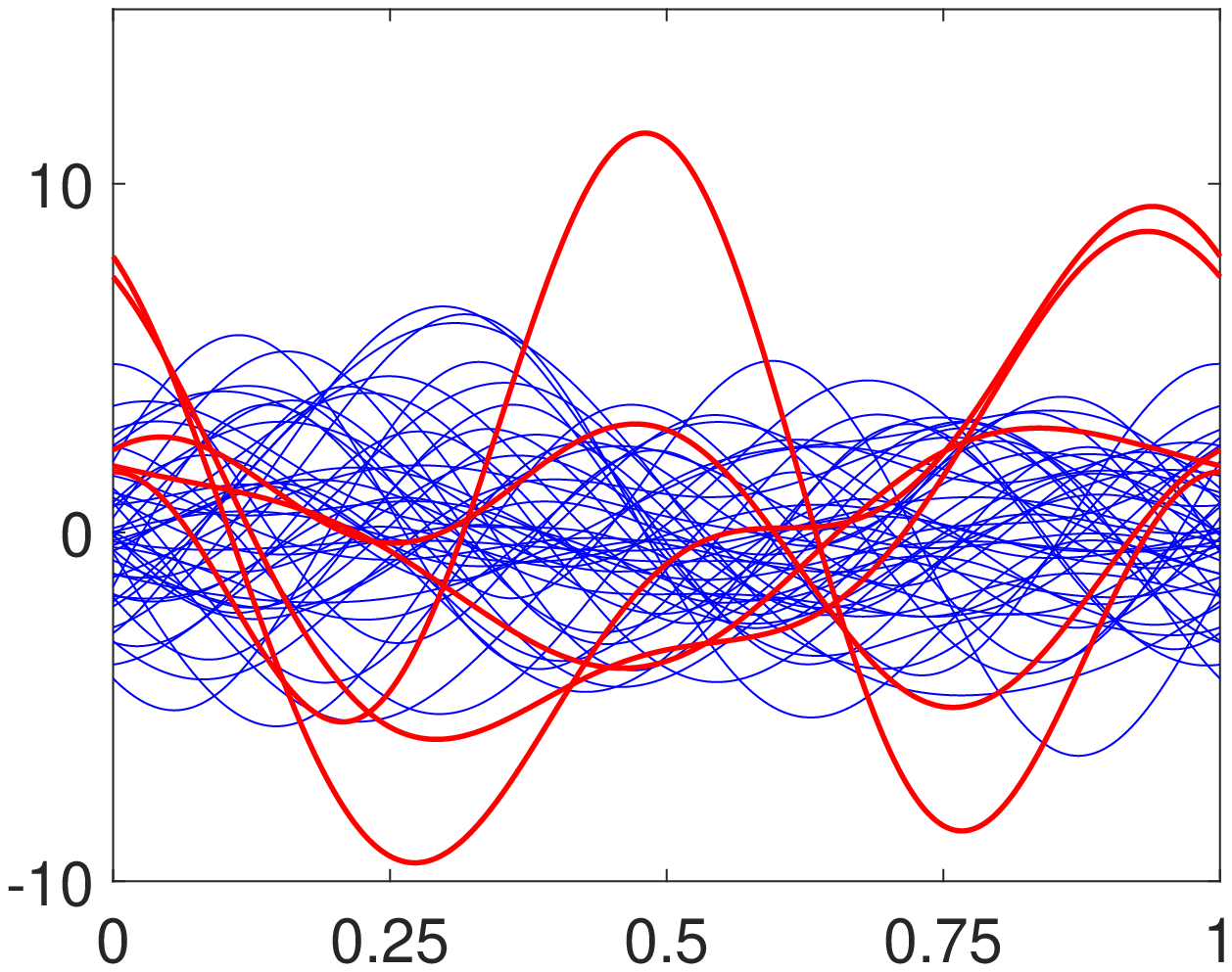}
\caption{50 functions with $P=4$}
\end{subfigure}%
\begin{subfigure}{.33\textwidth}
\centering
\includegraphics[height=4.2cm]{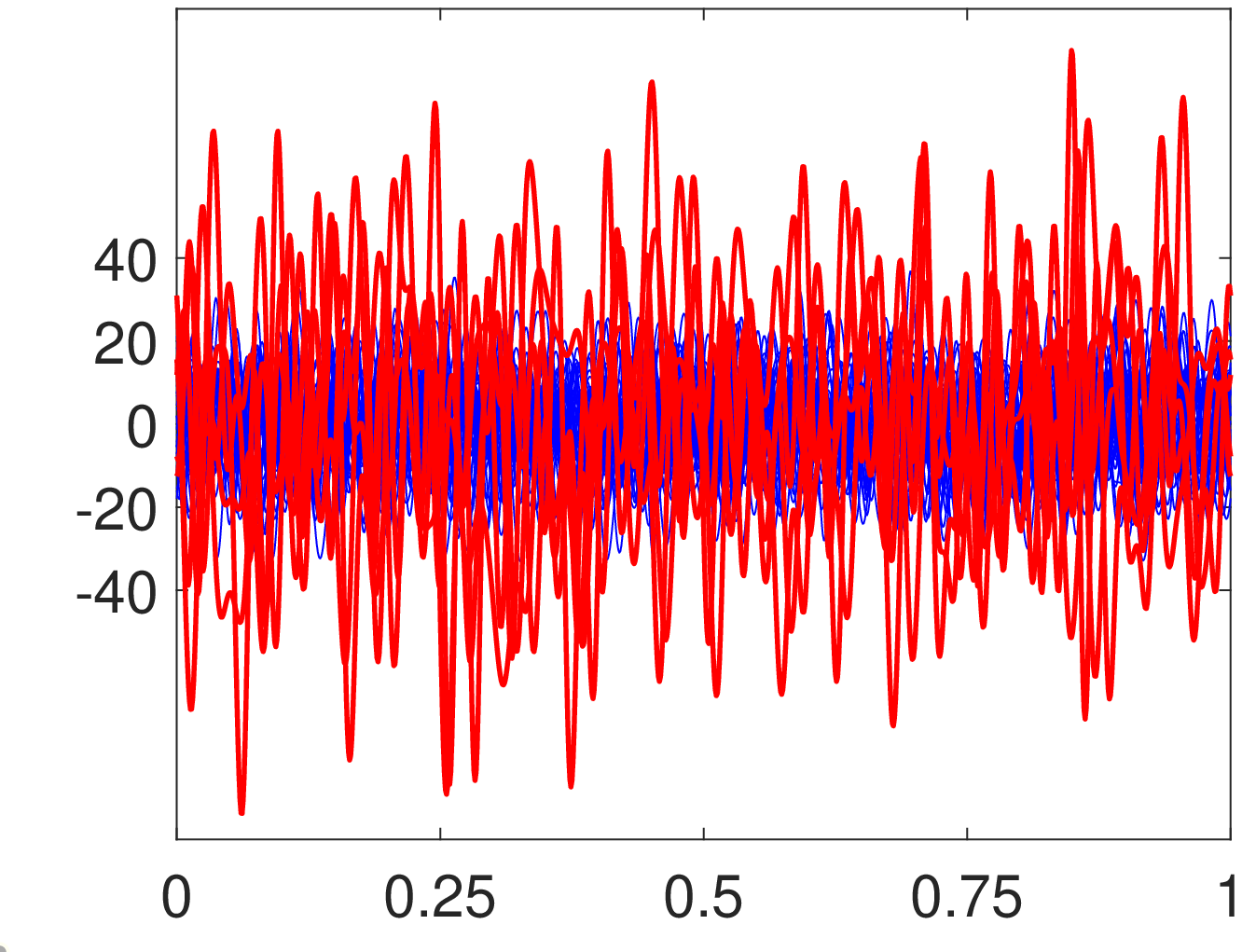}
\caption{50 functions with $P=100$}
\end{subfigure}%
\begin{subfigure}{.33\textwidth}
\centering
\includegraphics[height=4.2cm]{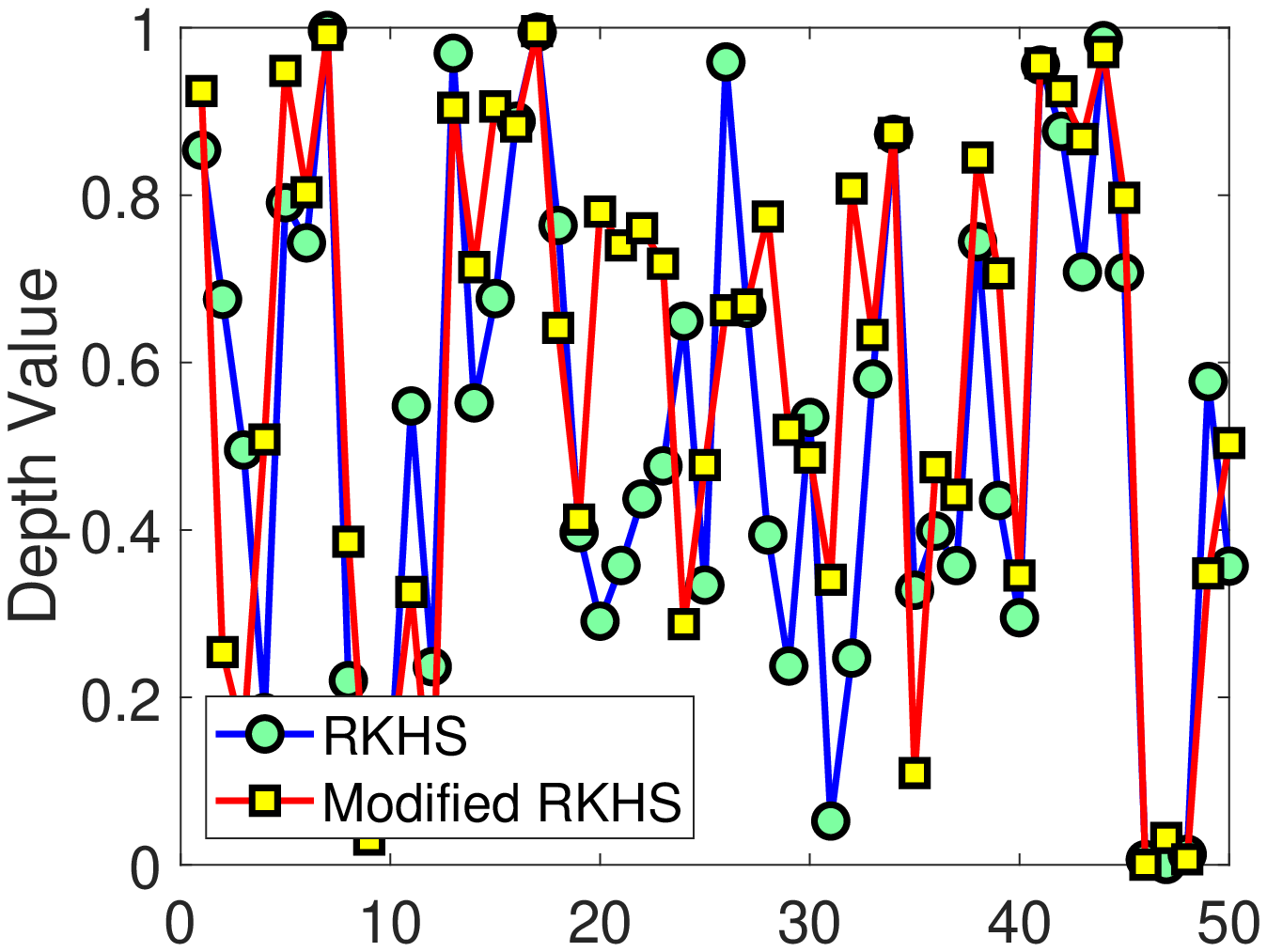}
\caption{Depths with $P=4$}
\end{subfigure}%

\begin{subfigure}{.33\textwidth}
\centering
\includegraphics[height=4.2cm]{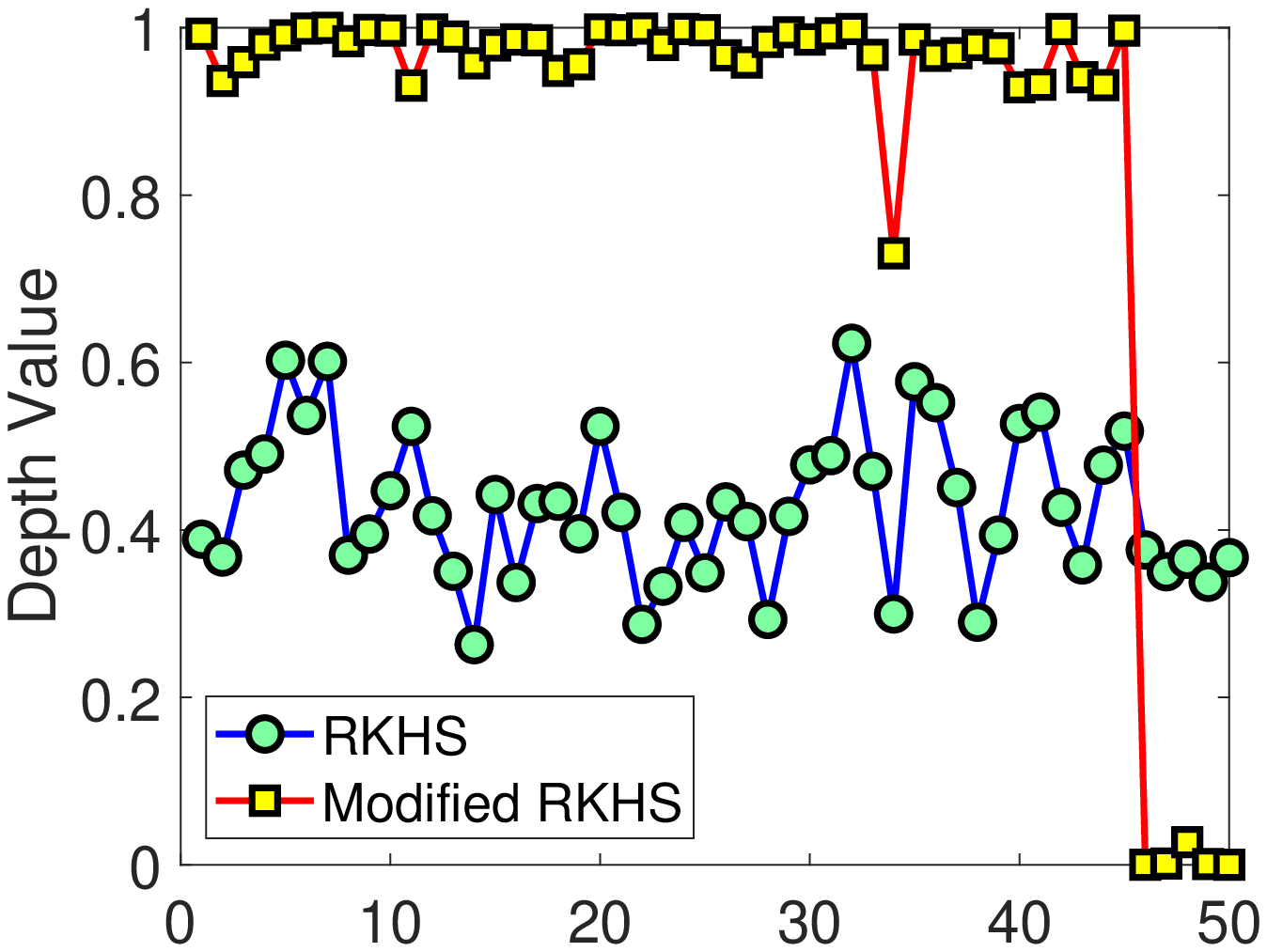}
\caption{Depths with $P=100$}
\end{subfigure}%
\begin{subfigure}{.33\textwidth}
\centering
\includegraphics[height=4.2cm]{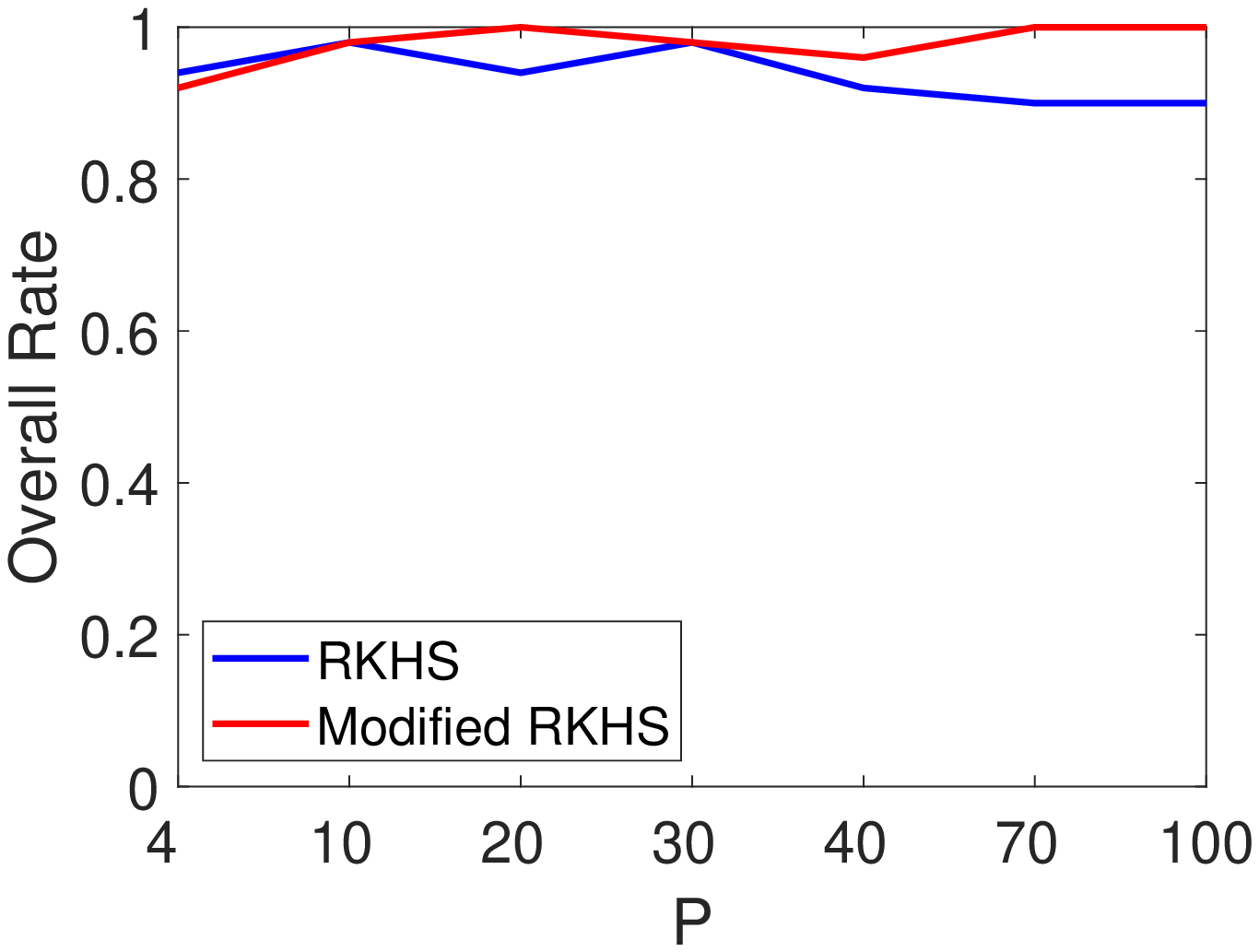}
\caption{Overall accuracy rate}
\end{subfigure}%
\begin{subfigure}{.33\textwidth}
\centering
\includegraphics[height=4.2cm]{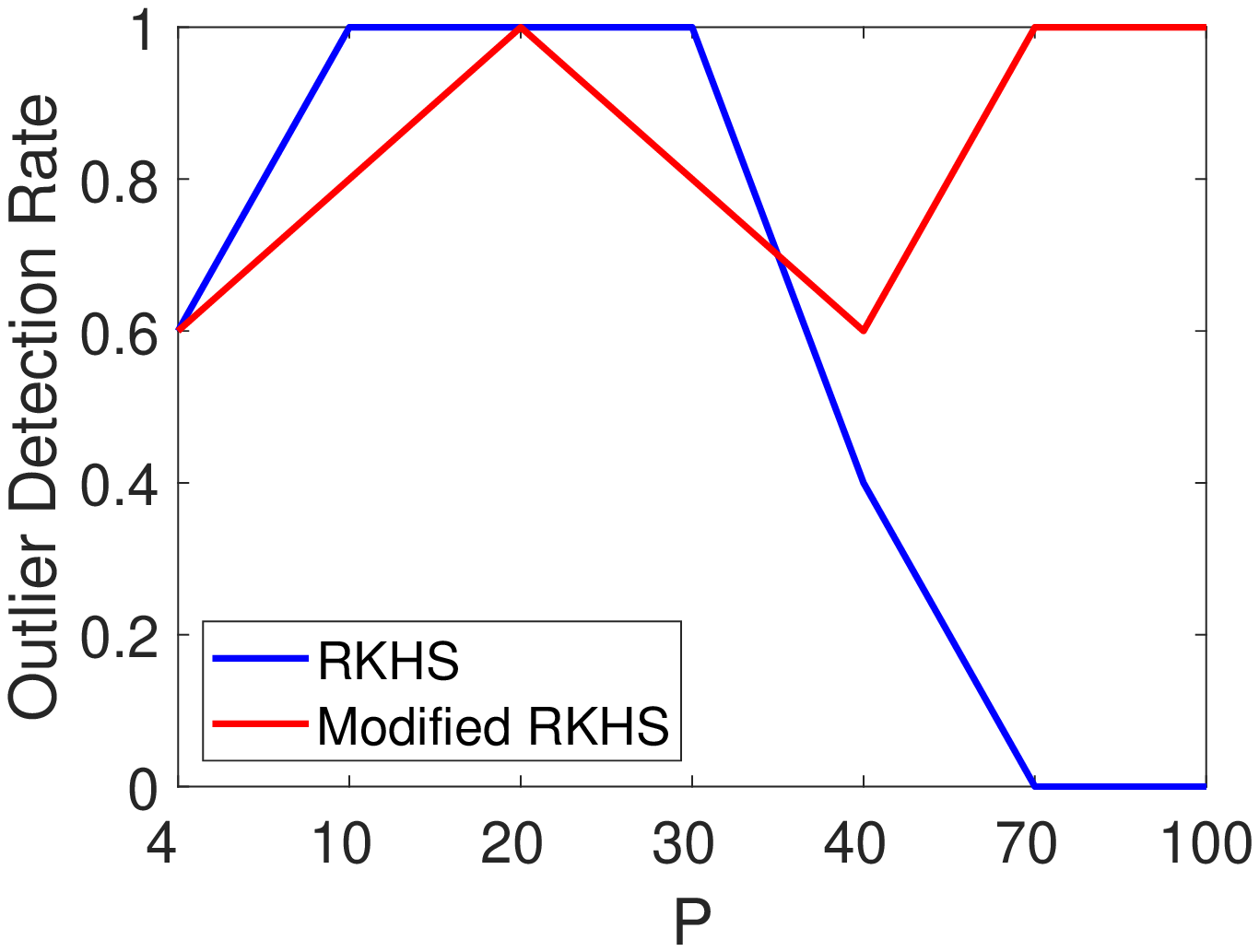}
\caption{Outlier detection rate}
\end{subfigure}%
\caption{Classification by depth values: {\bf (a)} 50 function samples for $P=4$, where the blue ones represent 45 functions in the main cluster and the red ones represent 5 outliers.  {\bf (b)} Same as (a) except $P=100$. {\bf (c)} Depth values of the 50 functions with $P=4$ using the RKHS norm (green circles over blue lines) and modified RKHS norm (yellow squares over red lines). {\bf (d)} Same as (c) except $P=100$. {\bf (e)} The classification accuracy for all 50 functions by using the RKHS norm (blue line) and modified RKHS norm (red line), where $P$ varies on seven different values 4, 10, 20, 30, 40, 70, and 100. {\bf (f)} Same as (e) except the accuracy on the 5 outlier functions.  
}
\label{fig:FG1}
\end{figure}

In particular, we show how the model-based depth can be used for the classification purpose, and compare the performance between RKHS norm and the modified one. Specifically, we select a sequence of orthonormal Fourier basis functions up to order $P$ on $[0,1]$ such that for $p = 1, 2, \cdots, P$, 
\begin{align*}
\phi_p(t)
= \left\{
\begin{array}{ccl}
\sqrt{2}\sin(\pi (p+1) pt)  &  & p \mbox { is odd}\\
\sqrt{2}\cos(\pi pt)  &  & p \mbox{ is even}
\end{array} \right. ,
\end{align*}
Then we generate $45$ functions as $f_i(t) = \sum_{p=1}^{P} a_{i,p} \phi_p(t), i = 1, \cdots, 45$ and $5$ functions as $f_i(t) = \sum_{p=1}^{P} b_{i,p} \phi_p(t), i = 46, \cdots, 50$, where independent coefficients $a_{i,p} \sim N(0,1)$ and $b_{i,p} \sim N(0,3)$.  Due to the different variance values on the coefficients, the first 45 functions are in the main cluster and the last 5 functions are outliers.  One special case when only $P = 4$ low frequency basis functions are used is shown in Figure \ref{fig:FG1}(a).  Because of the smaller coefficient variance, the first 45 functions are in a main cluster. In contrast, some of the last 5 function have much larger amplitude and are apparently outliers.   In another example, we use $P=100$ basis functions shown in Figure \ref{fig:FG1}(b).  As compared to when $P=4$, both main clusters functions and the 5 outliers have much higher frequency components.

We have shown that the RKHS induced norm can characterize the smoothness level in Equation \eqref{eq:equ1} and the modified RKHS norm can characterize the amplitude level ($\mathbb L^2$ norm) in Equation \eqref{eq:equ2}.   We at first use these two norms for the case when $P=4$ and the result on depth values are shown in Figure \ref{fig:FG1}(c).  Note that for the simulated 5 outliers, only 3 of them show large amplitude as compared to the main cluster, and therefore only these three have relatively lower depth values by using either RKHS norm or the modified norm.  In contrast, when $P=100$, the difference on amplitude for the main cluster and the 5 outliers are apparent.  This can be easily seen using the modified norm shown in Figure \ref{fig:FG1}(d).  As all high frequency basis functions can have large un-smooth level, the RKHS norm is not able to clearly differentiate 5 outliers from the main cluster.  This is also shown in Figure \ref{fig:FG1}(d).  

To measure the classification performance, we set a threshold of 0.1 on the depth value for all functions.  This is done for the number of basis components $P$ being 4, 10, 20, 30, 40, 70, or 100, which varies from highly smooth to highly nonsmooth observations.  The classification result on all 50 functions is shown in Figure \ref{fig:FG1}(e). In particular, we also show the detection on the 5 outliers in Figure \ref{fig:FG1}(f).   When $P$ is small, both norms produce reasonable classification accuracy around 95\% (a couple of errors in the outliers).  When $P$ gets larger, the modified RKHS can capture larger amplitude in the outliers and reach 100\% classification accuracy.  In contrast, all 50 functions have similar smoothing level which makes the RKHS norm not able to detect the outliers.

\subsection{Real Data Illustration}

\begin{figure}[hbt!]
\captionsetup[subfigure]{}

\begin{subfigure}{.33\textwidth}
\centering
\includegraphics[height=4.2cm]{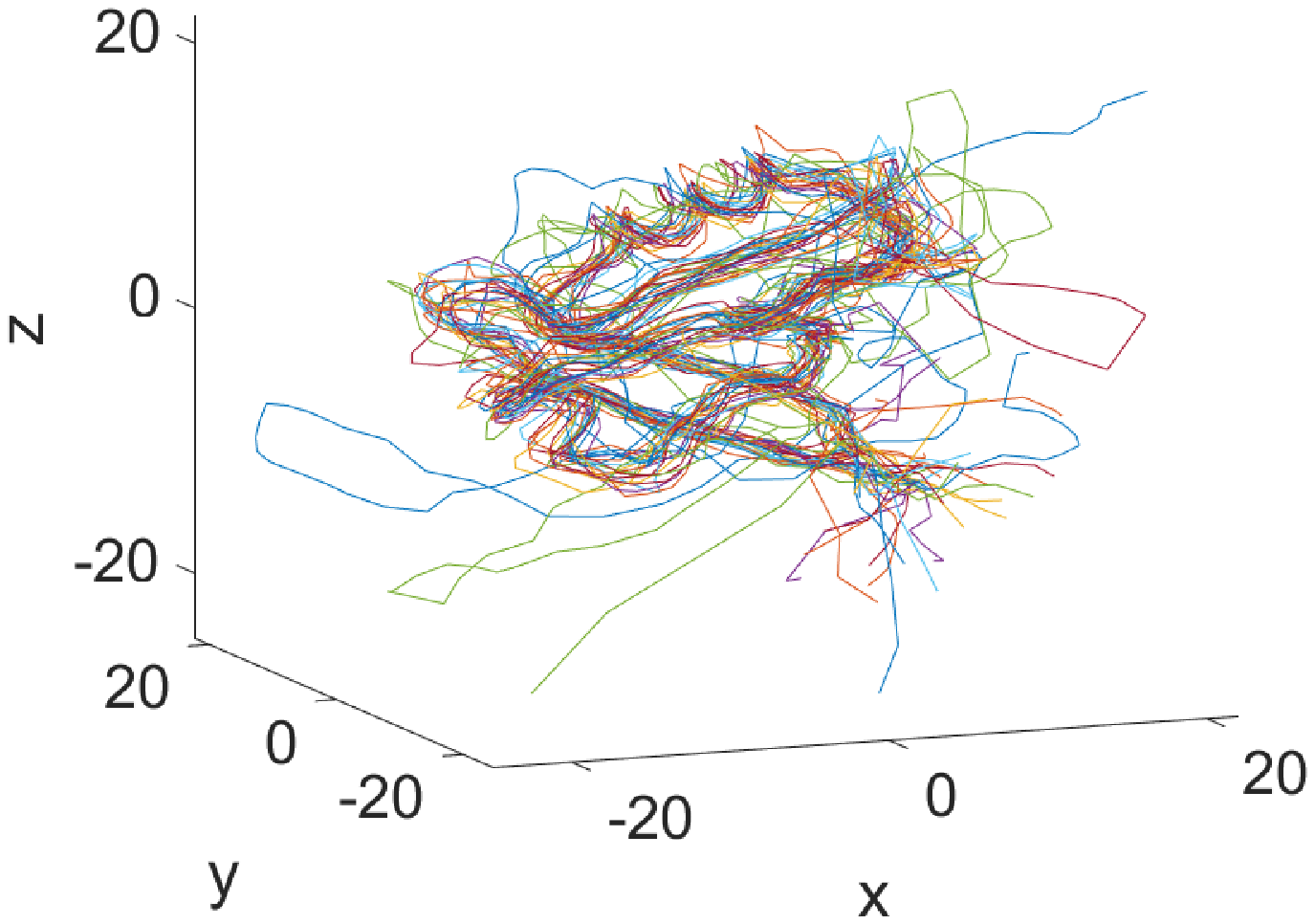}
\caption{3-D data}
\end{subfigure}%
\begin{subfigure}{.33\textwidth}
\centering
\includegraphics[height=4.2cm]{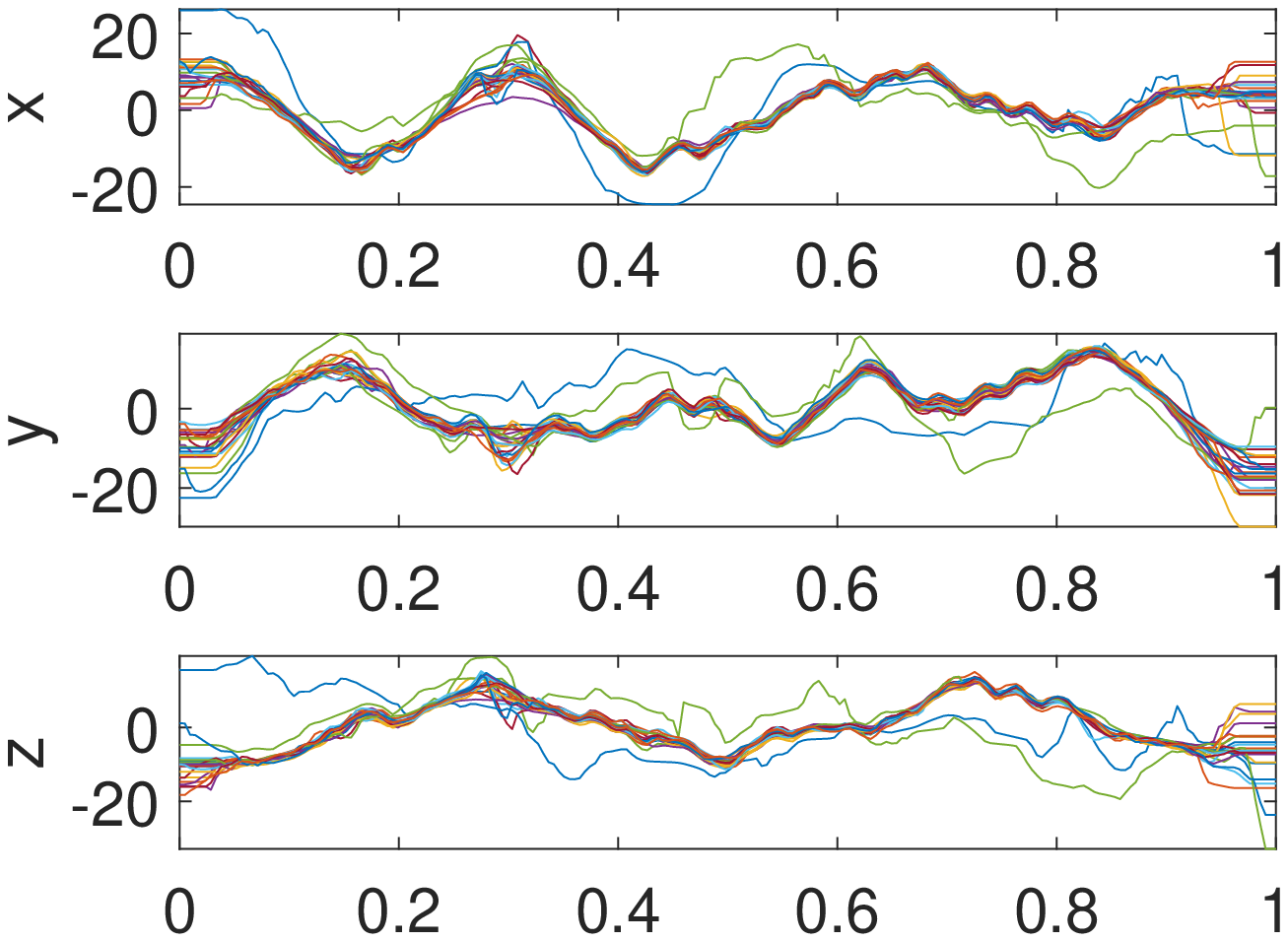}
\caption{Component-wise data}
\end{subfigure}%
\begin{subfigure}{.33\textwidth}
\centering
\includegraphics[height=4.2cm]{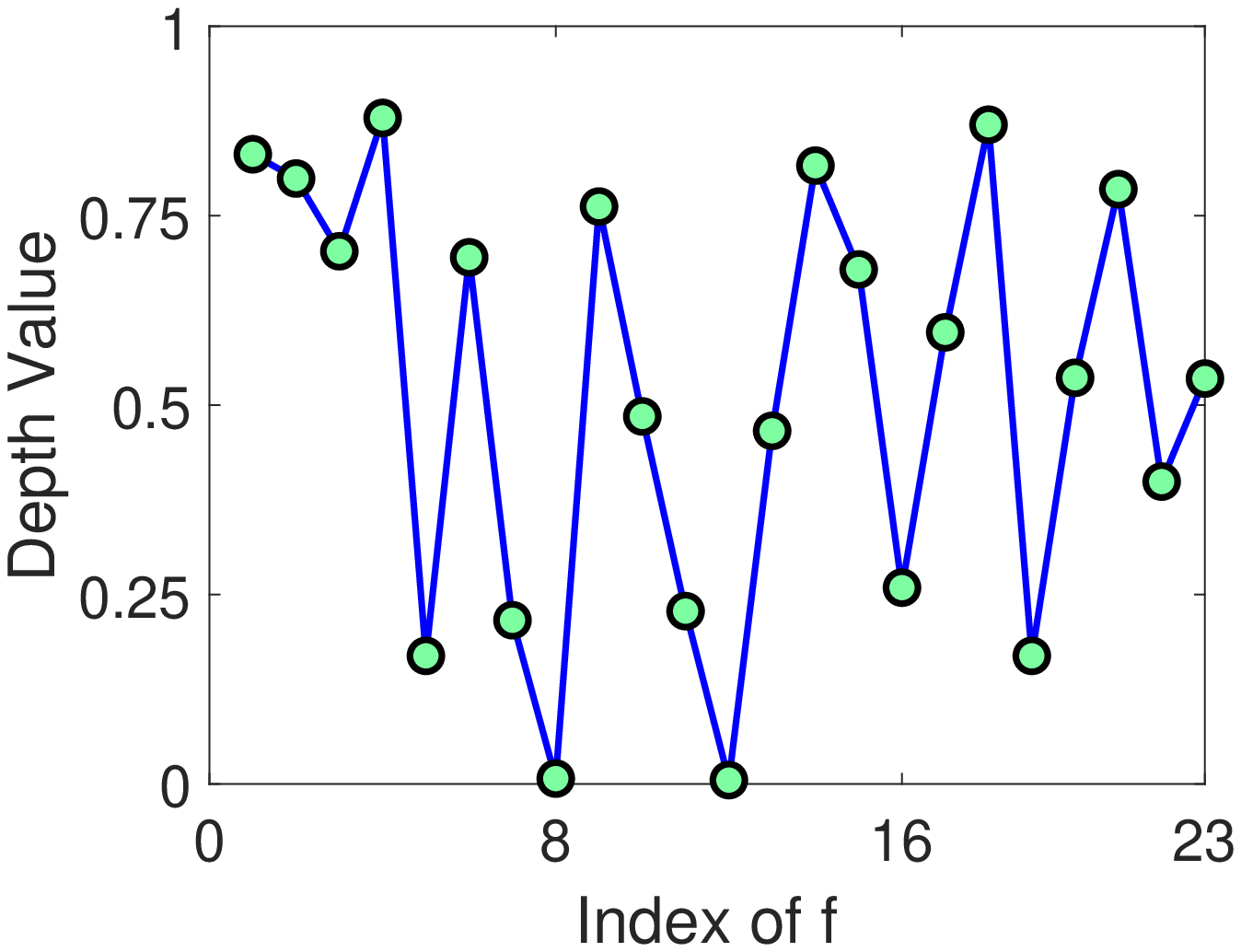}
\caption{Depth with $\zeta(f) = \|f \|_2$}
\end{subfigure}%

\begin{subfigure}{.33\textwidth}
\centering
\includegraphics[height=4.2cm]{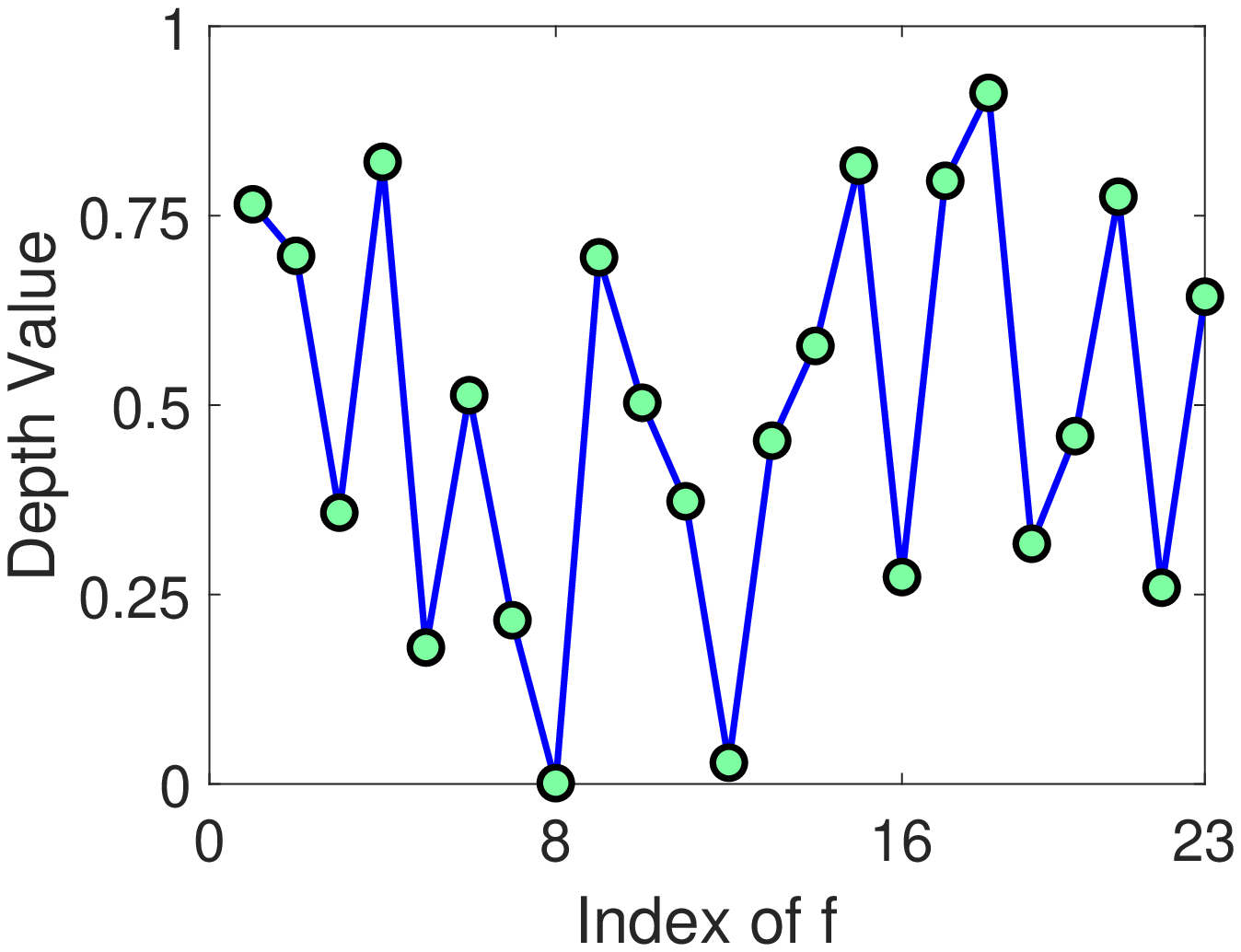}
\caption{Depth with $\zeta(f) = \|f' \|_2$}
\end{subfigure}%
\begin{subfigure}{.33\textwidth}
\centering
\includegraphics[height=4.2cm]{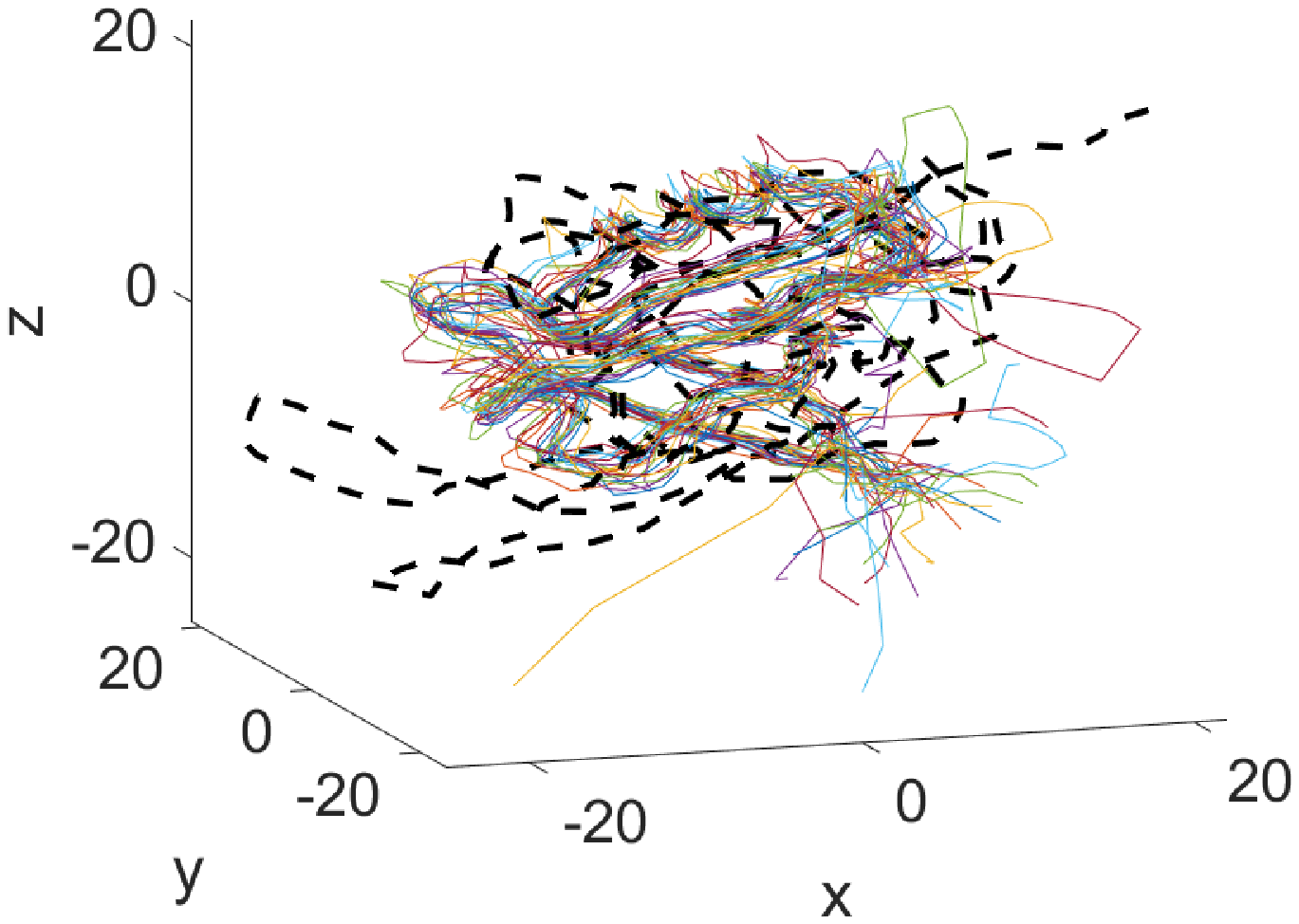}
\caption{Outliers in 3-D}
\end{subfigure}%
\begin{subfigure}{.33\textwidth}
\centering
\includegraphics[height=4.2cm]{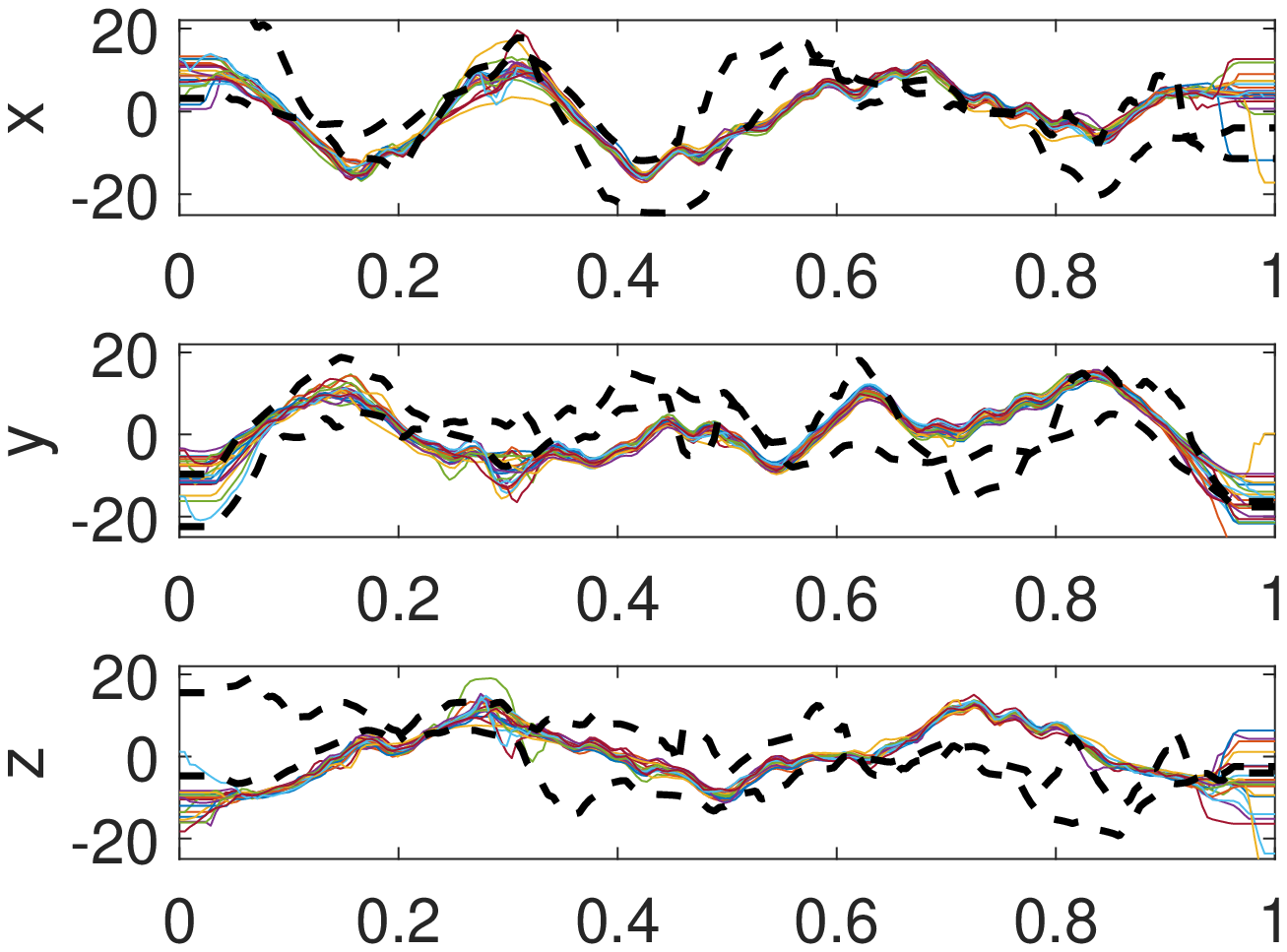}
\caption{Outlier components}
\end{subfigure}%
\caption{Real data example: {\bf (a)} 3-dimensional data of 23 PDZ domain observations. {\bf (b)} The 3-coordinate components of given observations. {\bf (c)} Depth values computed using the classical $\mathbb L^2$ norm. {\bf (d)} Depth values computed using the $\mathbb L^2$ norm on the first derivative functions. {\bf (e)} Detected outliers (black dash lines) in 3-dimension by using depth values. {\bf (f)} 3-coordinate components of the detected outliers (black dashed lines).  
}
\label{fig:RealData}
\end{figure}

 In this subsection, we apply our proposed method to detect outliers on a real dataset. The dataset is taken from the SCOP database \citep{murzin1995scop}. We take the subset of proteins with sample size 23 from PDZ domain using PISCES server \citep{wang2003pisces}. The data have been pre-processed as described in \citep{wu2013efficient}, and we get normalized data where the three componenets are properly rotated and aligned. This given data are shown in Figure \ref{fig:RealData}(a) as 3-dimensional curves and the three coponents are shown in Figure \ref{fig:RealData}(b).
 
This given data has been applied with two different norms; one is the classical $\mathbb L^2$ norm on 3-dimensional functions and the other is the $\mathbb L^2$ norm on the first derivative functions.  The depth values computed by these two different norms are shown in Fig \ref{fig:RealData}(c) and (d), respectively.  We note that the depth results are vey close to each for the two norms -- both methods indicate that the 8th and 12th protein sequences are outliers in our dataset by using a detection threshold $\alpha=0.05$.  The two outliers are shown in Figure \ref{fig:RealData}(e) and (f) as 3-dimensional curves and for 3 coordinate components, respectively. It is apparent that the depth values successfully detect the outliers in the given data. 

\begin{figure}[ht!]
\captionsetup[subfigure]{}

\begin{subfigure}{.25\textwidth}
\centering
\includegraphics[height=3.2cm]{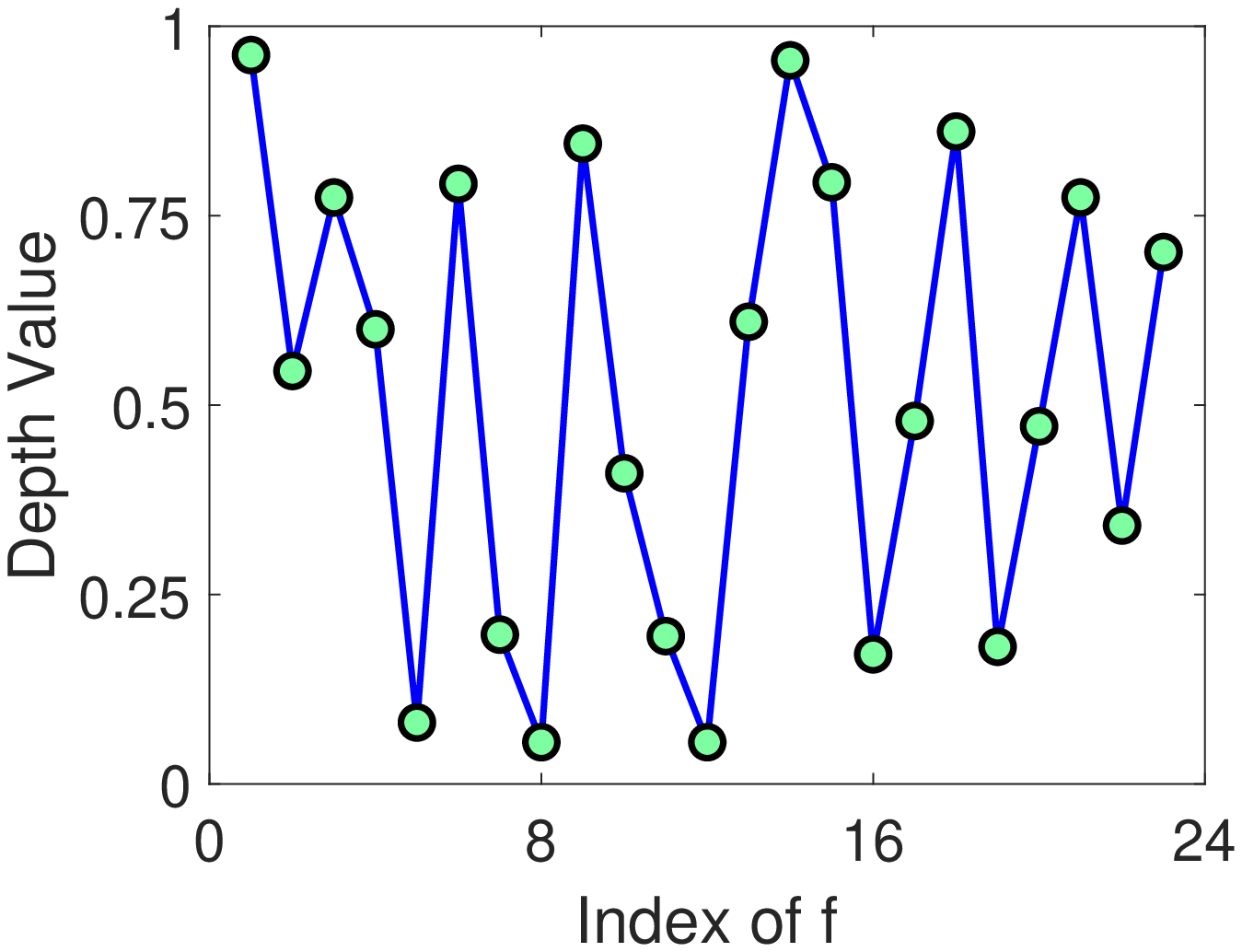}
\caption{RKHS norm}
\end{subfigure}%
\begin{subfigure}{.25\textwidth}
\centering
\includegraphics[height=3.2cm]{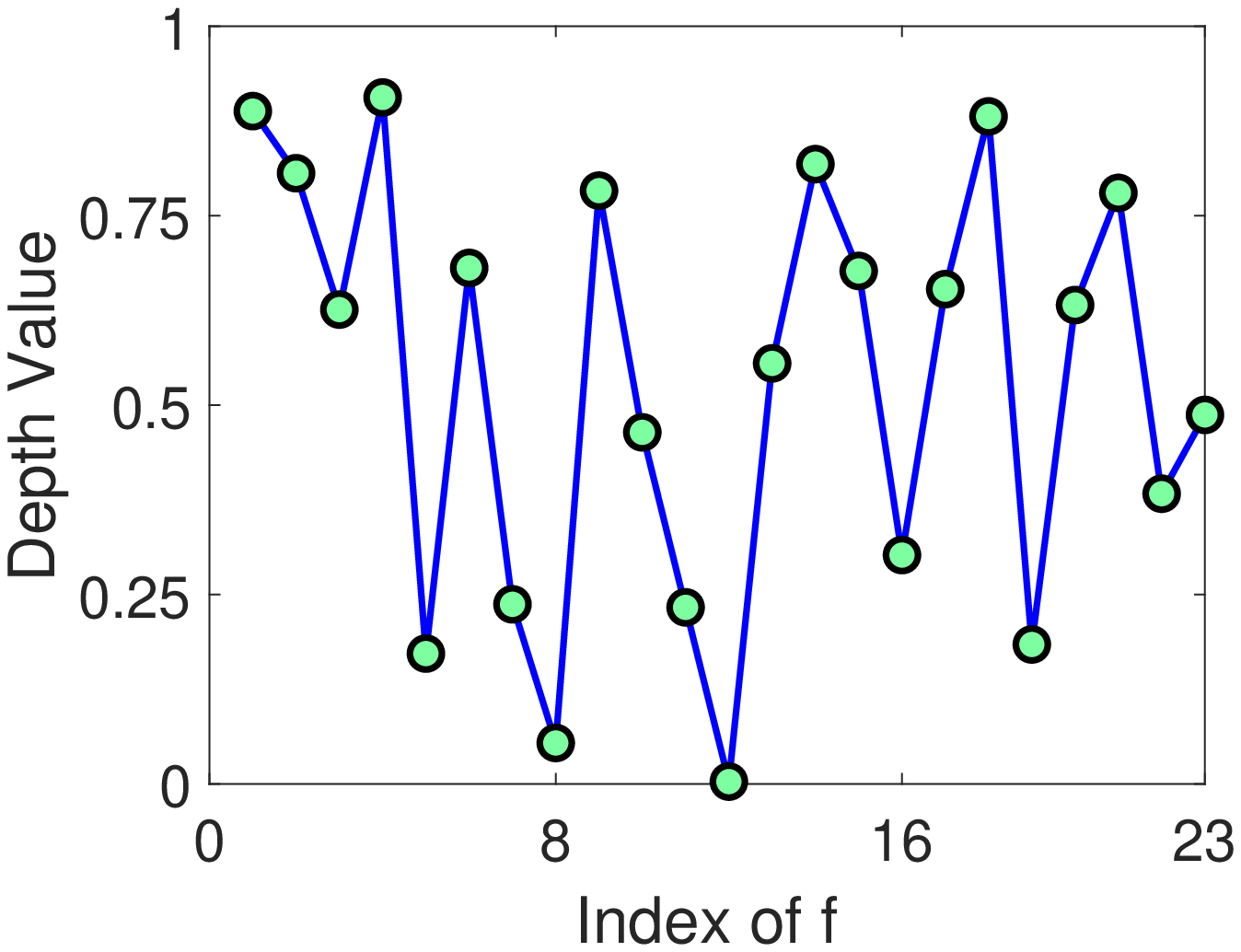}
\caption{Modified RKHS}
\end{subfigure}%
\begin{subfigure}{.25\textwidth}
\centering
\includegraphics[height=3.2cm]{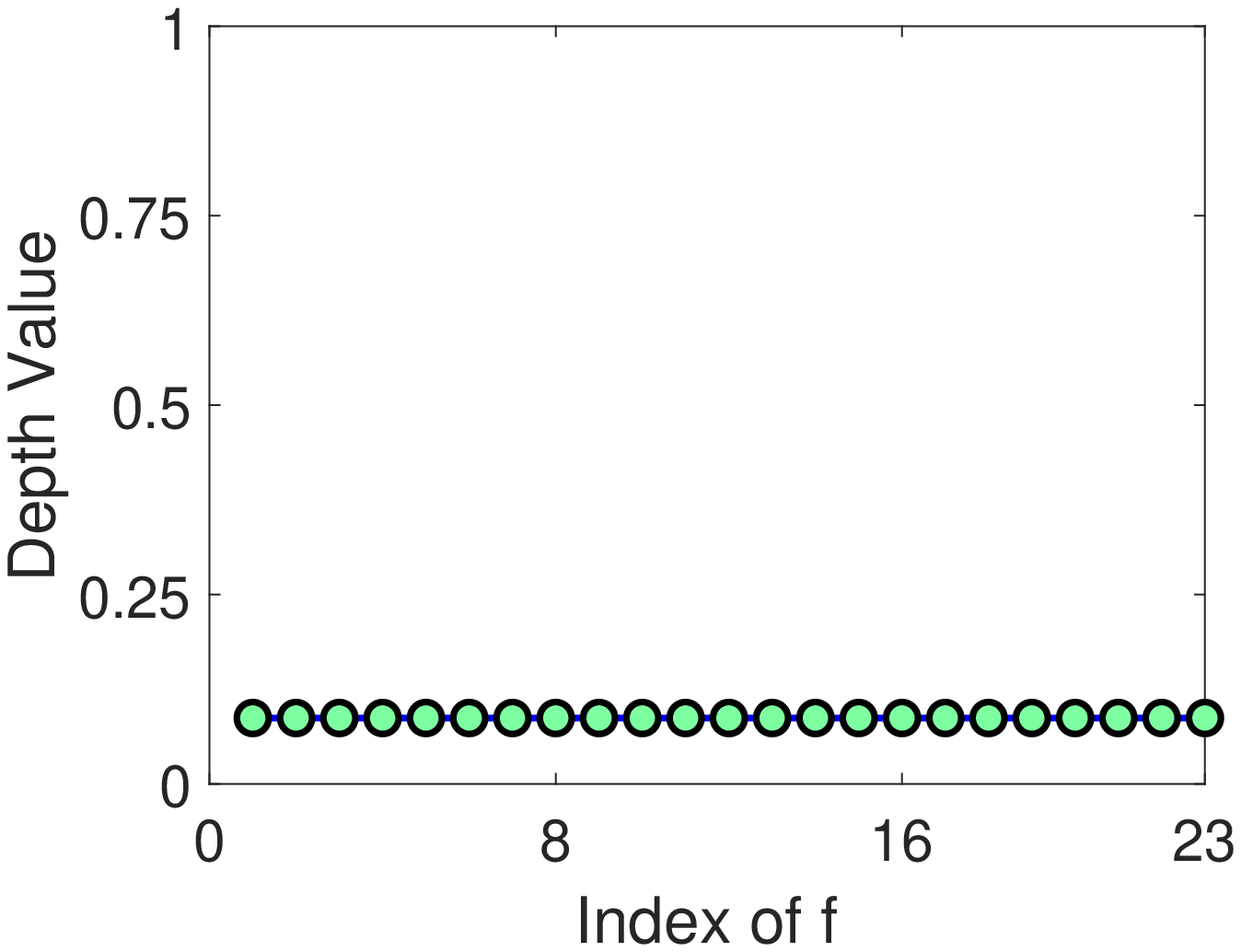}
\caption{Band depth}
\end{subfigure}%
\begin{subfigure}{.25\textwidth}
\centering
\includegraphics[height=3.2cm]{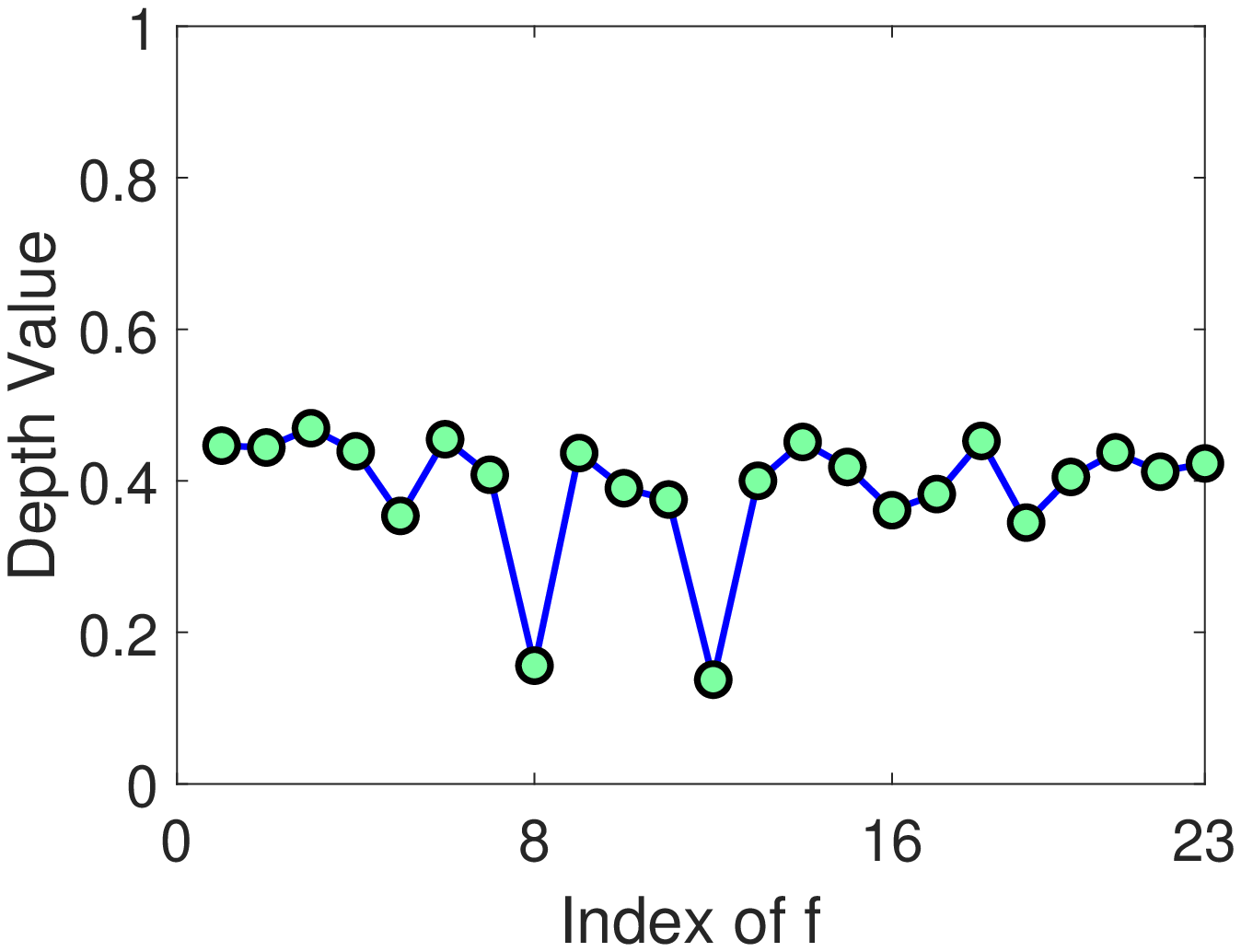}
\caption{Modified band depth}
\end{subfigure}%
\caption{Comparison: {\bf (a)} Depth values using RKHS induced norm. {\bf (b)} Depth values using modified RKHS norm with $a_p = 1/p$. {\bf (c)} Depth values computed using band depth method.  {\bf (d)} Depth values computed using modified band depth method.}
\label{fig:RealData_p2}
\end{figure}

For comparison, the depth values obtained by the RKHS norm in our framework are shown in Figure \ref{fig:RealData_p2}(a), where a lot of functions have low depth values and the two outliers cannot  be clearly identified.  Figure \ref{fig:RealData_p2}(b) shows the depth values computed by modified norm with $a_p = 1/p$ in our model based depth framework. It is clearly to see that the 8th and 12th functions have lowest depth values, though not as close to 0 as the two $\mathbb L^2$ norms in Figure \ref{fig:RealData}.  

In addition, we compare our approach to the well-known depth methods -- band depth and its modified version \citep{lopez2009concept}. The performance of band depth shown in Figure \ref{fig:RealData_p2}(c) is very poor, and it implies that there are great difficulties in applying the band depth due to large variation in the three coordinates. On the other hand, the result in Figure \ref{fig:RealData_p2}(d) shows that depth values obtained by modified band depth have a clear large gap between the two outliers and main portion of the data, consistent to the result in Figure \ref{fig:RealData}, whereas the depth values are distributed in a very narrow range.

\section{Summary and Future works}

In this article, we have proposed a new framework to define model-based statistical depth for functional as well as multivariate observations. Our definitions have two forms: norm-based and inner-product-based. Depending on the selection (of norms), the norm-based depth can have various center-outward ranks. For the inner-product depth, it is mainly the generalization of the multivariate halfspace depth. We then focus on using norms which are naturally defined with the generative model. That is, we use induced RKHS norm from the finite-dimensional covariance kernel in a second-ordered stochastic process. For an infinite-dimensional kernel, we have introduced a modified version to avoid the infinity value on the induced norm. For practical use, we propose efficient algorithms to compute the proposed depths. Through simulations and real data, we demonstrate the proposed depths reveal important statistical properties of given observations, such as median and quantiles. Furthermore, we establish the consistency theory on the estimators. 

Statistical depth is an extensively-studied area. However, all previous methods are either procedure-based or properties-based. To the best of our knowledge, this is the first model-based investigation. This paper introduced the basic framework, but the model is limited to covariance-based method. Due to the nature of covariance kernel, our framework have a tendency to deal with second-order stochastic process like Gaussian family well. We plan to work in a space where higher order statistics can also be important in the future. In addition, we have discussed four important properties for the proposed depths. As \cite{gijbels2017general} provided an elaboration on more desirable properties (such as receptivity and continuity) of statistical depths for functional data, our future work is to investigate whether our proposed framework would meet those properties.  Moveover, since we obtain median and quantiles by the proposed depths, we can also extend our method to construct boxplot visualization. Last but not least, we are seeking broader applications of the new framework in real world problems such as clustering, classification, and outlier detection.

%
%

\setcounter{theorem}{3}
\newpage

\bigskip
  \bigskip
  \bigskip
  \begin{center}
    {\LARGE\bf Supplement to Model-based Statistical Depth with Applications to Functional Data}
\end{center}
  \medskip

\spacingset{1.5}

\appendix

\section{Algorithms for general model-based depth}\label{sec:more_alg}

Suppose we have $n$ zero-mean independent sample functions $f_1, \cdots, f_n \in \mathcal F$ on $t \in [0,1]$, and our goal is to compute the model-based depth of any observed sample $f_{obs} \in \mathcal{F}$.   We first describe the \emph{norm-based depth} estimation algorithm as follows: 

\textbf{Algorithm II.}   (Input: observations $\{f_1, \cdots, f_n\}$, any observation $f_{obs}$, a threshold $\epsilon >0$, the center function $f_c$, and the selected norm $\|\cdot\|$, which means $\zeta(f,f_c)=\|f-f_c\|$ for any observation $f$.)  
\begin{enumerate}
    \item Compute the sample mean function $\bar {f}(t) = \frac{1}{n}\sum_{i=1}^n f_i(t)$, and empirical covariance kernel $\hat{K}(s,t) = \frac{1}{n}\sum_{i=1}^n [f_i(s)-\bar{f}(s)][f_i(t)-\bar{f}(t)]$;
    \item Eigen-decompose $\hat{K} = \sum_{p=1}^n \hat{\lambda}_{p,n} \hat{\phi}_{p,n}(s) \hat{\phi}_{p,n}(t)$;
    \item Choose a number $P$ if $\hat{\lambda}_{P+1}$ is the first eigenvalue such that $\hat{\lambda}_{P+1} < \epsilon$ for sufficiently small $\epsilon$; then $\hat{K}(s,t) = \sum_{p=1}^P \hat{\lambda}_{p,n} \hat{\phi}_{p,n}(s)\hat{\phi}_{p,n}(t)$;
    \item Compute $\hat{f}_{i,p} = \int_0^1 f_i(t) \hat{\phi}_{p,n}(t) dt$ for all $i = 1,\cdots,n$ and $p = 1,\cdots,P$, and compute $\hat{f}_{p} = \int_0^1 f_{obs}(t) \hat{\phi}_{p,n}(t) dt$;
    \item Re-sample (with replacement) a large number $N$ of coefficients $\{\hat{g}_{j,p}\}_{j=1}^N$ based on the coefficients of $\{\hat{f}_{i,p}\}$, and construct $g_j(t) = \sum_{p=1}^P \hat{g}_{j,p} \hat{\phi_p}(t)$;
    \item Estimate the sample depth of $f_{obs}$ w.r.t. $\{g_j\}$:
        \begin{align*}
            D_n(f_{obs};\{g_j\}_{j=1}^N) = \frac{1}{N}  \sum_{j=1}^N\textbf 1_{(\zeta(f_{obs},f_c) \leq \zeta( g_j , f_c))} = \frac{1}{N}  \sum_{j=1}^N\textbf 1_{(\| f_{obs} - f_c\| \leq \| g_j - f_c\|)},
        \end{align*} 
        where $\textbf 1_{(\cdot)}$ is the indicator function.
\end{enumerate}

Steps 1-4 are the first part in the algorithm. They aim to estimate the eigen-system of the covariance kernel via given observations. In particular,  the Karhunen Lo\`eve expansion \citep{ash1990information} is used in Step 2 to decompose the covariance kernel, and offer a method to reconstruct samples (the background on the Karhunen Lo\`eve expansion will be provided in Section \ref{sec:MBD}). Using a functional principal component analysis \citep{ramsay2005functional}, we retain the eigen-functions which explain meaningful variance in our system. Steps 5-6 are the second part of the algorithm. They estimate the depth value with the given norm, where we need re-sampling techniques and Monte Carlo approximations. This algorithm can be easily adapted to the multivariate data. In such case, the dimension of the data is already given and the principal component analysis and the multivariate metric can be directly applied. 

In general, computing a halfspace depth in $\mathbb R^d$ is a very challenging task.  So far, exact computations can be given only when $d=2$ \citep{rousseeuw1996algorithm} and $d=3$ \citep{rousseeuw1998computing}.  There are approximation algorithms when $d \geq 4$ \citep{zuo2018new}.  However, if the data distribution is a multivariate normal, our framework will result in an optimal solution similar to that obtained for the Gaussian process. 
For infinite dimensional GP, Lemma \ref{lem:ip} shows that 
the inner-produt-based depth can only be feasible for finite-dimensional space. Fortunately, when the random samples are from a finite-dimensional zero-mean Gaussian process, the depth has simple closed-form (see detail in Appendix A).  We adopt this special case and modify the above algorithm for halfspace depth as follows, where Steps 4-6 are simplified as follows: 
\begin{enumerate}
 \setcounter{enumi}{3}
        \item Compute $\hat{f}_{p} = \int_0^1 f_{obs}(t) \hat{\phi}_{p,n}(t) dt$ for $p = 1,\cdots,P$;
        \item Compute the induced RKHS norm $\| f_{obs} \|_{\mathbb H_{\hat{K}}}^2 = \sum_{p=1}^P \frac{\hat{f}_{p}^2}{\hat{\lambda}_{p,n}}$;
        \item Compute the depth as $ D_{ip}(f_{obs}) = 1 - \Phi(||f_{obs}||_{\mathbb H_{\hat{K}}})$, where $\Phi(x)$ denotes the cumulative distribution function of a standard normal random variable.
\end{enumerate}

\section{Applications of norm-based depth in finite-dimensional Process}
\label{sec:FDGP}
Finite-dimensional process is a commonly used stochastic process in practical applications. In particular, this include any finite-dimensional Gaussian Process (GP) and multivariate Gaussian distribution as special cases.
In this appendix, we simplify our model in Section~\ref{subsec:modified} into a zero-mean finite-dimensional process, which means that $K$ has a finite number $P$ of positive eigenvalues, and $P$ will be referred as the dimension of this process.  That is, $K(s,t) = \sum_{j=1}^P \lambda_j\phi_j(s)\phi_j(t)$. For convenience we denote this kernel as $K = K_P$.  One important benefit in this process is that the associated RKHS norm is always finite and can be directly used in our construction of norm-induced depth as described in Section~\ref{sec:examples}.

\subsection{RKHS norm induced depth for finite-dimensional process}\label{sec:depth_finite}

Suppose we have functional observation $f \in \mathbb L^2([0,1])$ from a zero-mean stochastic process with covariance kernel $K_P(s,t) = \sum_{p=1}^P \lambda_p \phi_p(s)\phi_p(t)$ on $[0, 1] \times [0, 1]$.  Then $f(t) = \sum_{p = 1}^P f_p \phi_p(t)$, where $f_1, \cdots, f_P$ are uncorrelated and $E(f_p) = 0, E(f_p^2) = \lambda_p, j = 1, \cdots, P $.  In particular, when the process is a Gaussian process, $f_1, \cdots, f_P$ are independent.  In this case, $\{X_p = f_p/\sqrt{\lambda_p}\}_{p=1}^P$ are i.i.d. samples from a standard normal distribution, and the squared induced norm $\| f \|_{\mathbb  H_K}^2 = \sum_{p=1}^P f_p^2/\lambda_p = \sum_{p=1}^P X_p^2$ follows a $\chi^2$ distribution with $P$ degrees of freedom, denoted as  $\chi^2(P)$. 

The computation of depth still depends on Definition 2: $ D(f_{obs},\, \mathbb P_{\theta}, \|\cdot\|, f_c) = \mathbb P_{\theta} \big [ \zeta(f,\,f_c) \geq \zeta(f_{obs},\, f_c)\big]$. The central function $f_c = 0$ is the mean function in our model; the criterion function $\zeta(f,\,g) = \| f -g\|_{H_K}$; $\mathbb P_{\theta}$ is a probability measure. 
We can now rewrite the definition of depth in the following form:
\begin{align}
   D(f_{obs},\, \mathbb P_{P}, \|\cdot\|_{\mathbb H_{K}}, 0)  &= \mathbb P_{P} \big [f: \|f\|_{\mathbb H_{K}} \geq \|f_{obs}\|_{\mathbb H_{K}} \big] \nonumber 
   \\ &= 1 - \mathbb P_{P} \big [f: \|f\|_{\mathbb H_{K}}^2 \leq \|f_{obs}\|_{\mathbb H_{K}}^2 \big] = 1 - F(\|f_{obs}\|_{\mathbb H_{K}}^2),
   \label{eq:chi} 
\end{align}
where $F(\cdot)$ denotes the cumulative distribution function (c.d.f.) of $\|f\|_{\mathbb H_{K}}^2$.  In the case of Gaussian process, this is a c.d.f. of  $\chi^2(P)$. Moreover, for any $\alpha \in [0,1]$, the $\alpha$-th depth contour is rewritten as
\begin{align*}
C(\alpha,\, \mathbb P_{P}, \|\cdot\|_{\mathbb H_{K}}, 0)  = \big\{ f\in \mathcal F:\, F(\|f_{obs}\|_{\mathbb H_{K}}^2) = 1 - \alpha\big\},
\end{align*}
and central region for this model is 
\begin{align*}
    R(\alpha,\, \mathbb P_{P}, \|\cdot\|_{\mathbb H_{K}}, 0) &= \big\{ f\in \mathcal F:\, F(\|f_{obs}\|_{\mathbb H_{K}}^2) \leq 1- \alpha\big\}.
\end{align*}

\begin{figure}[ht]
\captionsetup[subfigure]{}
\begin{subfigure}{.5\textwidth}
\centering
\includegraphics[height=4.5cm]{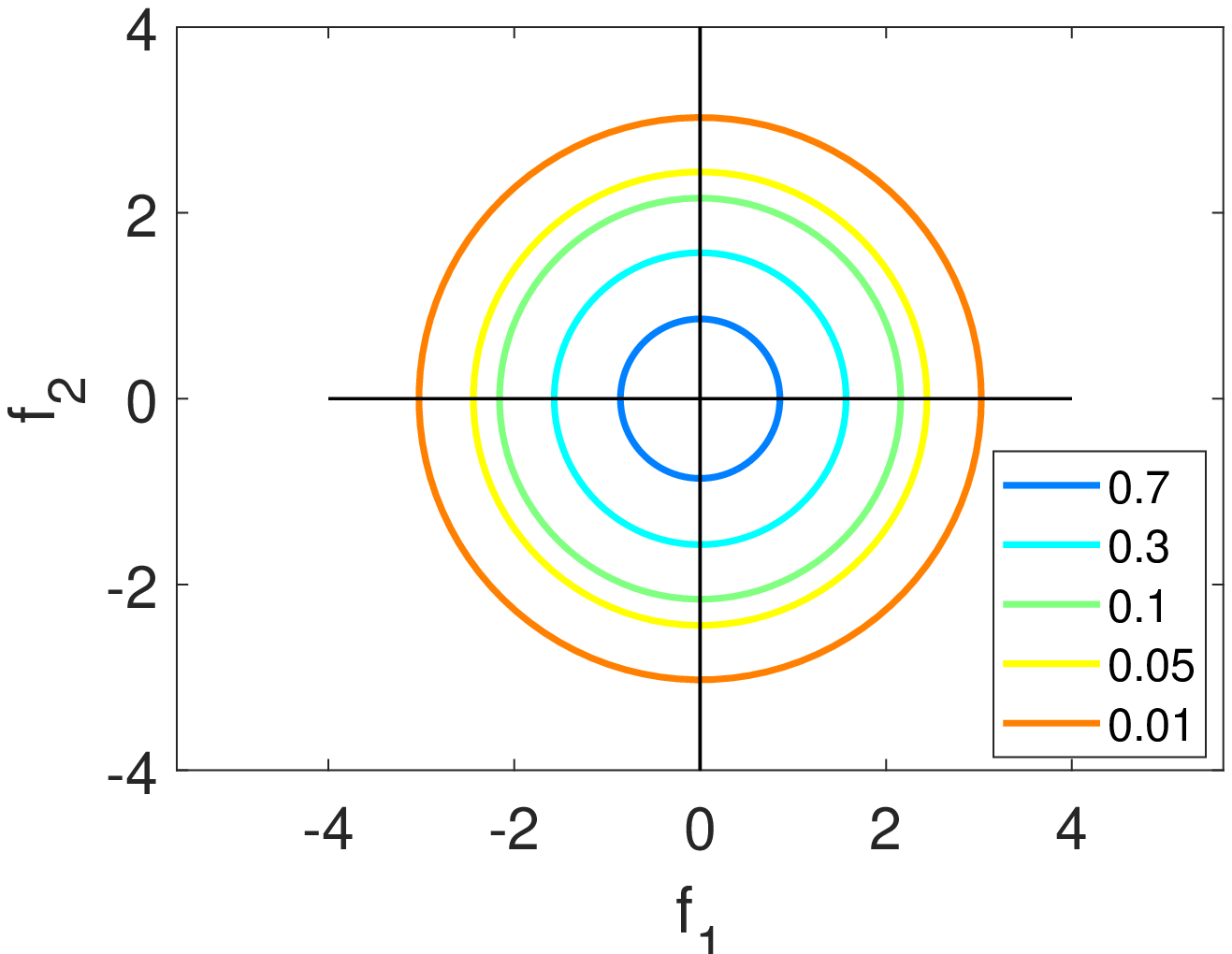}
\caption{$\Sigma = I_2$}
\end{subfigure}%
\begin{subfigure}{.5\textwidth}
\centering
\includegraphics[height=4.5cm]{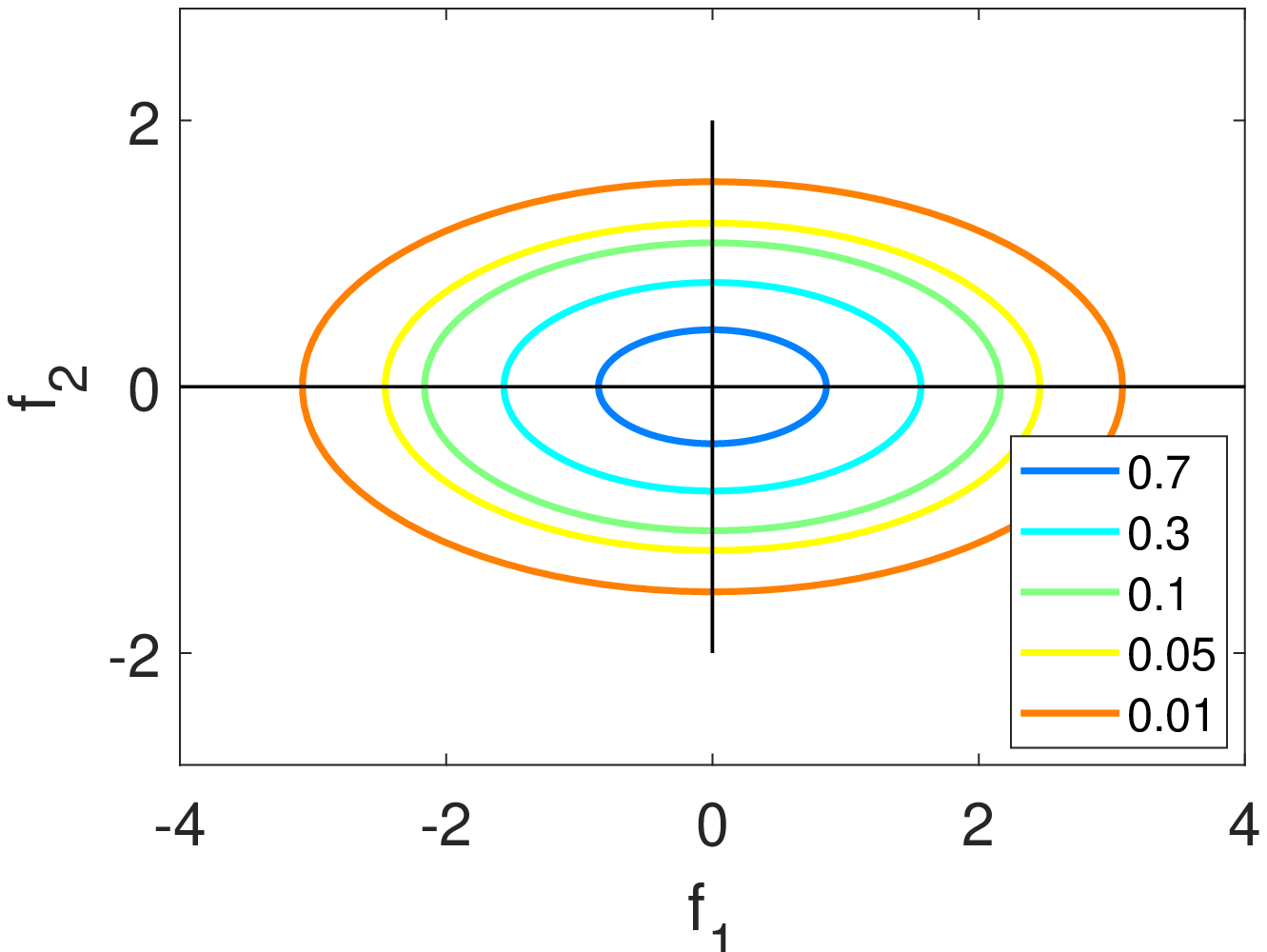}
\caption{$\Sigma$ = $diag(1, 0.25)$}
\end{subfigure}%
\caption{Illustration depth contours in the 2-D case where x-axis and y-axis represents the values of $f_1$ and $f_2$, respectively.  {\bf (a)} Depth contours at different levels for zero mean Gaussian Process with $(f_1, f_2) \sim N(0, I_2)$.  {\bf (b)} Same as (a) except that $(f_1, f_2) \sim diag(1, 0.25)$. }
\label{fig:DepCon}
\end{figure}

Based on the above derivation, it is easy to see that the depth contours defined via induced RKHS norm on a Gaussian process are $P$-dimensional ellipsoids, and the center of all ellipsoids is the origin in $\mathbb R^P$. For illustrative purpose, we let $P=2$ and $(f_1,f_2) \sim \mathcal N(0,\Sigma)$, with $\Sigma = diag(\lambda_1, \lambda_2)$. For any random samples $f(t) = \sum_{j = 1}^2 f_p \phi_p(t)$, we could use a point $(f_1,f_2) \in \mathbb R^2$ to represent random function $f(t)$, because the coefficients set for each $f(t)$ is unique with respect to the eigen-functions basis. 
In Figure \ref{fig:DepCon}, if we have any $(f_1,f_2)$ locating on the same ellipsoid, their corresponding random observations will have the same depth defined by the induced RKHS norm. In particular, when $\Sigma = I_2$, the depth contours are concentric circles.   Moreover, any random observations $f(t)$, whose coefficients $(f_1,f_2)$ locates inside of $\alpha$-th contour, will have a larger depth than $\alpha$. 

\subsection{Depth estimation procedure and algorithm}
\label{sec:algGP}
Similar to the infinite-dimensional case, we can derive algorithm to compute depth on a finite-dimensional stochastic process.  
Suppose we have $n$ independent random sample functions $\{f_1, \cdots, f_n \} \subseteq \mathcal F$ on $t \in [0,1]$, and $\mathcal{F}$ is a zero-mean $P$ dimensional stochastic Process. The following algorithm is to compute the depth based on $\mathcal{F}$ of any observed sample $f_{obs} \in \mathcal{F}$.  In practice when $P$ is unknown, we can set a small threshold $\epsilon$ to identify it. 

\textbf{Algorithm III.}  (Input: functional data $\{f_1, \cdots, f_n\}$, any observation $f_{obs}$,  and a threshold $\epsilon >0$.)  
    \begin{enumerate}
        \item Compute the sample mean function $\bar {f}(t) = \frac{1}{n}\sum_{i=1}^n f_i(t)$, and empirical covariance kernel $\hat{K}(s,t) = \frac{1}{n}\sum_{i=1}^n [f_i(s)-\bar{f}(s)][f_i(t)-\bar{f}(t)]$;
        \item Eigen-decompose $\hat{K} = \sum_{p=1}^n \hat{\lambda}_{p,n} \hat{\phi}_{p,n}(s) \hat{\phi}_{p,n}(t)$;
        \item Choose a number $P$ if $\hat{\lambda}_{P+1}$ is the first eigenvalue such that $\hat{\lambda}_{P+1} < \epsilon$ for sufficiently small $\epsilon$; then $\hat{K}(s,t) = \sum_{p=1}^P \hat{\lambda}_{p,n} \hat{\phi}_{p,n}(s)\hat{\phi}_{p,n}(t)$;
        \item Compute $\hat{f}_{i,p} = \int_0^1 f_i(t) \hat{\phi}_{p,n}(t) dt$ for all $i = 1,\cdots,n$ and $p = 1,\cdots,P$, and compute $\hat{f}_{p} = \int_0^1 f_{obs}(t) \hat{\phi}_{p,n}(t) dt$;
        \item For each $p \in {1,\cdots,P}$, re-sample (with replacement) a large number $N$ of coefficients $\{\hat{g}_{j,p}\}_{j=1}^N$ based on $\{ \hat{f}_{1,p}, \cdots, \hat{f}_{n,p} \}$;
        \item Construct $g_j(t) = \sum_{p=1}^{P} \hat{g}_{j,p} \hat{\phi}_{p,n}(t)$;
        \item Compute $||f_{obs}||_{\mathbb H_{\hat{K}}}^2 = \sum_{p = 1}^{P} \frac{ \hat{f}_{p}^2}{\hat{\lambda}_{p,n}}$, and $||g_j||_{\mathbb H_{\hat{K}}}^2 = \sum_{p = 1}^{P} \frac{ \hat{g}_{j,p}^{2}}{\hat{\lambda}_{p,n}}$;
        \item Estimate the depth of $f_{obs}$ using $\{g_j\}$:
            \begin{align*}
                D(f_{obs};\{g_j\}_{j=1}^N) = \frac{1}{N}  \sum_{j=1}^N 1_{\|f_{obs} \|_{\mathbb H_{\hat{K}}}^2 \leq \|g_j \|_{\mathbb H_{\hat{K}}}^2}.
            \end{align*}  
    \end{enumerate}
    

This algorithm is very similar to Algorithm II. The first 3 steps are to estimate the eigen-system of the covariance kernel via our observations.  As there are only finite number $P$ of positive eigenvalues, we can set small threshold to estimate $P$.  Steps 4-8 are to estimate the modified RKHS norm by resampling based on the eigen-decomposition on the covariance. 

An important special case is when the process is a Gaussian process.  In this case, we have pointed out the squared norm $\| f \|_{\mathbb H_{\hat{K}}}^2$ has a Chi-square distribution.  Therefore resampling will not be needed and the estimation of depth will be more robust and efficient.  Steps 4-8 can be simplified and modified to the following 3 steps: 
\begin{enumerate}
 \setcounter{enumi}{3}
        \item Compute $\hat{f}_{p} = \int_0^1 f_{obs}(t) \hat{\phi}_{p,n}(t) dt$ for all $i = 1,\cdots,n$ and $p = 1,\cdots,P$;
        \item Compute the induced RKHS norm $\| f_{obs} \|_{\mathbb H_{\hat{K}}}^2 = \sum_{p=1}^P \frac{\hat{f}_{p}^2}{\hat{\lambda}_{p,n}}$;
        \item Compute the depth as $ D = 1 - F(||f_{obs}||_{\mathbb H_{\hat{K}}}^2)$, where $F(x)$ denotes the cumulative distribution function of $\chi^2(P)$.
\end{enumerate}

\section{More simulation examples}\label{sec:more_sims}
\noindent \textbf{Simulation 5.} In this example, we illustrate the inner-product criterion in depth computation.  
We first select a sequence of orthonormal Fourier basis functions up to order $P=10$ on $[0,1]$ such that
\begin{align*}
\phi_p(t)
= \left\{
\begin{array}{ccl}
1  &  & p=1\\
\sqrt{2}\cos(\pi pt)  &  & p=2,4,6,8,10\\
\sqrt{2}\sin(\pi (p-1)t)  &  & p=3,5,7,9
\end{array} \right.
\end{align*}

\begin{figure}[ht]
\captionsetup[subfigure]{}
\begin{subfigure}{.5\textwidth}
\centering
\includegraphics[height=4.2cm]{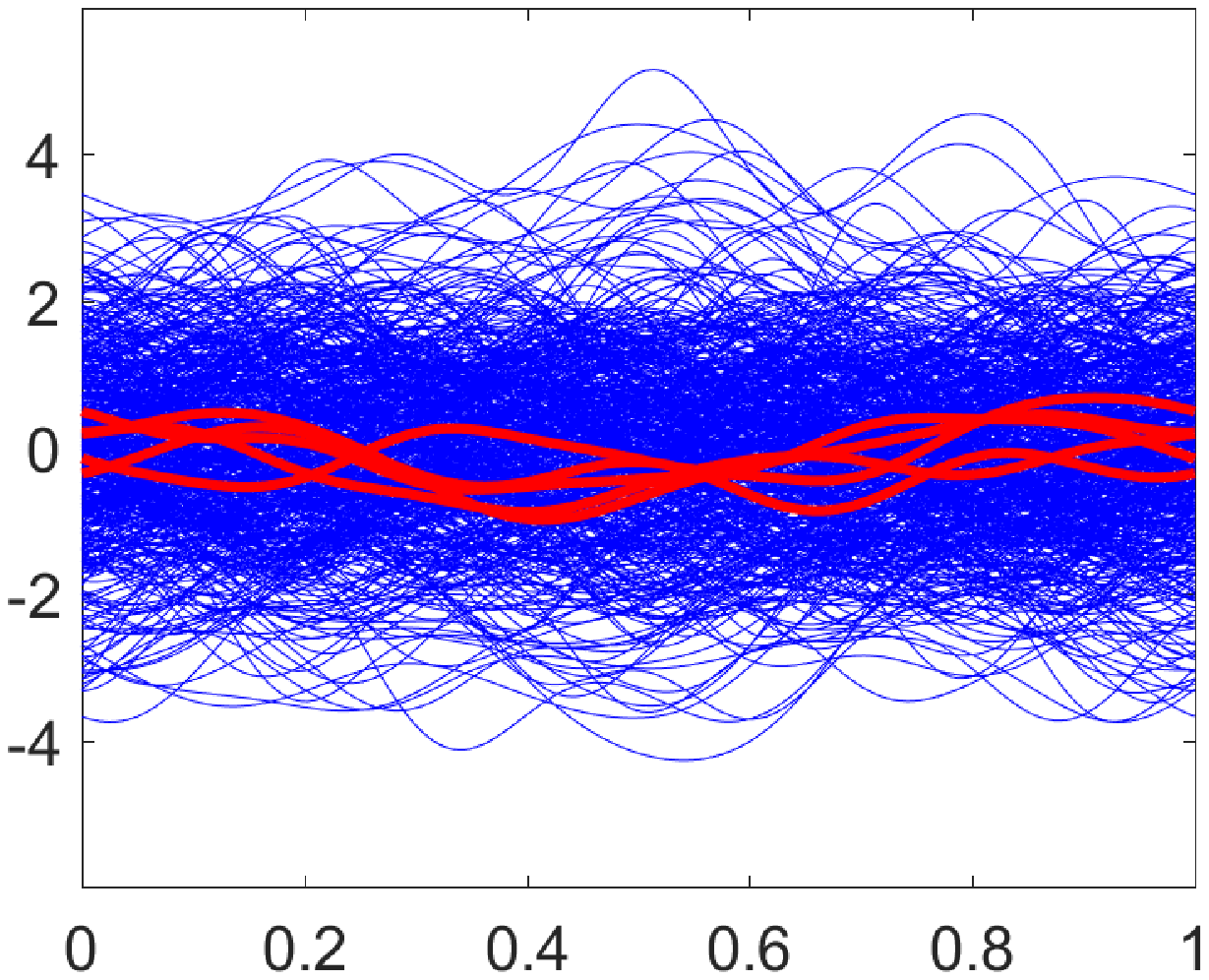}
\caption{Inner-product-based Depth}
\end{subfigure}%
\begin{subfigure}{.5\textwidth}
\centering
\includegraphics[height=4.2cm]{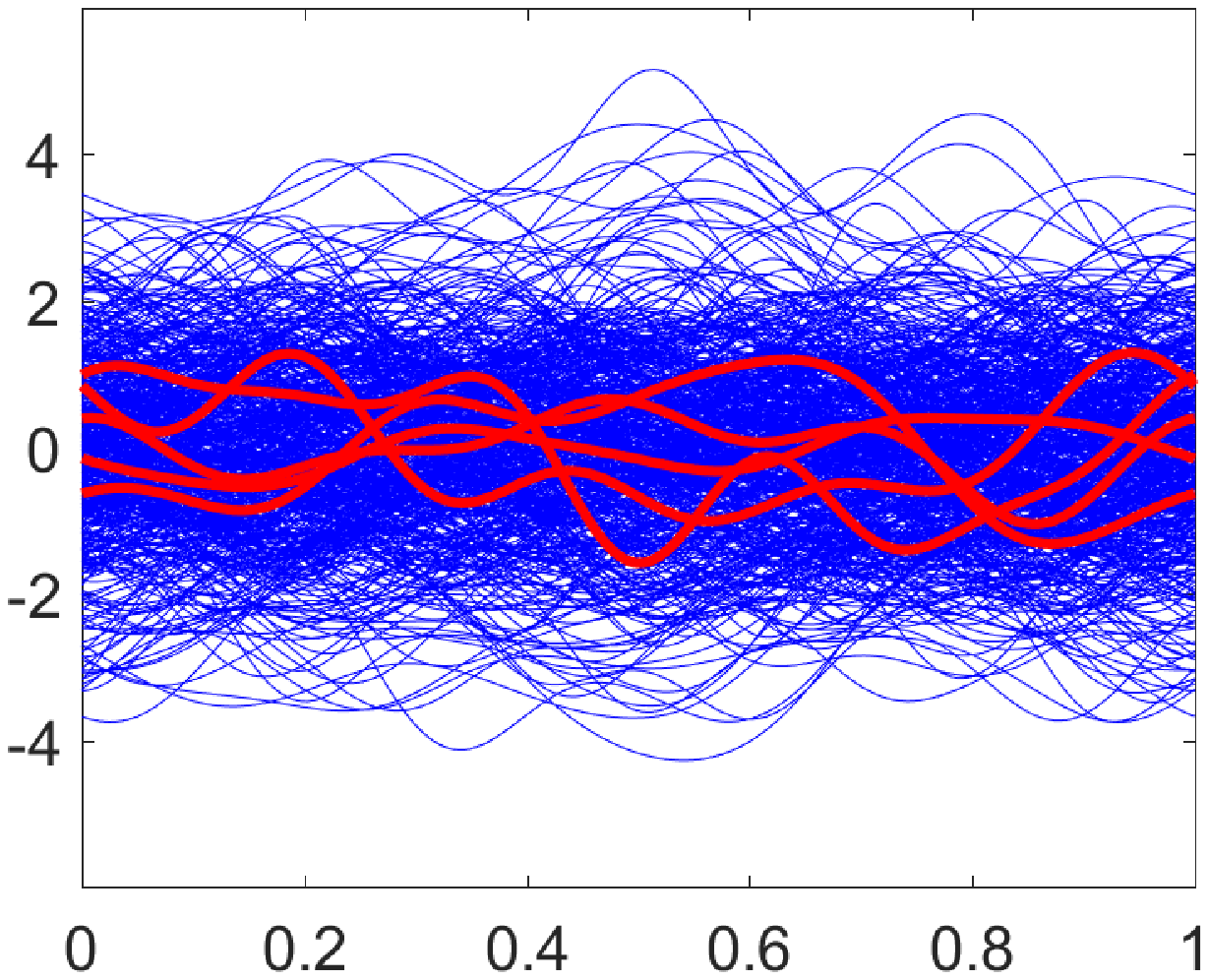}
\caption{Modified Half-region Depth}
\end{subfigure}%
\caption{
Simulation 3: {\bf (a)} All 500 simulated functions, where the red curves have the five deepest values obtained by the proposed method. {\bf (b)} Same as (a) except that the depth values are obtained by the modified half-region depth. }
\label{fig:DemoHR}
\end{figure}

Next we random generate $N=500$ coefficient vectors $\{ (a_{i,1},\cdots,a_{i,10}) \}_{i=1}^N$ following a multivariate normal distribution $\mathcal{N}(0, diag(1,((P-1)/P)^2, \cdots, (1/P)^2))$. Then we generate $N$ functions via linear combination $f_i = \sum_{p=1}^P a_{i,p} \phi_p$. We apply Algorithm I for inner-product depth discussed in the above section on this simulated data. We display these 500 functions in Figure \ref{fig:DemoHR}(a), where the five deepest curves are represented in bold red.  We see that these 5 red ones stay in the middle of the sample, which illustrate the effectiveness of the depth measurement. As a comparison, we also show the result obtained by modified half-region depth \citep{lopez2011half} and the result is shown in Figure \ref{fig:DemoHR}(b).  Visually, the five deepest functions displayed in Panel (a) seem to be more centralized near x-axis, and our method provide better center-outward rank than the modified half-region depth.

\noindent \textbf{Simulation 6.} In this example, we illustrate the norm-based depth on a multivariate data set and  compare the Monte Carlo estimate with sample average (as indicated in the beginning of this section). We at first generate $n=50$ random samples in $\mathbb R^2$ from multivariate normal distribution $\mathcal{N}(\mu, \Sigma)$, where 
$$\mu= \left ( \begin{array}{c} 0 \\ 0 \end{array} \right ) \mbox{ and } \Sigma = \left ( \begin{array}{cc} 1 &1/3 \\ 1/3 & 1/4 \end{array} \right ).$$  We choose a Mahalanobis distance as the criterion function, that is, for any $x \in \mathbb R^2$, $\zeta(x,\mu) = \sqrt{(x-\mu)^T\Sigma^{-1}(x-\mu)}$. Therefore, it is straightforward to derive the closed form of the depth function $D(x) = 1 - F(\zeta(x,\mu)^2)$, where $F(\cdot)$ denotes the cumulative distribution function of chi-square distribution with 2 degrees of freedom.

\begin{figure}[htb]
\captionsetup[subfigure]{}
\begin{subfigure}{.33\textwidth}
\centering
\includegraphics[height=4.2cm]{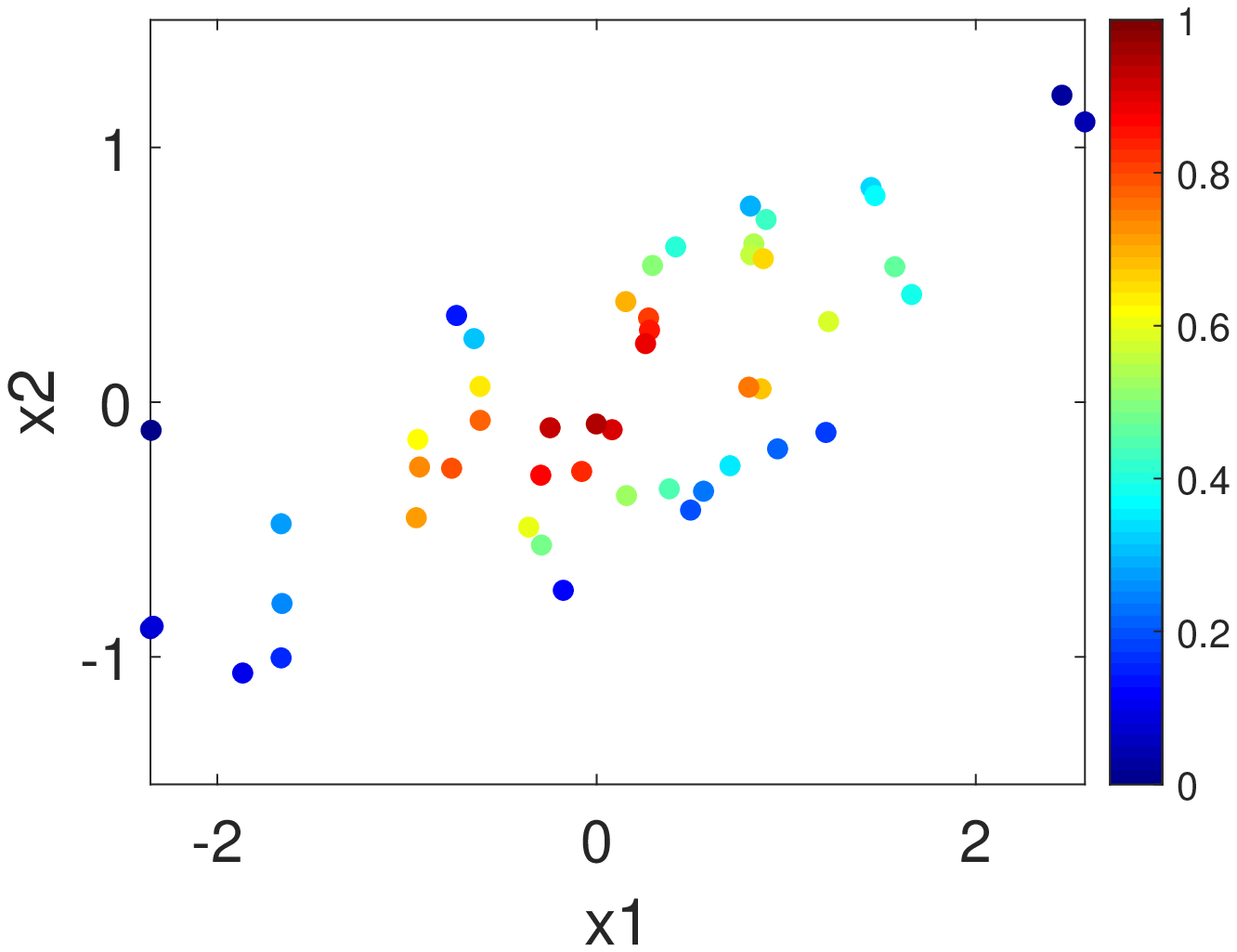}
\caption{Given observations}
\end{subfigure}%
\begin{subfigure}{.33\textwidth}
\centering
\includegraphics[height=4.2cm]{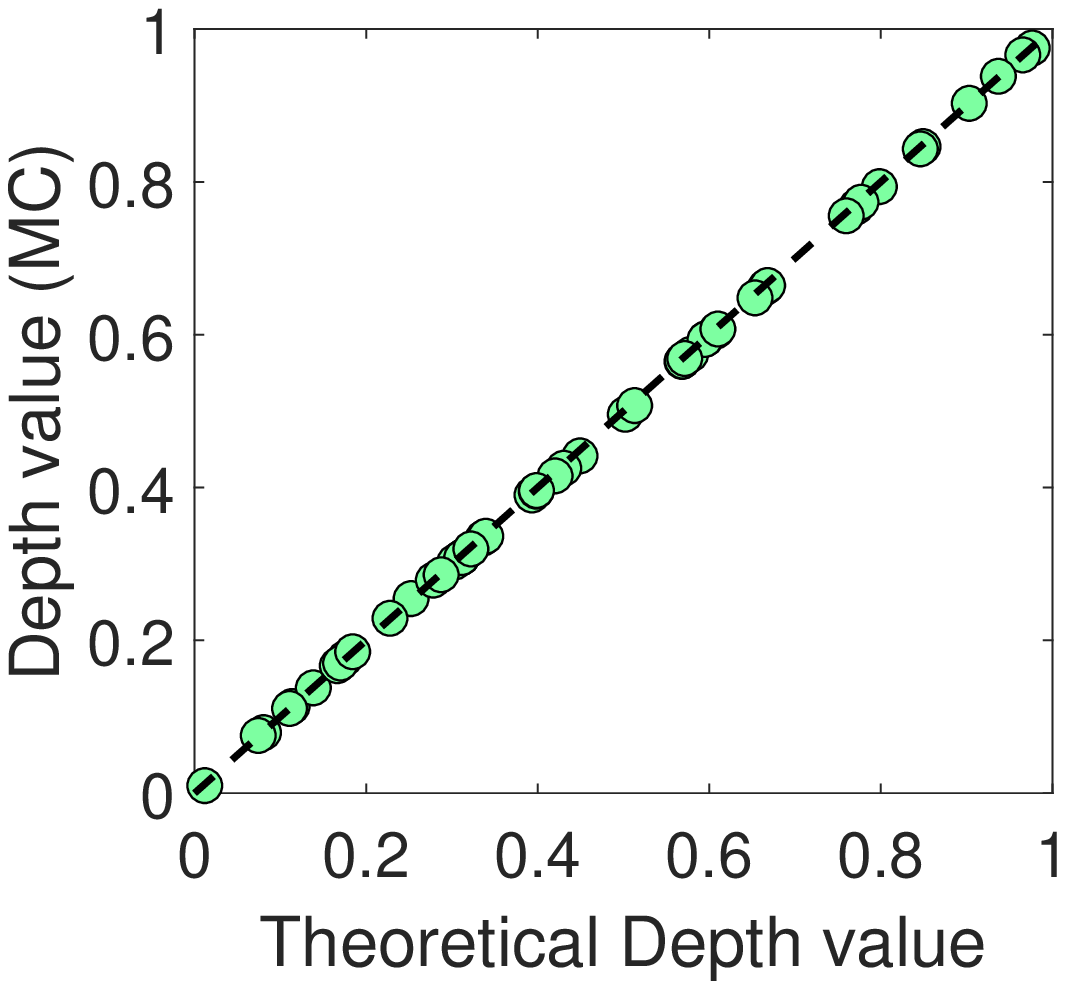}
\caption{Depth value comparison 1.}
\end{subfigure}%
\begin{subfigure}{.33\textwidth}
\centering
\includegraphics[height=4.2cm]{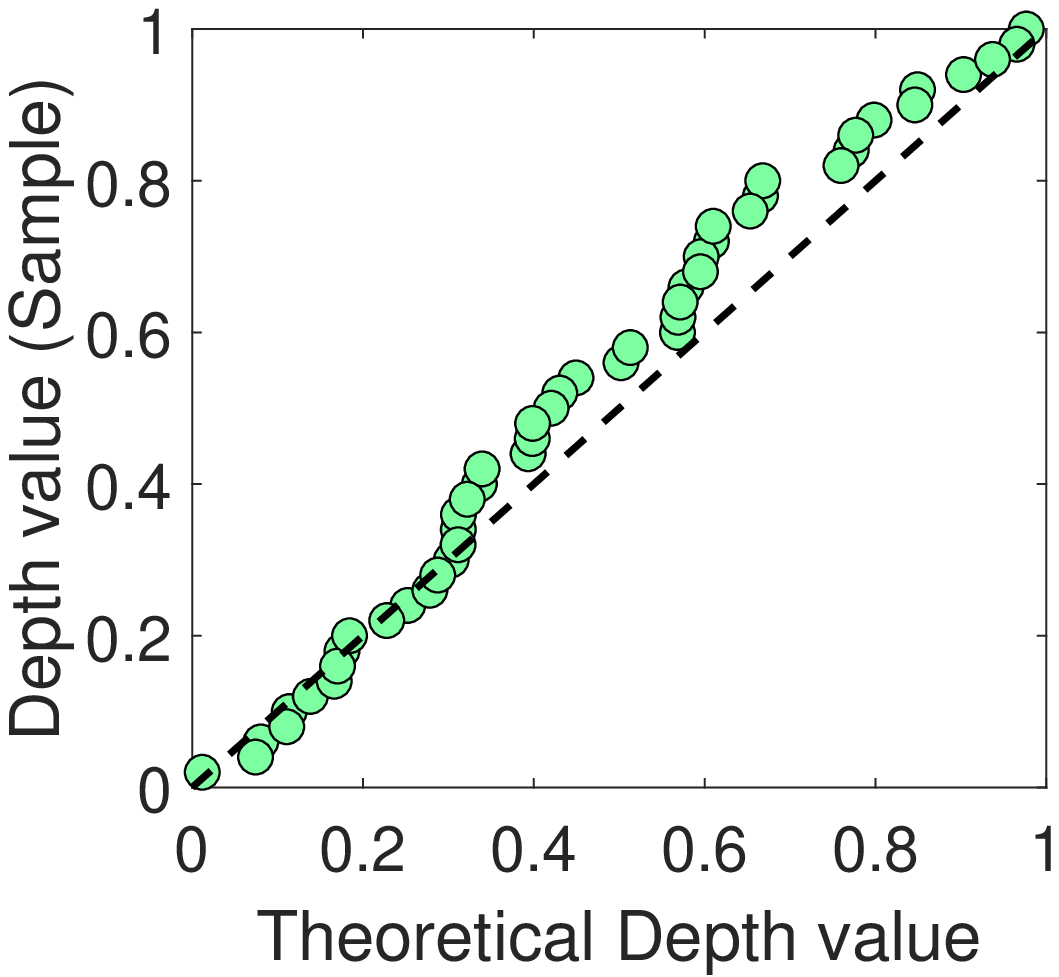}
\caption{Depth value comparison 2.}
\end{subfigure}%
\caption{Simulation 4: {\bf (a)} 50 points from multivariate normal density with color-labeled depth by Algorithm I. {\bf (b)} Depth value comparison:  closed-form depth function (x-axis) vs. Monte-Carlo-based Algorithm I (y-axis) {\bf (c)} Same as (b) except for sample average method in y-axis.}
\label{fig:DemoMul}
\end{figure}

We compute the depth value for each of these 50 points by Monte-Carlo-based Algorithm I, and then compare the result to the algorithm integrated with sample average of these points. We display these 50 points with color label of their depth values in Figure \ref{fig:DemoMul}(a). Note that the depth value using the Mahalanobis distance criterion ranges from 0 to 1 and the distirbution of these depth values approximately follow elliptic contours for a two dimensional normal distribution. Since we obtain the closed-form depth values, we can use them to compare the performance of Monte Carlo and sample average method. In Algorithm I, we generate 5000 re-sampling points in step 5. The results in Figure \ref{fig:DemoMul} (b)(c) show that the depth values computed by Algorithm I are very close to the theoretical ones, whereas the sample average method does not have the same level of accuracy.

\section{Depth estimation consistency in finite-dimensional data}\label{sec:consistency}
In this case, $\exists P \in \mathbb{N}$ such that $\lambda_P >0$ and $\lambda_p = 0,\ \forall\ p >P$ under the notation setup in Section 4 of the main paper. Then $K(s,t)=\sum_{p=1}^P \lambda_p \phi_p(s)\phi_p(t)$ and $\hat{K}(s,t) = \sum_{p=1}^n \hat{\lambda}_{p,n} \hat{\phi}_{p,n}(s) \hat{\phi}_{p,n}(t)$. Therefore, for any $f_{obs} \in \mathcal{F}$, we have the squared RKHS induced norm
\begin{equation}
    \| f_{obs} \|_{\mathbb{H}_K}^2 = \sum_{p=1}^P\frac{\langle f_{obs},\phi_p \rangle^2}{\lambda_p} < \infty,
    \label{eq:pnorm2}
\end{equation}
where $\langle \cdot, \cdot \rangle$ indicates the inner product operation in RKHS with $K$ as reproducing kernel. 

Based on the RKHS norm, the depth of $f_{obs}$ is given as follows: 
\begin{equation}
   d(f_{obs}) = D_n(f_{obs},\, \mathbb P, \|\cdot\|_{\mathbb{H}_K}, 0) = \mathbb P \big [f: \|f\|_{\mathbb{H}_K} \geq \|f_{obs}\|_{\mathbb{H}_K} \big]  = 1- F(\|f_{obs}\|_{{\mathbb{H}_K}}^2),
   \label{eq:pd2}
\end{equation}
where $F(x)$ denotes the cumulative distribution function of $\|f\|_{\mathbb{H}_K}^2$ for all $f \in \mathcal F$.  

As given in Algorithm II in Appendix G, the sample version of the squared modified norm is given as 
\begin{equation}
\|f_{obs}\|_{{\mathbb{H}_{\hat K}}}^2 = \sum_{p=1}^{n} \frac{\langle f_{obs}, \hat{\phi}_{p,n} \rangle^2}{\hat{\lambda}_{p,n}}.
\label{eq:snorm2}
\end{equation}
Similar to Case I, we adopt the sample version of the depth of $f_{obs}$
\begin{equation}
  d_n(f_{obs}) =  \mathbb P \big [f: \|f\|_{{\mathbb{H}_K}} \geq \|f_{obs}\|_{{\mathbb{H}_{\hat K}}} \big] = 1 - F(\|f_{obs}\|_{{\mathbb{H}_{\hat K}}}^2).
  \label{eq:sd2}
\end{equation}
We focus on proving $d_{n}(f_{obs})$ converges to $d(f_{obs})$ when $n$ is large.  This is shown in Theorem \ref{thm:finite} as follows.  In this case, neither a convergent weight series $\{a_p\}$ nor Assumption 1 is needed in the proof of consistency. 

\begin{theorem}
If the covariance kernel $K$ has only $P (\in \mathbb N)$ positive eigenvalues $\{\lambda_p\}$, then we have 
$$ \sup_{f_{obs} \in \mathcal{F}, ||f_{obs}|| \le 1} | \| f_{obs} \|_{\mathbb{H}_{\hat{K}}}^2  -  \| f_{obs} \|_{\mathbb{H}_K}^2| \overset{a.s.}{\longrightarrow} 0.$$ 
Moreover,  for any $f_{obs} \in \mathcal{F}$ 
$$\lim_{n \rightarrow \infty} d_{n}(f_{obs}) = d(f_{obs}).$$
\label{thm:finite}
\end{theorem}
\noindent {\bf Proof:}  As convergence almost surely implies convergence in distribution, it is apparent that we only need to prove the first convergence $\| f_{obs} \|_{\mathbb{H}_{\hat{K}}}^2 \overset{a.s.}{\longrightarrow} \| f_{obs} \|_{\mathbb{H}_K}^2$.

Based on the work done by \cite{dauxois1982asymptotic} and \cite{bosq2012linear}, when $n$ is large, we have $\hat{\lambda}_{p,n} \overset{a.s.}{\longrightarrow} \lambda_p > 0$ for $p \in \{ 1, \cdots, P\}$, while $\hat{\lambda}_{p,n} \overset{a.s.}{\longrightarrow} 0$ for $p \in \{ P+1, \cdots, n\}$\citep{tran2008introduction}.  

We denote $\breve{K}(s,t) = \sum_{p=1}^P \hat{\lambda}_{p,n} \hat{\phi}_{p,n}(s)\hat{\phi}_{p,n}(t)$, and we will get $\| \breve{K} - \hat{K} \| \overset{a.s.}{\longrightarrow} 0$ as $n \to \infty$. Besides, we have $\|\hat{K}-K \| \overset{a.s.}{\longrightarrow} 0 $\citep{dauxois1982asymptotic}, hence $\|\breve{K}-K \| \overset{a.s.}{\longrightarrow} 0$. 

If we denote $K^{-1}(s,t) = \sum_{p=1}^P \frac{\phi_p(s)\phi_p(t)}{\lambda_p} $ and $\breve{K}^{-1}(s,t) = \sum_{p=1}^P \frac{\hat{\phi}_{p,n}(s) \hat{\phi}_{p,n}(s)}{\hat{\lambda}_{p,n}}$, 
\begin{align*}
    \| \breve{K}^{-1} - K^{-1} \| &= \| \breve{K}^{-1} (\breve{K}-K) K^{-1}  \| \\
    &\leq \| \breve{K}^{-1}\| \|\breve{K}-K \| \|K^{-1} \| \\
    &= \frac{1}{\hat{\lambda}_{p,n}} \frac{1}{\lambda_p} \|\breve{K}-K \|\overset{a.s.}{\longrightarrow} 0.
\end{align*}

In Algorithm 2, the estimated depth is written as 
\begin{align*}
        \| f_{obs} \|_{\mathbb{H}_{\breve{K}}}^2 &= \sum_{p=1}^P\frac{\langle f_{obs},\hat{\phi}_{p,n} \rangle^2}{\hat{\lambda}_{p,n}}\\
        &= \sum_{p=1}^P\int_0^1\int_0^1 f_{obs}(s)f_{obs}(t) \frac{\hat{\phi}_{p,n}(s) \hat{\phi}_{p,n}(s)}{\hat{\lambda}_{p,n}} ds dt \\
        &= \int_0^1\int_0^1  f_{obs}(s)f_{obs}(t) \breve{K}^{-1}(s,t) ds dt
\end{align*}

Therefore, 
\begin{align*}
    | \ \| f_{obs} \|_{\mathbb{H}_{\breve{K}}}^2 -  \| f_{obs} \|_{\mathbb{H}_K}^2 \ | &= |\ \int_0^1\int_0^1  f_{obs}(s)f_{obs}(t) \breve{K}^{-1}(s,t) ds dt - \int_0^1\int_0^1  f_{obs}(s)f_{obs}(t) K^{-1}(s,t) ds dt   \ | \\
    &= |\ \langle f_{obs}, (\breve{K}^{-1} - K^{-1})f_{obs}    \rangle   \ | \\ 
    &\leq \|f_{obs} \|^2 \| \breve{K}^{-1} - K^{-1} \| \overset{a.s.}{\longrightarrow} 0.
\end{align*}
$\qed$

\section{Proof of Lemma \ref{lem:ip}} \label{sec:lemma:ip}
To better streamline the proof, we first prove the claimed result when the inner-product is induced from the reproducing kernel Hilbert space (RKHS) associated with the covariance function of the GP (which satisfies the Gram matrix condition of the lemma), and then extend the proof to a general inner product. Specifically, we show:
\begin{enumerate}
\item (basic form) If we take the induced RKHS (reproducing-kernel Hilbert space) inner-product $\left < \cdot, \cdot \right > $ using the covariance function $C$, then 
$$ D_{ip}(f_{obs}, \mathbb P_C, \left < \cdot, \cdot \right >, \mathcal F) = 0 $$
almost surely for $f_{obs} \in$ GP$(0, C)$
\item (general form) The above result will in fact hold for any inner-product on $\mathcal F$ that satisfies the condition in the lemma. 
\end{enumerate}

{\bf Proof:}  (Part 1)  Based on the result in Sec \ref{sec:inRKHS}, assume the covariance function $C(\cdot, \cdot)$ in a Gaussian process GP$(0, C)$ has infinite number of positive eigenvalues.  Then the covariance can be represented as 
$C(s,t) = \sum_{p=1}^\infty\lambda_p\phi_p(s)\phi_p(t).$
For any $f_{obs} \in GP(0, C)$, let $ f_{obs,p} = \int_0^1 f_{obs}(s)\phi_p(s)ds$.  We have $f_{obs}(t) = \sum_{p=1}^\infty f_{obs,p} \phi_p(t)$.  Hence, the induced RKHS norm 
$$\|f_obs\|_{\mathbb H_C} = \sum_{p=1}^\infty \frac{f_{obs,p}^2}{\lambda_p} = \infty \ \ \  (a.s.)$$
For any integer $P>0$ and function $f\in\mathcal F$, we let $f^P$ represent the finite cutoff of $f$ at the $P$-th order.  That is, $f^P(t) = \sum_{p=1}^P f_p \phi_p(t)$.  Let $ \mathcal G_P$ denote the finite-dimensional space expanded by $\{\phi_p(t)\}_{p=1}^P$.  Using the result in Appendix A, the inner-product depth 
\begin{equation}
D_{ip} (f^P_{obs},  \mathbb P_C, \left < \cdot, \cdot \right >, \mathcal G_P) = 1 - \Phi(\|f^P_{obs}\|_{\mathbb H_C}). 
\label{eq:Dip}
\end{equation}
Note that $\|f^P_{obs}\|_{\mathbb H_C} = \sum_{p=1}^P \frac{f_{obs,p}^2}{\lambda_p}  \rightarrow \infty  (a.s.)$ as $P \rightarrow \infty$.  Then $1 - \Phi(\|f^P_{obs}\|_{\mathbb H_C}) \rightarrow 1 - \Phi(\infty) = 1 - 1 = 0 \ \ (a.s.)$  Finally, we have 
$$D_{ip} (f_{obs},  \mathbb P_C, \left < \cdot, \cdot \right >, \mathcal F)  \le \inf_P D_{ip} (f^P_{obs},  \mathbb P_C, \left < \cdot, \cdot \right >, \mathcal G_P)   \rightarrow 0. \ \ \ (a.s.)$$

(Part 2)  We see that the proof of Part 1 mainly relies on the result in Appendix A (Equation \eqref{eq:Dip}), where we use the induced RKHS inner-product. Let $f$ be a realization from the Gaussian process GP$(0,C)$. Here we just need to show that using the new inner-product, such equation will still hold. Again, we consider the finite cut-off of $f^P_{obs}$ at the $P$-th order, and will show that Equation~\eqref{eq:Dip} remains valid with the new inner product $\langle \cdot,\cdot\rangle$. Therefore, we suppress the superscript $P$ in proving this equation in the rest of this part. Under this notation, we can write
\begin{align*}
f(t) &= \sum_{p=1}^P f_p \phi_p(t) \in \mathbb H_K, \\
g(t) &= \sum_{p=1}^P g_p \phi_p(t) \in \mathbb H_K, \\
f_{obs}(t) &= \sum_{p=1}^P f_{obs,p} \phi_p(t) \neq {\bf 0} \in \mathbb H_K,
\end{align*}
where $f_p$ are independent normal random variables with $\mathbb Ef_p=0$ and $Var f_p = \lambda_p, p = 1, \cdots, P$.

For this new inner-product $\left < \cdot, \cdot \right >$ on $\mathcal F$, let $r_{ij} = \left < \phi_i, \phi_j \right >$ for $1\le i, j \le P$.  We also denote
\begin{align*}
 X = \left < f - f_{obs}, g \right > = \sum_{i=1}^P \sum_{j=1}^P (f_i - f_{obs,i})g_j r_{ij}
=- \sum_{i=1}^P \sum_{j=1}^P f_{obs,i} g_j r_{ij} + \sum_{i=1}^P \sum_{j=1}^P f_i g_j r_{ij}.
\end{align*}
It is straightforward to know that  $X$ is normally distributed with $\mathbb EX =- \sum_{i=1}^P \sum_{j=1}^P f_{obs,i} g_j r_{ij}:\, = \mu_g$ and $Var X =  \sum_{i=1}^P (\sum_{j=1}^P g_j r_{ij})^2 \lambda_i :\, = \sigma_g^2$.  Now we can compute the probability 
\begin{align*}
    \mathbb P_{\theta} \big [ \left < f - f_{obs}, g \right > \ge 0\big] = \mathbb P_{\theta} \big [ X \ge 0\big] = \mathbb P_{\theta} \big [ \frac{X-\mu_g}{\sigma_g} \ge -\frac{\mu_g}{\sigma_g}\big] = 1- \mathbb P_{\theta} \big [ \frac{X-\mu_g}{\sigma_g} \le -\frac{\mu_g}{\sigma_g}\big].
\end{align*}
With the normal assumption, $\mathbb P_{\theta} \big [ \left < f - f_{obs}, g \right >_{mod} \ge 0\big] = 1 - \Phi(-\frac{\mu_g}{\sigma_g})$ where $\Phi$ is the c.d.f. of a standard normal random variable $\frac{X-\mu_g}{\sigma_g}$ (it does not depend on $g$). To minimize the probability with respect to $g$, we need to maximize $-\mu_g/\sigma_g$, or $\mu_g^2/\sigma^2_g$.    

By the Cauchy inequality, we have
\begin{align*}
\frac{\mu_g^2}{\sigma^2_g} &=   \frac{(\sum_{i=1}^P \sum_{j=1}^P f_{obs,i} g_j r_{ij})^2}{\sum_{i=1}^P (\sum_{j=1}^P g_j r_{ij})^2 \lambda_i }  = 
  \frac{(\sum_{i=1}^P \frac{f_{obs,i}}{\sqrt{\lambda_i}} \sqrt{\lambda_i} \sum_{j=1}^Pg_j r_{ij})^2}{\sum_{i=1}^P (\sum_{j=1}^P g_j r_{ij})^2 \lambda_i }  \\
  & \le   \frac{\sum_{i=1}^P (\frac{f_{obs,i}}{\sqrt{\lambda_i}})^2 \sum_{i=1}^P(\sqrt{\lambda_i} \sum_{j=1}^Pg_j r_{ij})^2}{\sum_{i=1}^P (\sum_{j=1}^P g_j r_{ij})^2 \lambda_i } 
= \sum_i \frac{f_{obs,i}^2}{\lambda_i}
\end{align*}
The equality holds if and only if there exists $c>0$ such that $c \frac{f_{obs,i}}{\sqrt{\lambda_i}} = \sqrt{\lambda_i} \sum_{j=1}^P g_j r_{ij}, i = 1, \cdots, P$.  That is, 
$$\sum_{j=1}^P g_j r_{ij} = c\frac{f_{obs,i}}{\lambda_i}, \ \ i=1,2,\ldots,P.$$
Under the condition on the inner-product in the lemma, this set of linear equations always admits a unique solution.
By plugging-in this solution, the maximum of $-\frac{\mu_{g}}{\sigma_{g}}$ is obtained at 
\begin{align*}    
    -\frac{- \sum_{i=1}^P f_{obs,i} \sum_{j=1}^P  g_j r_{ij}}{\sqrt{ \sum_{i=1}^P (\sum_{j=1}^P g_j r_{ij})^2 \lambda_i }} 
    = \frac{\sum_{i=1}^P f_{obs,i} c\frac{f_{obs,i}}{\lambda_i}}{\sqrt{ \sum_{i=1}^P (c\frac{f_{obs,i}}{\lambda_i})^2 \lambda_i }} 
    =\sqrt{\sum_{i=1}^P \frac{f_{obs,i}^2}{\lambda_i}} =  \|f_{obs}\|_{\mathbb H_C}.
\end{align*}
Finally, the depth of $f_{obs}$ is still given in the following form:
$$ D_{ip}(f_{obs},\, \mathbb P_\theta, \left < \cdot, \cdot \right > , \mathcal F)  = 1 - \Phi (\|f_{obs}\|_{\mathbb H_C}).$$

\section{Optimal solution of depth with inner-product-based criterion} \label{sec:opt_col}

We assume that $f \in \mathcal{F} (\subset \mathbb L^2([0,1]))$ is random realizations from one zero-mean Gaussian process with covariance kernel $K$ in a finite Karhunen Lo\`eve expansion $K(s,t) = \sum_{p=1}^P\lambda_p\phi_p(s)\phi_p(t), s, t \in [0, 1]$. As discussed in Sec. \ref{sec:inRKHS}, the realizations from the Gaussian process form an RKHS $\mathbb H_K$.  
Let 
\begin{align*}
f(t) &= \sum_{p=1}^P f_p \phi_p(t) \in \mathbb H_K, \\
g(t) &= \sum_{p=1}^P g_p \phi_p(t) \in \mathbb H_K, \\
f_{obs}(t) &= \sum_{p=1}^P f_{obs,p} \phi_p(t) \neq {\bf 0} \in \mathbb H_K,
\end{align*}
where $f_p$ are independent normal random variables with $\mathbb Ef_p=0$ and $Var f_p = \lambda_p, p = 1, \cdots, P$.

Using the inner-product, we denote
\begin{align*}
 X = \left < f - f_{obs}, g \right >_{\mathbb H_K} = \sum_{p=1}^P \frac{ (f_p - f_{obs,p})g_p}{\lambda_p } 
 = - \sum_{p=1}^P \frac{f_{obs,p}g_p}{\lambda_p} + \sum_{p=1}^P \frac{ f_p g_p}{\lambda_p }.
\end{align*}
It is straightforward to know that  $X$ is normally distributed with $\mathbb EX = - \sum_{p=1}^P \frac{f_{obs,p}g_p}{\lambda_p}  :\, = \mu_g$ and $Var X = \sum_{p=1}^P \frac{g_p^2}{\lambda_p}  :\, = \sigma_g^2$.  Now we can compute the probability 
\begin{align*}
    \mathbb P_{\theta} \big [ \left < f - f_{obs}, g \right >_{\mathbb H_K} \ge 0\big] = \mathbb P_{\theta} \big [ X \ge 0\big] = \mathbb P_{\theta} \big [ \frac{X-\mu_g}{\sigma_g} \ge -\frac{\mu_g}{\sigma_g}\big] = 1- \mathbb P_{\theta} \big [ \frac{X-\mu_g}{\sigma_g} \le -\frac{\mu_g}{\sigma_g}\big].
\end{align*}
With the normal assumption, $\mathbb P_{\theta} \big [ \left < f - f_{obs}, g \right >_{\mathbb H_K} \ge 0\big] = 1 - \Phi(-\frac{\mu_g}{\sigma_g})$ where $\Phi$ is the c.d.f. of a standard normal random variable $\frac{X-\mu_g}{\sigma_g}$ (it does not depend on $g$). To minimize the probability with respect to $g$, we need to maximize $-\mu_g/\sigma_g$, or $\mu_g^2/\sigma^2_g$.    

Let $$a_p = \frac{f_{obs,p}}{\lambda_p}, b_p = \frac{1}{\sqrt{\lambda_p}}.$$  Then use the Cauchy inequality,
\begin{align*}
\frac{\mu_g^2}{\sigma^2_g} &=  \frac{(\sum_p a_p g_p)^2}{\sum_p b_p^2 g_p^2} = \frac{(\sum_p \frac{a_p}{b_p} b_p g_p)^2}{\sum_p b_p^2 g_p^2}  \le \frac{\sum_p (\frac{a_p}{b_p})^2 \sum_p (b_p g_p)^2}{\sum_p b_p^2 g_p^2}  =  \sum_{p=1}^P (\frac{a_p}{b_p})^2. 
\end{align*}
The equality holds if and only if there exists $c>0$ such that $c\frac{a_p}{b_p} = b_p g_p, p = 1, \cdots, P$.  That is, 
$$g_p = c\frac{a_p}{b_p^2} = c  \cdot \frac{f_{obs,p}}{\lambda_p} \cdot \lambda_p =  c f_{obs,p}.$$
With the constraint $||g||_{\mathbb H_K} = 1$, 
$$ 1 =  \left < g, g \right >_{\mathbb H_K} = \sum_{p=1}^P \frac{g_p^2}{\lambda_k } = \sum_{p=1}^P \frac{c^2 f_{obs,p}^2}{\lambda_p} =  c^2 \sum_{p=1}^P \frac{f_{obs,p}^2 }{\lambda_p} = c^2\|f_{obs}\|_{\mathbb H_K}^2.   $$ 
Therefore, $c^2 = \frac{1}{\|f_{obs}\|_{\mathbb H_K}^2}$ and $g_p = \frac{f_{obs,p}}{\|f_{obs}\|_{\mathbb H_K}}$. 
We have found the optimal solution  
$$g^{\ast}(t) = \mbox{arginf}_{g \in \mathcal F, ||g||_{\mathbb H_K} = 1} \mathbb P_{\theta} \big [ \left < f - f_{obs}, g \right >_{\mathbb H_K} \ge 0\big] = \sum_{p=1}^P \frac{f_{obs,p}}{\|f_{obs}\|_{\mathbb H_K}}\phi_p(t) = \frac{f_{obs}(t)}{\|f_{obs}\|_{\mathbb H_K}}. $$ 
With this optimal $g^*$, 
\begin{align*}
    -\frac{\mu_{g^*}}{\sigma_{g^*}} = -\frac{- \sum_{p=1}^P \frac{f_{obs,p}g_p}{\lambda_p}}{\sqrt{\sum_{p=1}^P \frac{g_p^2}{\lambda_p}}} = \frac{\sum_{p=1}^P \frac{f_{obs,p}}{\lambda_p}\cdot\frac{f_{obs,p}}{\|f_{obs}\|_{\mathbb H_K}}}{\sqrt{\sum_{p=1}^P \frac{f_{obs,p}^2}{\|f_{obs}\|_{\mathbb H_K}^2}\cdot\frac{1}{\lambda_p}}} = \|f_{obs}\|_{\mathbb H_K}.
\end{align*}
Finally, the depth of $f_{obs}$ is given in the following form:
$$ D_{ip}(f_{obs}) := D_{ip}(f_{obs},\, \mathbb P_\theta, \left < \cdot, \cdot \right > , \mathcal F)  = 1 - \Phi (\|f_{obs}\|_{\mathbb H_K}).$$

\section{Proof of Lemma \ref{lem:prop}} \label{sec:lem_prop}

{\bf Proof:}  1) norm-based depth in general form:
\begin{itemize}
\item
P-2:  
By definition, the general depth $D_n(f_{obs},\, \mathbb P_\theta, \| \cdot \|, f_c)$ is strictly decreasing with respect to $\|f_{obs}- f_c\|$.  As $\|f_{obs}-f_c\| \ge \|f_{c}-f_c\| =0$, we have $D_n(f_{obs},\, \mathbb P_\theta, \| \cdot \|, f_c)  \ge D_n(f_{c},\, \mathbb P_\theta, \| \cdot \|, f_c). $
\item
P-3: 
For any $\alpha \in (0,1)$, $\|f_c+\alpha(f_{obs}-f_c) - f_c\| = \alpha \|f_{obs}-f_c\| \le \|f_{obs}-f_c\|$.  By Definition \ref{def:depthng},  $D_n(f_c+\alpha(f_{obs}-f_c),\, \mathbb P_{\theta}, s, f_c) \ge D_n(f_{obs},\, \mathbb P_\theta, \| \cdot \|, f_c) .$
\item P-4: Obvious. 
 \end{itemize}

 2) norm-based depth in specific form:
\begin{itemize}
\item
P-1: 
$D_n(af_{obs}+h,\, \mathbb P_{\theta, aF+h}, \| \cdot \|, af_c+h) = \mathbb P_{\theta} \big [f: \|af+h-(af_c+h)\| \geq \|af_{obs}+h- (af_c+h)\| \big] = \mathbb P_{\theta} \big [f: \|f-f_c\| \geq \|f_{obs}- f_c\| \big] = D_n(f_{obs},\, \mathbb P_{\theta, F}, \| \cdot \|, f_c) $
\item
P-2:  
$D_n(f_{obs},\, \mathbb P_\theta, \| \cdot \|, f_c) =\mathbb P_{\theta} \big [f: \|f-f_c\| \geq \|f_{obs}- f_c\| \big] 
\le \mathbb P_{\theta} \big [f: \|f-f_c\| \geq 0 \big]  = D_n(f_{c},\, \mathbb P_\theta, \| \cdot \|, f_c) 
$
\item
P-3: 
For any $\alpha \in (0,1)$, $D_n(f_c+\alpha(f_{obs}-f_c),\, \mathbb P_{\theta}, s, f_c) = \mathbb P_{\theta} \big [f: \|f-f_c\| \geq \|f_c+\alpha(f_{obs}-f_c) - f_c\| \big]  = \mathbb P_{\theta} \big [f: \|f-f_c\| \geq \alpha\|(f_{obs}-f_c)\| \big]  \ge \mathbb P_{\theta} \big [f: \|f-f_c\| \geq \|f_{obs}-f_c\| \big] = D_n(f_{obs},\, \mathbb P_\theta, \| \cdot \|, f_c) .$  
\item
P-4:
$D_n(f_{obs},\, \mathbb P_\theta, \| \cdot \|, f_c) =\mathbb P_{\theta} \big [f: \|f-f_c\| \geq \|f_{obs}- f_c\| \big] 
\le \mathbb P_{\theta} \big [f: \|f\| \geq \|f_{obs}\|- 2\|f_c\| \big] \to 0$ (as $\|f_{obs}\| \to \infty$). 
\end{itemize}

3) inner-product-based depth:
 \begin{itemize}
\item

P-1':
\begin{eqnarray*}
& & D_{ip}(af_{obs}+h,\, \mathbb P_{\theta, aF+h}, \left < \cdot, \cdot \right >, \mathcal{G}) \\
&=& \inf_{g \in \mathcal G, ||g|| = 1} \mathbb P_{\theta} \big [f \in \mathcal F: \left < af+h, g \right > \ge \left < af_{obs}+h, g \right > \big]  \\
&=&  \inf_{g \in \mathcal G, ||g|| = 1} \mathbb P_{\theta} \big [f \in \mathcal F: a\left < f, g \right > \ge a \left < f_{obs}, g \right > \big] \\
&=&  D_{ip}(f_{obs},\, \mathbb P_{\theta, F}, \left < \cdot, \cdot \right >, \mathcal{G}) 
\end{eqnarray*}


\item
P-2':
It is straightforward to prove this property followed by Assumption 1. For any $g\in \mathcal G$, it is easy to verify that the set $\big \{f \in \mathcal F: \left < f - f_c, g \right > \ge 0 \big\}$ is a closed halfspace that contains $f_c$.  By Assumption 1, $D_{ip}(f_{c},\, \mathbb P_{\theta, F}, \left < \cdot, \cdot \right >, \mathcal{G})
= \inf_{g \in \mathcal G, ||g|| = 1} \mathbb P_{\theta} \big [f \in \mathcal F: \left < f - f_c, g \right > \ge 0 \big] \ge 1/2$.  Assume $h (\neq f_c) \in \mathcal F$ satisfies that  $D_{ip}(h,\, \mathbb P_{\theta, F}, \left < \cdot, \cdot \right >, \mathcal{G}) > 1/2$.  Then for any $g \in \mathcal G, \mathbb P_{\theta} \big [f \in \mathcal F: \left < f - h, g \right > \ge 0 \big] > 1/2$. Hence, $\mathbb P_\theta$ is also H-symmetirc about $h$, contradicting to Assumption 1 that $f_c$ is unique. Therefore,  $D_{ip}(f_c,\, \mathbb P_{\theta}, \left < \cdot, \cdot \right >, \mathcal{G}) = \sup_{f_{obs} \in \mathcal{F}}D_{ip}(f_{obs},\, \mathbb P_{\theta}, \left < \cdot, \cdot \right >, \mathcal{G})$. 
\item
P-3':
For any $f_{obs} (\neq f_c) \in \mathcal F$, we need to prove that for any $\alpha \in (0,1)$, 
$$\inf_{g \in \mathcal G, ||g|| = 1} \mathbb P_{\theta} \big [f \in \mathcal F: \left < f - f_{obs}, g \right > \ge 0 \big] \le \inf_{g \in \mathcal G, ||g|| = 1} \mathbb P_{\theta} \big [f \in \mathcal F: \left < f - (f_c+\alpha(f_{obs}-f_c)), g \right > \ge 0 \big].$$ 
In fact, note that $f_c \in \big \{f \in \mathcal F: \left < f - f_{obs}, g \right > \ge 0 \big \} \Leftrightarrow f_c \in \big \{f \in \mathcal F: \left < f - (f_c+\alpha(f_{obs}-f_c)), g \right > \ge 0 \big \}$.
By Assumption 1, we only need to consider $g$ such that the halfspace does not contain $f_c$.  Therefore, 
\begin{eqnarray*}
& &\inf_{g \in \mathcal G, ||g|| = 1} \mathbb P_{\theta} \big [f \in \mathcal F: \left < f - f_{obs}, g \right > \ge 0 \big]  \\
&=& \inf_{g \in \mathcal G, ||g|| = 1, \left < f_c - f_{obs}, g \right > < 0} \mathbb P_{\theta} \big [f \in \mathcal F: \left < f - f_{obs}, g \right > \ge 0 \big]  \\
&\le& \inf_{g \in \mathcal G, ||g|| = 1, \left < f_c - f_{obs}, g \right > < 0} \mathbb P_{\theta} \big [f \in \mathcal F: \left < f - f_{obs}, g \right > \ge  (1-\alpha)\left < f_{c}-f_{obs}, g \right > \big] \\
&=& \inf_{g \in \mathcal G, ||g|| = 1, \left < f_c - f_{obs}, g \right > < 0} \mathbb P_{\theta} \big [f \in \mathcal F: \left < f - (f_c+\alpha(f_{obs}-f_c)), g \right > \ge 0 \big] \\
&\le& \inf_{g \in \mathcal G, ||g|| = 1} \mathbb P_{\theta} \big [f \in \mathcal F: \left < f - (f_c+\alpha(f_{obs}-f_c)), g \right > \ge 0 \big].
\end{eqnarray*}
\item
P-4':
\begin{eqnarray*}
& & D_{ip}(f_{obs},\, \mathbb P_{\theta, F}, \left < \cdot, \cdot \right >, \mathcal{G})
= \inf_{g \in \mathcal G, ||g|| = 1} \mathbb P_{\theta} \big [f \in \mathcal F: \left < f - f_c, g \right > \ge \left < f_{obs} - f_c, g \right > \big] \\
 &=&  \inf_{g \in \mathcal G, ||g|| = 1} \mathbb P_{\theta} \big [f \in \mathcal F: \left < f, g \right > \ge \left < f_{obs}, g \right > \big] 
 \le \mathbb P_{\theta} \big [f \in \mathcal F: \left < f, f_{obs} \right > \ge \left < f_{obs}, f_{obs} \right > \big] \\
 &\le&  \mathbb P_{\theta} \big [f \in \mathcal F:  \sqrt{\left < f, f \right > \left < f_{obs}, f_{obs} \right >} \ge \left < f_{obs}, f_{obs} \right > \big]  \mbox{ (Cauchy inequality) } \\
 &=&  \mathbb P_{\theta} \big [f \in \mathcal F: \|f\| \ge \|f_{obs}\| \big]  \to 0 \mbox { (as $\|f_{obs}\| \to \infty$)}
  \end{eqnarray*}
\end{itemize}

\section{Proof of Theorem \ref{thm:infinite}} \label{sec:thm_infinite}

{\bf Proof:} 
Throughout the proof, we use letter $C$ to denote some constant whose meaning may change from line to line.
According to Lemma 14 in \cite{tran2008introduction}, we have $\mathbb{E}\|\hat{K} - K \|_{\infty}^2 \leq C\,n^{-1}$. Therefore, by Markov's inequality
\begin{align*}
    \mathbb{P}(\|\hat{K} - K \|_\infty \geq \frac{\log n}{\sqrt{n}}) \leq (\frac{\sqrt{n}}{\log n})^2 \mathbb{E}(\|\hat{K} - K \|_\infty^2) \leq C(\log n)^{-2}\to 0.
\end{align*}
Let $\mathcal A$ denote the event $\{\|\hat{K} - K \|_\infty \leq \frac{\log n}{\sqrt{n}}\}$.   Then, $\mathbb P(\mathcal A) \to 1$ as $n\to\infty$.

Recall that in Algorithm I, we set $\hat{\lambda}_{p,n} = 0$ if $\hat{\lambda}_{p,n}$ is less than a threshold $\delta_n$ satisfying $\delta_n \to 0$ and $\delta_n \geq C(\frac{\sqrt{n}}{\log n })^{-\frac{\beta}{2\beta+1}}$. Then for a sufficiently large $n$, we have $\delta_n \geq 2\frac{\log n}{\sqrt{n}}$. Consequently, the $M_n = \operatorname*{arg\ max}_m \{ \lambda_m \geq \delta_n\}$ as defined in Equation \eqref{eq:snorm} satisfies $M_n \to \infty$ as $n\to\infty$. In addition, using Assumption 1, we have
\begin{align*}
    C_1 M_n^{-\beta} \geq \lambda_{M_n} \geq \delta_n \geq C(\frac{\sqrt{n}}{\log n })^{-\frac{\beta}{2\beta+1}},\quad 
    C_2 (M_n+1)^{-\beta} \leq \lambda_{M_n+1} <\delta_n,
\end{align*}
implying $C_1'\,\delta_n^{-1/\beta}\leq M_n \leq C_2'\,\delta_n^{-1/\beta}\leq  C(\frac{\sqrt{n}}{\log n })^{\frac{1}{2\beta+1}}$.
In addition, under event $\mathcal A$, we have, by Weyl's theorem and the definition of $M_n$, that
\begin{align*}
    \operatorname*{arg\ max}_{1\leq p\leq M_n} |\hat{\lambda}_{p,n} - \lambda_p | \leq \|\hat{K} - K \|_\infty \leq \frac{\log n}{\sqrt{n}} \leq \frac{\delta_n}{2} \leq \frac{\lambda_{M_n}}{2} \leq \frac{\lambda_p}{2},
\end{align*}
where  in the last step we have used the fact that $\lambda_p$ is a nonincreasing sequence.
Consequently, $\hat{\lambda}_{p,n} \geq \frac{\lambda_p}{2}$ holds for each $p = 1,\cdots, M_n$. 

By Proposition 16 in \cite{tran2008introduction} and our Assumption 1 on the eigenvalues, we obtain that for each $p=1,\ldots,M_n$,
\begin{align*}
    \|\hat{\phi}_{p,n}-\phi_p \| &\leq \frac{C}{\min\{\lambda_{p-1}-\lambda_{p},\,\lambda_p-\lambda_{p+1}\}}\|\hat{K} - K \|_\infty\\
    &\leq \frac{C}{p^{-(\beta+1)}}\|\hat{K} - K \|_\infty \leq C p^{\beta+1} \frac{\log n}{\sqrt{n}} \leq C M_n^{\beta+1}\frac{\log n}{\sqrt{n}}.
\end{align*}
By combining this with the bound on $M_n$, we obtain
\begin{align*}
    \|\hat{\phi}_{p,n}-\phi_p  \| \leq C\,\delta_n^{-\frac{\beta+1}{\beta}}\, \frac{\log n}{\sqrt{n}}\leq C\,\frac{n^{\frac{\beta+1}{4\beta+2}-\frac{1}{2}}}{(\log n)^{(\beta+1)/(2\beta+1)}} \log n= C n^{\frac{-\beta}{4\beta+2}}(\log n)^{\frac{\beta}{2\beta+1}} \to 0.
\end{align*}
By the Cauchy-Schwarz inequality, for any $p \in \{1,\cdots, M_n\}$,
\begin{align*}
    |\langle f_{obs}, \hat{\phi}_{p,n} \rangle^2 - \langle f_{obs}, \phi_p \rangle^2| &= |\langle f_{obs}, \hat{\phi}_{p,n}+\phi_p \rangle \langle f_{obs}, \hat{\phi}_{p,n}-\phi_p \rangle |\\
    &\leq \|f_{obs} \|^2(\|\hat{\phi}_{p,n}\|+\|\phi_p \|)\|\hat{\phi}_{p,n}-\phi_p \|\\
    &= 2\|f_{obs} \|^2\|\hat{\phi}_{p,n}-\phi_p \|.
\end{align*}
Combing the last two displays, we obtain
$$|\langle f_{obs}, \hat{\phi}_{p,n} \rangle^2 - \langle f_{obs}, \phi_p \rangle^2| \leq C \|f_{obs} \|^2\,p^{\beta+1} \frac{\log n}{\sqrt{n}} \leq C \|f_{obs} \|^2\,n^{\frac{-\beta}{4\beta+2}}(\log n)^{\frac{\beta}{2\beta+1}},$$
and for each $p=1,2,\ldots,M_n$,
\begin{align*}
    \Big|\frac{\langle f_{obs}, \hat{\phi}_{p,n} \rangle^2 - \langle f_{obs}, \phi_p \rangle^2}{\lambda_p}\Big|  &\leq  C \|f_{obs} \|^2\,p^{2\beta+1} \frac{\log n}{\sqrt{n}} \leq C \|f_{obs} \|^2\,M_n^{2\beta+1} \frac{\log n}{\sqrt{n}}\leq C\|f_{obs} \|^2.
\end{align*}

Now we are ready to prove Equation \eqref{eqn:conv1}.  By Assumption \ref{assump:ub}, it is easy to verify that the series $\sum_{p=1}^\infty \frac{\langle f_{obs},\phi_p\rangle ^2}{\lambda_p} a_p^2$ is uniformly convergent for any $\|f_{obs}\|_b \le 1$ (as for $N$ sufficiently large, $\sum_{p \ge N} \frac{\langle f_{obs},\phi_p\rangle ^2}{\lambda_p} a_p^2 \leq N^{-2\alpha}\,\sum_{p \ge N} \frac{\langle f_{obs},\phi_p\rangle ^2}{\lambda_p} b_p^2 \le N^{-2\alpha}\,\|f_{obs}\|_b)$.  Therefore, according to Assumption 2, for each $N\geq1$, we have $\sum_{p=N+1}^\infty \frac{\langle f_{obs},\phi_p\rangle^2}{\lambda_p} a_p^2 < N^{-2\alpha}$ and $\sum_{p=N+1}^\infty a_p^2 < N^{-2\alpha} \sum_p b_p^2 \leq C\,N^{-2\alpha}$.
According to the error bounds on $\hat{\lambda}_{p,n}$ and $\langle f_{obs}, \hat{\phi}_{p,n} \rangle$, we have that under event $\mathcal A_n$,
\begin{align*}
  | \frac{\langle f_{obs}, \hat{\phi}_{p,n} \rangle^2}{\hat{\lambda}_{p,n}} - \frac{\langle f_{obs}, {\phi}_{p} \rangle^2}{{\lambda}_{p}}| &<C \|f_{obs} \|^2\,\Big(\delta_n+N^{2\beta+1} \frac{\log n}{\sqrt{n}}\Big),\quad p=1,\ldots,N,\\
    \Big|\sum_{p=N+1}^{M_n \land n} \frac{\langle f_{obs}, \hat{\phi}_{p,n} \rangle^2 - \langle f_{obs}, \phi_p \rangle^2}{\lambda_p} a_p^2\Big| &\leq C\,\|f_{obs}\|^2 \sum_{p=N+1}^{M_n \land n} a_p^2 \leq C\,\|f_{obs}\|^2\,N^{-2\alpha}.
\end{align*}
Therefore, we obtain
\begin{align*}
    \sum_{p=N+1}^{M_n \land n} \frac{\langle f_{obs}, \hat{\phi}_{p,n} \rangle^2}{\hat{\lambda}_{p,n}} a_p^2 &\leq 2 \sum_{p=N+1}^{M_n \land n} \frac{\langle f_{obs}, \hat{\phi}_{p,n} \rangle^2}{{\lambda}_{p}}a_p^2\\
    &\leq 2\, \Big|\sum_{p=N+1}^{M_n \land n} \frac{\langle f_{obs}, \hat{\phi}_{p,n} \rangle^2 - \langle f_{obs}, \phi_p \rangle^2}{\lambda_p} a_p^2 \Big|+\sum_{p=N+1}^{M_n \land n} \frac{\langle f_{obs}, {\phi}_{p} \rangle^2}{{\lambda}_{p}}a_p^2 \leq C\,N^{-2\alpha},
\end{align*}
where the first inequality is due to $\hat{\lambda}_{p,n}\geq \lambda_p/2$ for all $p\leq M_n$.

Putting pieces together, we can conclude that 
\begin{align*}
    \big|\|f_{obs}\|_{\hat{mod}}^2 - \|f_{obs}\|_{{mod}}^2 \big| &= \Big|\sum_{p=1}^{M_n \land n} \frac{\langle f_{obs}, \hat{\phi}_{p,n} \rangle^2}{\hat{\lambda}_{p,n}} a_p^2 - \sum_{p=1}^\infty \frac{\langle f_{obs}, {\phi}_{p} \rangle^2}{{\lambda}_{p}}a_p^2\Big|\\
    &\leq \Big|\sum_{p=1}^N\Big(\frac{\langle f_{obs}, \hat{\phi}_{p,n} \rangle^2}{\hat{\lambda}_{p,n}} - \frac{\langle f_{obs}, {\phi}_{p} \rangle^2}{{\lambda}_{p}}\Big)\, a_p^2\Big|+ \sum_{p=N+1}^\infty \frac{\langle f_{obs}, {\phi}_{p} \rangle^2}{{\lambda}_{p}}a_p^2 + \sum_{p=N+1}^{M_n \land n} \frac{\langle f_{obs}, \hat{\phi}_{p,n} \rangle^2}{\hat{\lambda}_{p,n}} a_p^2\\
    &< C\Big(N^{-2\alpha} + \delta_n+N^{2\beta+1} \frac{\log n}{\sqrt{n}}\Big).
\end{align*}
By choosing $N=\Big(\frac{\log n}{\sqrt{n}}\Big)^{-(2\alpha+2\beta+1)}$, we have $\big|\|f_{obs}\|_{\hat{mod}}^2 - \|f_{obs}\|_{{mod}}^2 \big| \leq n^{-\kappa}$ for $\kappa = 2\alpha/(2\alpha+2\beta+1)>0$ under event $\mathcal A$.
$\qed$

\section{Proof of Theorem~\ref{thm:sa} and Theorem~\ref{thm:mc}}\label{sec:thm_samc}
{\bf Proof of Theorem~\ref{thm:sa}:}
According to the proof of theorem~\ref{thm:infinite}, there exists some event $\mathcal A$ whose probability tending to one as $n\to\infty$, such that under this event
\begin{align*}
\big|\|g\|_{\hat {mod}} - \|g\|_{mod} \big| \leq C\,n^{-\kappa}\,\|g\|_b
\end{align*}
for all $g$ such that $\|g\|_b\leq \infty$ (note that $\|g\|_{mod}$ is always dominated by $\|g\|_b$ according to Assumption 2). Under this condition, we have the following inclusion relationships
\begin{align*}
\big\{\|g_p\|_{mod} \geq \|f_{obs}\|_{mod}-Cn^{-\kappa}\,(\|g_p\|_b+\|f_{obs}\|_b)\big\} &\subset \big\{\|g_p\|_{\hat {mod}} \geq \|f_{obs}\|_{\hat mod}\big\} \\
&\subset \big\{\|g_p\|_{mod} \geq \|f_{obs}\|_{mod}+Cn^{-\kappa}\,(\|g_p\|_b+\|f_{obs}\|_b)\big\}.
\end{align*}
According to Assumption 3 and a standard tail probability bound for the max of sub-Gaussian random variables, we have $\mathbb P(\max_{p=1,\ldots,n}\|g_p\|_b \leq C\sigma\sqrt{\log n}) \geq 1-n^{-1}$ for some constant $C>0$. Let $\mathcal B$ to denote this event. Then, under event $\mathcal A\cap \mathcal B$, we have
\begin{align*}
U_n=n^{-1}\sum_{p=1}^n 1_{\|g_p\|_{mod} \geq \|f_{obs}\|_{mod}+\varepsilon_n}\leq
\frac{1}{n} \sum_{p=1}^n 1_{\|g_p\|_{\hat {mod}} \geq \|f_{obs}\|_{\hat mod}}\leq V_n= \frac{1}{n} \sum_{p=1}^n 1_{\|g_p\|_{mod} \geq \|f_{obs}\|_{mod}-\varepsilon_n},
\end{align*}
where $\varepsilon_n=Cn^{-\kappa}\sqrt{\log n}$. By Markov inequality, we have
\begin{align*}
\mathbb P\Big(\Big|U_n - \big(1-F((\|f_{obs}\|_{ {mod}}+\varepsilon_n)^2)\Big)| \leq \frac{\log n}{\sqrt{n}}\Big)  \geq 1-\frac{C}{\log^2 n},\\
\mathbb P\Big(\Big|V_n - \big(1-F((\|f_{obs}\|_{ {mod}}-\varepsilon_n)^2)\Big)| \leq \frac{\log n}{\sqrt{n}}\Big)  \geq 1-\frac{C}{\log^2 n}.
\end{align*}
Let $\mathcal C$ denote the intersection of the two events inside above probabilities, and $\mathcal E=\mathcal A\cap \mathcal B\cap\mathcal C$. Then $\mathbb P(\mathcal E)\to 1$ as $n\to\infty$, and under this event $\mathcal E_n$, we have
\begin{align*}
1-F((\|f_{obs}\|_{ {mod}}+\varepsilon_n)^2)\leq \frac{1}{n} \sum_{p=1}^n 1_{\|g_p\|_{\hat {mod}} \geq \|f_{obs}\|_{\hat mod}}\leq 1-F((\|f_{obs}\|_{ {mod}}-\varepsilon_n)^2).
\end{align*}
This implies the claimed result by using the fact that $F$ is a continuous function and $\varepsilon_n\to 0$ as $n\to \infty$.

\bigskip

\noindent
{\bf Proof of Theorem~\ref{thm:mc}:}
By the Markov inequality, given the data $D$, the conditional probability
\begin{align*}
\mathbb P\Big( \Big|\frac{1}{N} \sum_{p=1}^N 1_{\|g_p\|_{\hat {mod}} \geq \|f_{obs}\|_{\hat {mod}}} -\big(1- F_n(\|f_{obs}\|_{\hat {mod}})\big)\Big| \leq \frac{\log N}{\sqrt{N}}\,\Big|\, D\Big)\geq 1-C/(\log N)^2,
\end{align*}
where the randomness in $\mathbb P$ is due to the Monte Carlo sampling, and for any $t>0$, 
\begin{align*}
F_n(t) = \mathbb P\Big(\sum_{p=1}^{M_n} a_p^2Z_p^2\leq t^2\,\Big|\, D\Big),
\end{align*}
only dependent on $M_n$ (defined in Equation \eqref{eq:snorm}), is the probability that a weighted sum of squares of the first $M_n$ standard normal random variables $\{Z_p\}_{p=1}^{\infty}$ are less than or equal to $t$. By taking expectation with respect to $D$ on both side, we can further obtain
\begin{align*}
\mathbb P\Big( \Big|\frac{1}{N} \sum_{p=1}^N 1_{\|g_p\|_{\hat {mod}} \geq \|f_{obs}\|_{\hat {mod}}} -\big(1- F_n(\|f_{obs}\|_{\hat {mod}})\big)\Big| \leq \frac{\log N}{\sqrt{N}}\Big)\geq 1-C/(\log N)^2,
\end{align*}
where now the randomness in $\mathbb P$ is due to both the randomness in data $D$ and the randomness in the Monte Carlo sampling.
 In addition, function $F$ in the desired limit is 
\begin{align*}
F(t) = \mathbb P\Big(\sum_{p=1}^{\infty} a_p^2Z_p^2\leq t^2\Big).
\end{align*}
According to Theorem~\ref{thm:infinite}, we have $\big|\|f_{obs}\|_{\hat {mod}} -\|f_{obs}\|_{{mod}}\big| \leq C\, n^{-\kappa}$ with probability tending to one as $n\to\infty$. Therefore, due to the continuity of $F$ in $t$, it remains to show that for each $t\in \mathbb R$,
\begin{align*}
F_n(t) \to F(t)\quad  \mbox{in probability as }n\to\infty.
\end{align*}
In fact, according to Assumption 2 and the fact that $M_n\to \infty$ as $n\to \infty$, we have
\begin{align*}
\mathbb E\Big[\sum_{p=M_n+1}^{\infty} a_p^2Z_p^2\Big] = \sum_{p=M_n+1}^{\infty} a_p^2 \leq M_n^{-2\alpha}\sum_{p=M_n+1}^\infty b_p \to 0
\end{align*}
as $n\to\infty$. This implies the convergence in probability of $\sum_{p=1}^{M_n} a_p^2Z_p^2$ to $\sum_{p=1}^{\infty} a_p^2Z_p^2$ as $n\to\infty$. Then the desired convergence of $F_n$ to $F$ is a consequence of the fact that convergences in probability imply convergences in distribution.


\newpage
\bibliographystyle{agsm}
\bibliography{references}

\end{document}